\documentclass[a4paper,fleqn]{cas-sc}
\usepackage[
type={CC},
modifier={by-nc-nd},
version={4.0},
]{doclicense}
\usepackage{float}
\usepackage[numbers]{natbib}
\usepackage{lineno}
\modulolinenumbers[5]
\usepackage{rotating}
\usepackage{graphicx}
\usepackage{subcaption}
\usepackage{amsmath,amsfonts}
\usepackage{comment}
\usepackage[dvipsnames]{xcolor}
\usepackage{array}
\usepackage{stfloats}
\usepackage{url}
\usepackage{epstopdf}
\usepackage{lscape}
\usepackage{booktabs}
\usepackage{adjustbox}
\usepackage{pifont}
\usepackage[para]{threeparttable}
\usepackage{tablefootnote}
\usepackage{longtable}
\usepackage{wrapfig}
\def\tsc#1{\csdef{#1}{\textsc{\lowercase{#1}}\xspace}}
\tsc{WGM}
\tsc{QE}
\tsc{EP}
\tsc{PMS}
\tsc{BEC}
\tsc{DE}
\begin{document}
	\let\WriteBookmarks\relax
	\def\floatpagepagefraction{1}
	\def\textpagefraction{.001}
	
	\shorttitle{Routing Attack and Defense in RPL: A Survey}
	
	
	
	
	\title [mode = title]{A Comprehensive Survey on RPL Routing-based Attacks, Defences and Future Directions in Internet of Things}
	
	\tnotetext[1]{https://doi.org/10.1016/j.compeleceng.2025.110071}
	
	
	
	
	
	

	%
	\author[1,5]{Anil Kumar Prajapati}
	\ead{2018rcp9156@mnit.ac.in, anilkumar.prajapati@jaipur.manipal.edu}
	
	\credit{Data curation, Writing - Original draft preparation}
	
	\affiliation[1]{organization={Malaviya National Institute of Technology Jaipur},
		addressline={JLN Marg, Malviya Nagar}, 
		city={Jaipur},
		country={India}}
	
	\author[1]{Emmanuel~S.~Pilli}
	\ead{espilli.cse@mnit.ac.in (**Corresponding Author)}
	\credit{Conceptualization of this study, Methodology, Supervision, Resources}
	\cormark[2]
	\author[1]{Ramesh~Babu~Battula}
	\ead{rbbattula.cse@mnit.ac.in}
	
	\credit{Conceptualization of this study, Methodology, Supervision}
	
	\author[2]{Vijay Varadharajan}
	\ead{vijay.varadharajan@newcastle.edu.au}
	\credit{Writing - Review \& Editing}
	\affiliation[2]{organization={School of Information and Physical Sciences, The University of Newcastle },
		city={Callaghan NSW},
		country={Australia}}
	
	\author[3]{Abhishek~Verma}
	\ead{abhiverma866@gmail.com}
	\credit{Writing - Review \& Editing, Formal Analysis}
	\affiliation[3]{organization={Department of Information Technology, Babasaheb Bhimrao Ambedkar University},
		city={Lucknow},
		country={India}}
	
	\author[4]{R. C.~Joshi}
	\ead{rcjoshi.geu@gmail.com}
	\credit{Supervision}
	\affiliation[4]{organization={Department of Electronics \& Computer Engineering, Indian Institute of Technology Roorkee},
		city={Roorkee},
		country={India}}
	
	\affiliation[5]{organization={Department of Data Science and Engineering, Manipal University},
		city={Jaipur},
		country={India}}
	
	
	\begin{abstract}
		The Internet of Things (IoT) is a network of digital devices like sensors, processors, embedded and communication devices that can connect to and exchange data with other devices and systems over the internet. IoT devices have limitations on power, memory, and computational resources. Researchers have developed the IPv6 Over Low-power Wireless Personal Area Network (6LoWPAN) protocols to provide wireless connectivity among these devices while overcoming the constraints on resources. 6LoWPAN has been approved subsequently by the Internet Engineering Task Force (IETF). The IETF Routing Over Low-power and Lossy Networks (ROLL) standardized the Routing Protocol for LLNs known as RPL (IETF RFC 6550), which is part of the 6LoWPAN stack. However, IoT devices are vulnerable to various attacks on RPL-based routing. This survey provides an in depth study of existing RPL-based attacks and defense published from year 2011 to 2024 from highly reputed journals and conferences. By thematic analysis of existing routing attacks on RPL, we developed a novel attack taxonomy which focuses on the nature of routing attacks and classifies them into 12 major categories. Subsequently, the impact of each attack on the network is analyzed and discussed real life scenarios of these attacks. Another contribution of this survey proposed a novel taxonomy for classification of defense mechanisms into 8 major categories against routing attacks based on type of defense strategy. The detailed analysis of each defense mechanism with real life applicability is explained. Furthermore, evaluation tools  such as testbeds and simulators for RPL-based attack and defense are discussed and critically analyzed in terms of real world applicability. Finally, open research challenges are presented on the basis of research gaps of existing literature along with research directions for practitioners and researchers. We believe our study will give actionable insights and solid foundation for researchers to expand effective defense solutions against emerging RPL routing attacks in IoT networks.
		
	\end{abstract}
	
	\doclicenseThis
	
	\maketitle
	
	\section{Introduction} \label{sec_introduction}
\begin{figure}[h]
		\centering  
		\includegraphics[width=.25\linewidth]{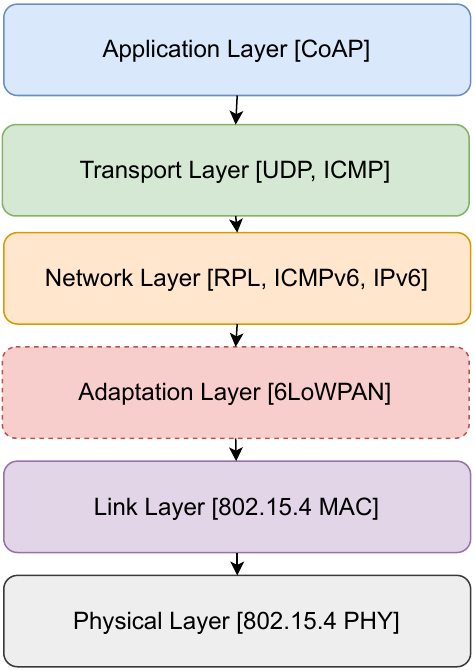}
		\caption{6LoWPAN Protocol Stack}
		\label{iotpstack}
	\end{figure}

	The Internet of Things (IoT) is critical for connecting and exchanging information with other things (physical objects, sensors, and computing devices) and facilitating communication or data transfer over wireless links without human intervention. IoT can be conceptualized as a system in which sensors collect data, gateways transmit it, and back-end systems make intelligent decisions. The Internet of Things has numerous applications in agriculture, healthcare, industry, markets, vehicles, transportation, and smart homes \cite{al2015internet}. The various architectures have been proposed for IoT based on user perspective \cite{itu2012overview}. The popular architecture are 3-layered \cite{krvco2014designing}, middle-ware \cite{chaqfeh2012challenges} and service oriented \cite{wu2010research}. The 3-layer based architecture provide abstraction on IoT devices and has perception, network and application layer. The perception layer play the role of data collection and processing from IoT devices. The network layer provides the secure connection and transmission of collected data from perception layer to application layer. The application layer consist the user interfaces which provides the services to IoT users \cite{alotaibi2023securing}. \newline
	\indent The main constraints for IoT networks are Low-power and lossy links, resulting in low throughput and high packet drop rate. Due to these resource constraints, the ROLL working group created the Routing Protocol for Low-power and Lossy Networks (RPL). Though RPL has security mechanisms, it is still vulnerable to emerging attacks. Researchers have developed a significant number of security solutions during the development of the RPL protocol. Hence there is still scope for security researchers to design a coherent defense mechanism for up-and-coming attacks\cite{granjal2015security}. 
	The rapid expansion of IoT devices has made it imperative to assign unique identifiers to each device; this issue has been addressed with the introduction of the IPv6 protocol. Another primary concern for IoT devices is ensuring efficient and secure routing from external attacks \cite{neshenko2019demystifying}\cite{butun2019security}. The 6LoWPAN \cite{kushalnagar2007ipv6}, an IPv6-enabled low-power wireless personal area network, provides the protocol stack for Low-power IoT devices and addresses both unique addressing and secure routing issues. The 6LoWPAN protocol stack was developed by the IETF and is outlined in RFC 4944 \cite{montenegro2007transmission}. 
	The 6LoWPAN protocol stack is depicted graphically in Fig. \ref{iotpstack}. The 6LoWPAN physical and link layers are designed for Low-power personal area networks and are compliant with the IEEE standard 802.15.4. Layer 2.5, the adaption layer, primarily supports the IPv6 requirements for the Maximum Transmission Unit (MTU) through header compression, fragmentation, and reassembly. The network layer provides routing decisions and includes IPv6, ICMPv6, and RPL \cite{winter2012rpl}. The transport layer employs the UDP and ICMP protocols, while the application layer uses CoAP.  
	
\begin{figure}[h]
		\centering
		\includegraphics[width=.5\linewidth]{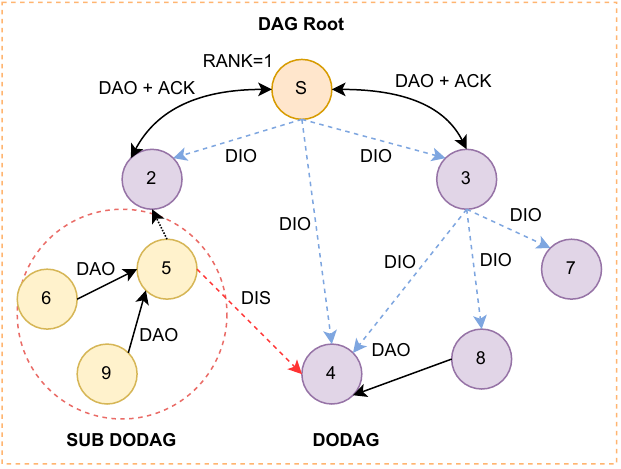}
		\caption{Illustration of RPL DODAG}
		\label{fig:rpldodag}
	\end{figure}

	\textcolor{black}{RPL, which stands for Routing Protocol for Low-Power and Lossy Networks, plays a pivotal role in IoT networks by offering specialized optimization features tailored to the needs of low-power devices, lossy links, and scalable connectivity. These key benefits significantly contribute to the overall efficiency and reliability of IoT systems: Firstly, RPL excels in efficient resource usage by addressing the unique requirements of low-power devices while ensuring seamless and dependable communication over lossy links. This capability not only enhances the overall performance of IoT networks but also extends the lifespan of resource-constrained devices. Additionally, the scalability of RPL is noteworthy, as it enables the protocol to effectively manage large-scale IoT deployments while remaining adaptable to dynamic network topologies. This adaptability is crucial for ensuring uninterrupted connectivity and smooth operation across diverse applications and industries such as smart homes, industrial IoT, smart cities, agriculture, and healthcare. Moreover, RPL offers application-specific optimization strategies, allowing for customizable routing decisions based on metrics like energy consumption, latency, and link quality. This fine-grained control facilitates efficient data transmission and supports a wide range of applications, including multicast scenarios for tasks such as firmware updates and group communication. Furthermore, RPL's interoperability with IPv6 and its status as an IETF standard ensure seamless integration and compatibility with various devices and systems in the IoT ecosystem. This standardization contributes to a cohesive and interconnected network fabric that promotes efficient data exchange and communication.
		Lastly, the inclusion of robust security features in RPL, such as encryption and secure key management protocols \cite{gong2023slim,tu2023eake,badshah2024usaf} safeguards sensitive data and applications within IoT environments. This robust security framework enhances the overall trustworthiness of IoT deployments and protects against potential threats and vulnerabilities. In essence, RPL emerges as a cornerstone technology in the realm of IoT communication, offering a comprehensive suite of features that optimize performance, reliability, scalability, and security across diverse IoT applications.}\\
	\indent The primary principle of RPL is to support the Objective Function of specific applications specified in terms of minimizing energy, latency, or constraints. RPL is a distance vector protocol for routing defined by the Direction-Oriented Directed Acyclic Graph (DODAG), which divides the network into sub-directed acyclic graphs, each of which contains constrained devices; all the traffic is directed towards the root node known as DAG root, which is represented by an IPv6 Border Router or Gateway. RPL instances are DODAGs having multiple DODAGs. The Objective Function (OF) helps nodes in DODAG in selecting and optimizing routes based on various routing metric constraints such as Expected Transmission Count (ETX), link quality, latency, and color. Objective Function Zero (OF0) \cite{thubert2012objective} and Minimum Rank with Hysteresis Objective Function (MRHOF) \cite{gnawali2012minimum} are the two standard objective functions defined for RPL. The OF has a Rank-based mechanism to identify the best parent, the position of nodes relative to the root nodes, or the nodes' distance from the root in DODAG. In DODAG, the lowest rank node best suits the parent. RPL-specific control messages keep the DODAG up to date. RPL control messages are represented by the ICMPv6 protocol. DODAG Information Solicitation (DIS), DODAG Information Object (DIO), Destination Advertisement Object (DAO), and Destination Advertisement Object Acknowledgment (DAO-ACK) are the RPL control messages that manage DODAG construction and maintenance regularly by exchanging messages between DODAG nodes as shown in Fig. \ref{fig:rpldodag}.
	
	\indent The Trickle Timer defined in RFC 6206 \cite{levis2011trickle} is a dedicated algorithm described in RPL for DODAG maintenance primarily responsible for synchronizing DIO messages. RPL aids Point-to-Point (P2P), Multipoint-to-Point (MP2P), and Multipoint-to-Multipoint (MP2MP) traffic and functions in two modes: storing (each node maintains the routing table) and non-storing (no routing information is stored). 
	There are three modes of security supported by RPL. The first one is preinstalled mode in which RPL control message are secured with symmetric keys. The second one is authenticated mode in which node authentication is required at the time of joining and this operation is performed using preinstalled keys. The third one is unsecured mode which resembles RPL with no security enabled. Many RPL features, such as self-configuration, neighbor discovery, DODAG construction and maintenance mechanisms, route construction processes, etc., are defined in RFC 6550 \cite{winter2012rpl}.
	The principal contributions of this survey are:
	\begin{enumerate}
		\item A novel taxonomy for RPL-based routing attacks not addressed in prior research and analyze their impact on Resource Consumption, Routing Decisions, and Performance Metrics. 
		\item A novel classification for RPL defense mechanisms based on different approaches, i.e., RPL-specification based, Machine Intelligence based, Trust and Threshold etc.
		\item Extensively discussion on the various existing RPL-based simulators, testbeds, and datasets. 
	\end{enumerate}

	The paper is organized as follows: Section \ref{sec_relatedwork} accounts for various existing and related surveys and highlights the originality and significance of our survey. Section \ref{sec_classficationofroutingattack} discusses RPL-based attacks, classifies them, and analyses their impact. Section \ref{sec_defensclassfication} provides a detailed discussion of the defense mechanisms for RPL-based routing attacks. The evaluation tools and testbeds are discussed in Section \ref{sec_evalution_dataset}. We outline the existing and open research challenges for further investigation in Section \ref{sec_research_challenges}. Section \ref{sec_conclusion} concludes the paper by summarizing the usefulness of the survey. The list of abbreviations of key terms used in survey is shown in Table \ref{tab:labbr}.
	
	\begin{table}[!h]
		\centering
		\color{black}
		\caption{\textcolor{black}{List of Abbreviations}}
		\label{tab:labbr}
		\begin{tabular}{ll|ll}
			\toprule
			\textbf{Abbreviation} 	&	\textbf{Stands For}	&	\textbf{Abbreviation} 	&	\textbf{Stands For}	\\
			\midrule
			IoT	&	Internet of Things	&	 6LoWPAN 	&	IPv6 Over Low-power Wireless Personal Area Network \\ IETF &	Internet Engineering Task Force	&	ROLL &	Routing Over Low-power and Lossy Networks	\\
			LLN	&	Low Power and Lossy Network	& RPL	& Routing Protocol for Low Power and Lossy Network \\
			IPv6 & Internet Protocol version 6	&	 ICMPv6	&	Internet Control Message Protocol version 6	\\
			UDP	&	User Datagram Protocol	&	CIA	&	Confidentiality, Integrity, and Availability	\\
			CoAP & Constrained Application Protocol	&	 DODAG 	&	 Destination Oriented Directed Acyclic Graph\\
			OF & Objective Function & ETX & Expected Transmission Count \\
			OF0		&	Objective Function
			Zero &	MRHOF	&	 Minimum Rank with Hysteresis Objective Function 	\\
			DIS  & DODAG Information Solicitation & DIO & DODAG Information Object \\ DAO & Destination Advertisement
			Object  & DAO-ACK & Destination Advertisement Object Acknowledgment \\
			P2P &  Point-to-Point &  MP2P & Multipoint-to-
			Point \\
			IDS	&	Intrusion Detection System	&	DTSN	&	 DAO Trigger Sequence Number \\
			PDR & Packet Delivery Ratio &
			E2ED & End-to-End Delay \\
			ML & Machine Learning & DL & Deep Learning \\
			CLI & Command Line Interface & GUI & Graphical User Interface \\
			\bottomrule
		\end{tabular}
	\end{table}

	\section{Related Work} \label{sec_relatedwork}
	\subsection{Existing Surveys} \label{subsec_existing_survey}
	Researchers have published several surveys on intrusion detection systems and mitigation methods for RPL based routing attacks in IoT networks, predominantly starting from 2010 and is still in its early stages. Wallgren \textit{et al.} \cite{wallgren2013routing}  were the first to identify vulnerabilities in RPL-based routing attacks and proposed a lightweight defense mechanism known as the heartbeat protocol against selective forwarding attacks. Pongle \textit{et al.} \cite{pongle2015survey} described an attack survey on the 6LowPAN and RPL and intrusion detection mechanisms. They also discussed research trends for RPL and 6LoWPAN intrusion detection techniques. Airehrour \textit{et al.}. \cite{airehrour2016secure} published a survey covering the routing protocol used for IoT and classifying possible attacks based on CIA (Confidentiality, Integrity, and Availability). They investigated some open issues and challenges for secure routing and also discussed trust models proposed by other researchers. Mayzaud \textit{et al.} \cite{mayzaud2016taxonomy} were the first to propose a taxonomy for RPL attack classification based on three aspects: topology, resource, and traffic. They categorized the attacks as internal or external, active or passive, and described attack prerequisites, impact assessment, proposed mitigation techniques, and assessment of their overheads. Kim \textit{et al.} \cite{kim2017challenging} investigated RPL in terms of experimentation and simulation using ContikiOS or TinyOS and testbeds and proposed future implementations of RPL optional functionalities. Zarpelao \textit{et al.} \cite{zarpelao2017survey} proposed a classification of intrusion detection systems for IoT in terms of detection method, placement, validation strategy, and security threats. Mangelkar \textit{et al.} \cite{mangelkar2017comparative} provided a comparative study of attacks and classified them based on CIA in relation to attack type and name. They also discussed the countermeasures proposed by various researchers. Kamgueu \textit{et al.} \cite{kamgueu2018survey}  proposed a survey of RPL topology optimization, security, and mobility. However, their work did not address defense mechanisms for routing attacks. Raoof \textit{et al.} \cite{raoof2018routing} proposed an extensive survey of routing attacks in RPL and their mitigation. This work categorized routing attacks based on their origin into RPL-specific and WSN-inherited attacks. They are the first to classify attack mitigation techniques based on extensive research. In terms of mitigation, they included a summary and insights for each attack in their survey. Their survey also outlined some research challenges and future directions. Verma \textit{et al.} \cite{verma2020security} published a survey that described an improved taxonomy for RPL-based routing attacks and a classification of defense mechanisms in terms of secure protocols and IDS. They also identified potential areas to investigate for future RPL security research. Almusaylim \textit{et al.} \cite{almusaylim2020proposing} proposed a review of defense mechanisms based on version and rank attacks. Simoglou \textit{et al.} \cite{simoglou2021intrusion} studied the design requirements, best practices, research gaps, and guidelines for designing an intrusion detection system for RPL security from 2003 to 2020. Pasikhani \textit{et al.} \cite{pasikhani2021intrusion} provided an assessment of the impact of each attack, statistical analysis for the defense mechanism, and a simulator for RPL along with investigating existing IDS. Seyfollahi \textit{et al.} \cite{seyfollahi2021review} investigated the machine learning algorithms used to design an IDS for securing RPL. Bang \textit{et al.} \cite{bang2022assessment} proposed a novel classification and mitigation mechanism for routing attacks based on RPL control messages. 
	
	\begin{table}[!h]
		\centering
		\scriptsize
		\color{black}
		\caption{Comparison of Existing Surveys on RPL Routing attacks and Defense Mechanisms}
		\label{tbl_survey_comr}
		\begin{tabular}{p{0.8cm}p{1.2cm}p{3cm}p{5cm}p{4cm}}
			\toprule
			\textbf{Ref}	&	\textbf{Research Span}	&	\textbf{Scope}	&	\textbf{Technical Coverage}	&	\textbf{Limitations} \\
			\midrule
			\cite{wallgren2013routing}	&	2010-2013	&	RPL-based routing attacks and its counter measures.	& RPL-based routing attacks implementation on ContikiOS and highlights the placement of IDS for elimination of malicious node.	&	No impact analysis and classification of attacks. 	\\
			\midrule
			\cite{pongle2015survey}	&	2011-2014	&	Routing attacks on RPL and 6LowPAN	&	New routing attacks on RPL protocol with IDS techniques and research trends.	& No discussion about classification of Attacks, Defense.\\
			\midrule
			\cite{airehrour2016secure}	&	2011-2015	&	Secure communication for IoT networks and routing protocols.	&	Classification of Routing attacks based on confidentiality, availability, integrity and open research challenges with trust models.	&	No classification of Defense Mechanism.	\\ \midrule
			\cite{mayzaud2016taxonomy}	&	2011-2016	&	Classification of RPL-based Routing attacks in IoT networks.	&	Proposed a taxonomy for classification of RPL-based attacks on the basis of topology, resources and traffic. 	&	No discussion about the classification of mitigation and future research direction. \\ \midrule
			\cite{kim2017challenging}	&	2012-2017	&	Protocol for IPV6 based routing on Low Power and Lossy Networks	&	Discussion about the investigation and simulation of RPL using ContikiOS and TinyOS.	&	No classification for routing attacks and No detailed discussion about simulators. 	\\ \midrule
			\cite{mangelkar2017comparative}	&	2012-2016	&	Study for RPL-based routing attacks and existing defense solutions.	&	Comparative analysis of  Attack Type and Names  based on CIA principals were used for classification of routing attacks  and possible counter measures.	&	Not performed any classification of countermeasure and No discussion about RPL attacks.	\\ \midrule
			\cite{kamgueu2018survey}	&	2012-2017	&	Focus on RPL improvement on basis of mobility, security and optimization.	&	Survey address the RPL protocols enhancement considering security, mobility and optimization 	&	Not discussed attacks/defense classification, impact analysis and research gaps. \\ \midrule
			\cite{raoof2018routing}	&	2011-2018	&	Routing Attacks and its Mitigation in RPL-based IoT Networks.	&	Proposed a classification of routing attacks on RPL specifics and WSN inherited, Issues and Research Challenges. & Not discussed any RPL attack datasets and practical implementation.	\\ \midrule
			\cite{verma2020security}	&	2011-2019	&	Security aspects of RPL-based routing in 6LoWPAN network in IoT	&	Extended attack taxonomy based on traffic, resources and topology and defense taxonomy focusing on secure protocol and IDS. Open Research Issues and Challenges along with cross layer defense solution for RPL. 	&	Not consider the attacks impact analysis and No discussion about simulation tools, testbed, attack datasets	\\ \midrule
			\cite{almusaylim2020proposing}	&	2011-2019	&	Securing RPL from Specific routing attacks in IoT	&	Provides discussion for defense solution for version and rank attack in RPL-based IoT network with minor impact analysis.	&	Not focused on attacks classification as well as future research goals.	\\ \midrule
			\cite{simoglou2021intrusion}	&	2013-2020	&	Focusing on building Intrusion Detection System for Securing RPL from routing attacks.	&	Design requirement, best practices and gaps to build the IDS system for RPL. IDS requirement and guidelines for robust IDS.	&	No consideration for attack and defense taxonomy as well as impact of routing attacks.	\\ \midrule
			\cite{pasikhani2021intrusion}	&	2011-2020	&	Reviewing Intrusion Detection System for  RPL-based routing attacks in 6LowPAN network.	&	Detailed discussion about IDS for RPL, routing attack impact analysis, brief about simulator tools and research directions. 	&	Not considered the RPL-based routing attacks datasets and testbed facility for RPL .	\\ \midrule
			\cite{bang2022assessment}	&	2011-2020	&	RPL-based routing attacks and its mitigation  focusing RPL control messages.	&	Classify the RPL-based routing attacks based on RPL control messages (DAO, DIS, DIO, DAO-ACK) and mitigation according to defense method &	No discussion about the attack dataset and testbed for RPL.	\\ \midrule
			\cite{al2022systematic}	&	2016-2021	&	Focusing Machine and Deep Learning Based defense solution for RPL-based routing attacks in 6LoWPAN.	&	Detailed discussion about future research direction of RPL along with critical analysis of existing study. Very short compassion for existing RPL attacks dataset.  	&	Not considered the conventional security solution of RPL, critical analysis of existing routing attack datasets and testbeds.	\\ \midrule
			\cite{albinali2024towards}	&	2013-2023	&	RPL-based routing Attacks and Mitigation for IoT network.	&	Taxonomy for classification of RPL attacks based in attack vector and launching methods. 	&	Not performed impact of RPL attacks, no discussion about the RPL-attack dataset and testbeds. 	\\ \midrule
			\cite{alfriehat2024rpl}	&	2011-2024	&	Detection approach for RPL-based routing attacks in IoT networks. 	&	Classification of attack and methods for safeguarding the RPL for routing attacks.  	&	Not considered the RPL routing attacks datasets, testbeds and attack impact analysis.	\\\midrule
			This Survey	&	2011-2024	&	Focusing RPL Attacks and Defense solutions with critical analysis, Detailed analysis of routing attacks datasets, RPL testbeds, simulator tools, future research directions.	&	Proposed a Novel Taxonomy for RPL attacks and defense classification also detailed impact analysis of each attack. Detailed investigation on RPL-based routing attack datasets, simulator tools for building RPL-based IoT network and discussion of real time testbed for experimental study of RPL. 	&	We are not considered the demonstration of RPL attacks and mitigation. \\
			\bottomrule
		\end{tabular}
	\end{table}
	
	\clearpage
	This study included statistical analysis and research challenges. Table \ref{tbl_survey_comr} displays existing surveys on RPL-based routing attacks and defense mechanisms from 2011 to 2024. The table illustrates several shortcomings of this research domain and inspires our efforts, which led to a novel taxonomy for routing attacks and defense mechanisms for RPL. Our research introduces novel ideas not covered in previous studies, such as RPL-specific simulators, datasets, and testbeds.

	\subsection{Survey Issues} \label{subsec_researchquestion}
	Our survey addresses the following issues: 
	\begin{itemize}
		\item What are the most common routing attacks discovered by researchers that effectively make the RPL protocol insecure?  
		\item What are the routing attacks covered by researchers in the literature so far?
		\item What parameters are affected by routing attacks, and how do they affect routing decisions? 
		\item What are the various defense mechanisms developed for securing RPL? 
		\item What parameters are used to assess the performance of the defense mechanisms? 
		\item What simulators, datasets, and testbeds are available for RPL research work, and to what extent can researchers use them?   
		\item What are the routing attacks that researchers have not addressed, and hence the new defense mechanisms that need to be developed? 
	\end{itemize}
	\section{Classification of RPL-Based Attacks}
	\label{sec_classficationofroutingattack}
\begin{figure*}[h]
		\centering
		\includegraphics[width=1\linewidth]{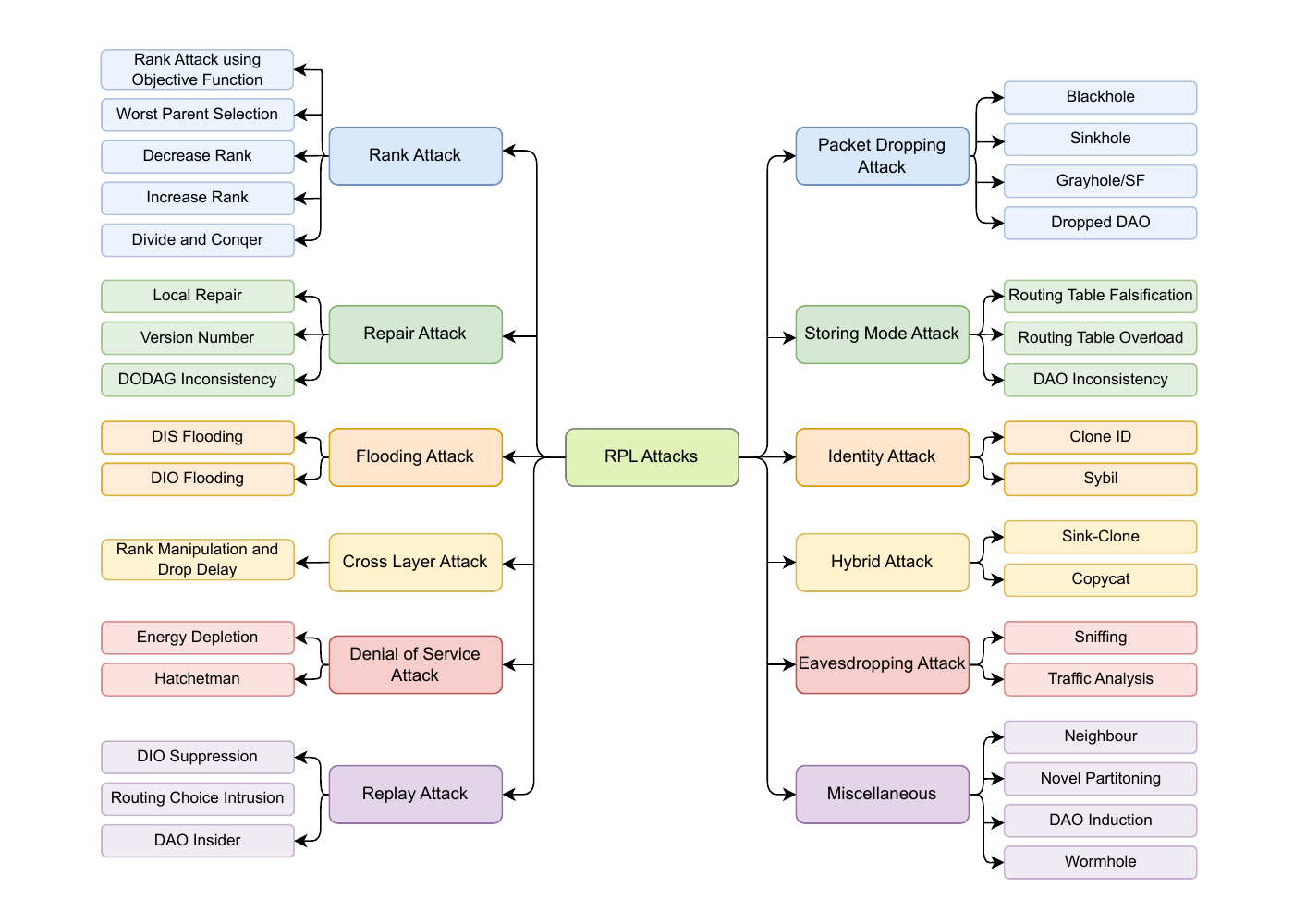}
		\caption{Taxonomy for Classification of RPL Protocol-Based Attacks}
		\label{tx_attackclass}
	\end{figure*}

	Based on our review of the existing literature, we present a new taxonomy for the classification of RPL-based routing attacks in Fig. \ref{tx_attackclass}. The new proposed attack taxonomy is based on the nature of RPL attacks, while in literature \cite{mayzaud2016taxonomy} proposed based on attack goals, \cite{raoof2018routing} based on attack origin, \cite{airehrour2016secure} on CIA properties, and \cite{bang2022assessment} based on RPL control messages. The statistics shown in Table \ref{tbl_attck_identified} are derived from the previous survey papers present in the literature and clearly show that our survey covers a significant number of unaddressed RPL-based routing attacks in the taxonomy.
	\textcolor{black}{As mentioned in Table \ref{tbl_attck_identified}, it has been observed that most of the existing survey papers have focused on maximum 22 attacks, whereas our survey is more exhaustive and includes all the recent attacks proposed in the literature. The new range of attacks included in the new taxonomy are Induced Blackhole, Coordinated Blackhole, Buffer Reservation, DAO Induction, Multicast, and Spam DIS Flooding, Dropped DAO, Energy Depletion, Hatchetman, Novel Partitioning, Divide and Conquer, Hybrid Attack (Copycat and Sink-clone) and Cross Layer Attack (Rank Manipulation and Drop Delay).}
	\begin{table}[!htbp]
		\caption{Routing Attack Vertical for RPL}
		\centering
		\begin{tabular}{l c}
			\toprule
			\textbf{Researcher} & \textbf{Inclusion of Attacks}\\
			\midrule
			Mayzud \textit{et al.} \cite{mayzaud2016taxonomy} & 16 \\
			Raoof \textit{et al.} \cite{raoof2018routing} & 16 \\
			Verma \textit{et al.} \cite{verma2020security} & 22 \\
			Pasikhani \textit{et al.} \cite{pasikhani2021intrusion} & 20 \\
			Bang \textit{et al.} \cite{bang2022assessment} & 17 \\
			\textbf{This Survey} & \textbf{37}\\
			\bottomrule
		\end{tabular}
		\label{tbl_attck_identified}
	\end{table}
	\newline \indent
	\textcolor{black}{The attacks reviewed in the new taxonomy (33 attack categories) are grouped into 12 major categories, as shown in Fig. \ref{tx_attackclass}. Existing surveys, such as Verma et al. \cite{verma2020security}, focus on attack classification based on the primary target of the attack, i.e., resources, topology, or traffic. While this provides a broader viewpoint for analyzing attack behavior, many recent attacks target RPL in multiple ways. For instance, hybrid attacks not only target topology but also resources. Another issue is that several attacks appear very similar, e.g., DAO insider, Dropping DAO, and DAO induction attacks. Placing such a broad range of attacks into existing taxonomies is challenging and may limit the clear understanding of RPL security for new researchers. To address this issue, we have proposed a new taxonomy that focuses on the nature of the attack, such as whether it exploits rank or the local repair mechanism of RPL. Additional categories are created based on whether the attack involves DoS, flooding, replay, packet dropping, or eavesdropping. Furthermore, we have introduced separate categories for hybrid attacks, identity-related attacks, storing mode attacks, and miscellaneous attacks. We believe this taxonomy better reflects the actual nature of these attacks and can help researchers advance their work in a clearer and more structured manner.}        
	
	Subsections \ref{subsec_rankattack} to \ref{subsec_miscattack} provide an overview of these routing attacks. However, we encourage the reader to refer to the papers cited under each attack for a detailed discussion of the implementation of RPL attacks.
	Algahtani \cite{algahtani2021reference} \textit{et al.} lays down detailed steps of procedure to implement RPL-based routing attacks in Contiki-NG.

	
	\begin{figure}[h!]
		\centering
		\begin{subfigure}[t]{0.45\linewidth}
			\centering
			\includegraphics[width=\linewidth]{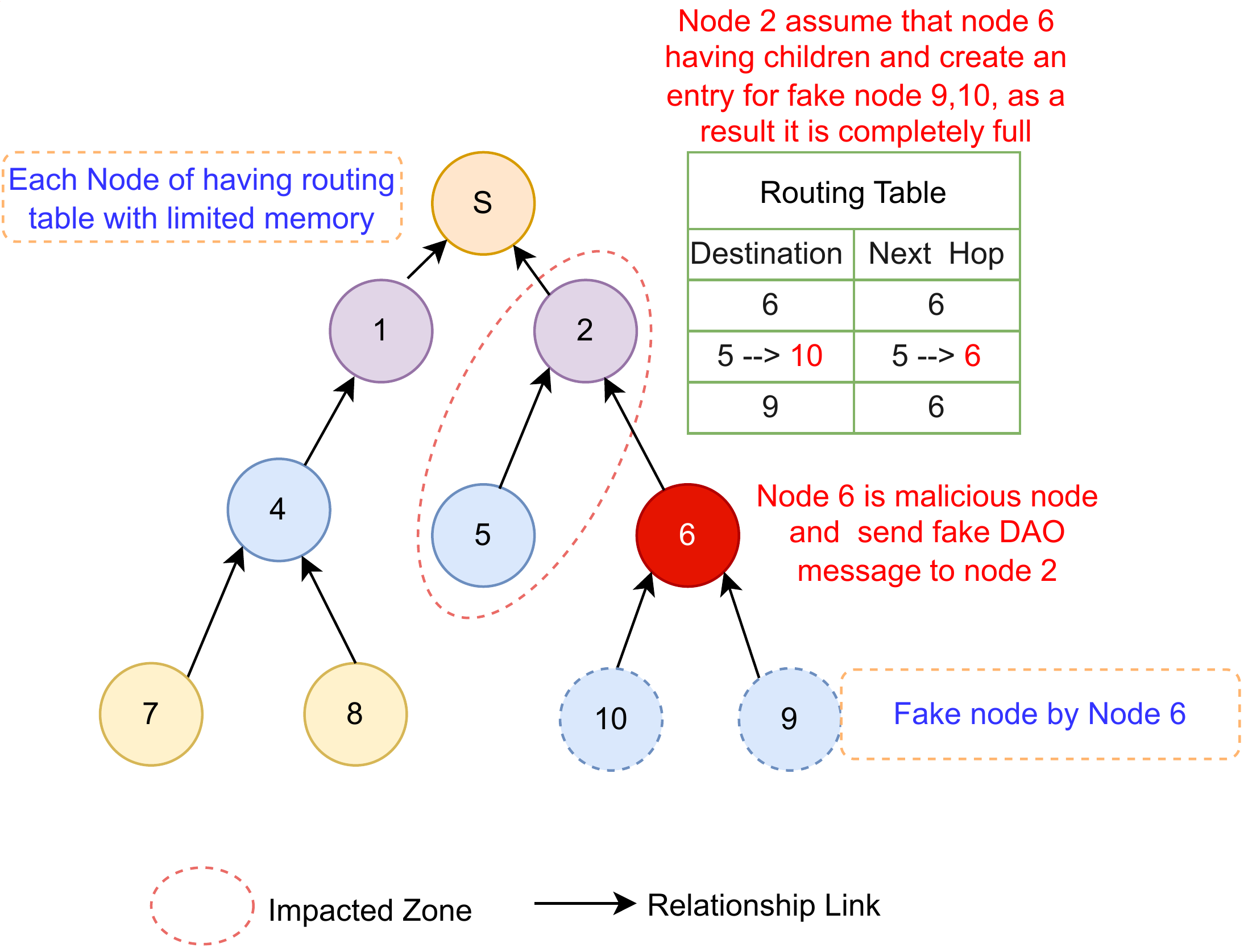}
			\caption{Routing Table Overload Attack}
			\label{fig:routingtableoverlaod}
		\end{subfigure}
		\begin{subfigure}[t]{0.45\linewidth}
			\centering
			\includegraphics[width=\linewidth]{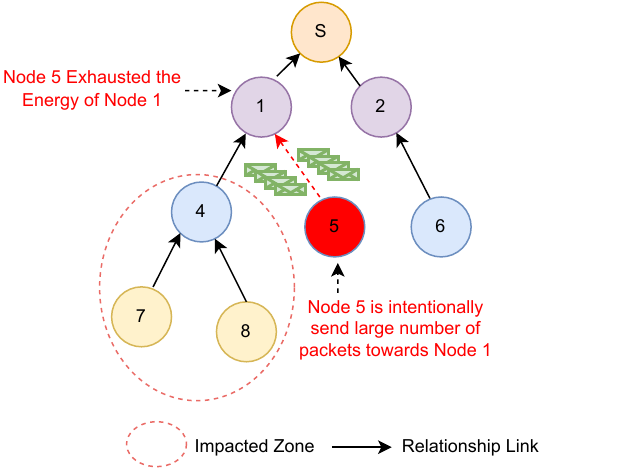}
			\caption{Energy Depletion Attack}
			\label{fig:energydepletion}
		\end{subfigure}
		\\
		\begin{subfigure}[t]{0.45\linewidth}
			\centering
			\includegraphics[width=\linewidth]{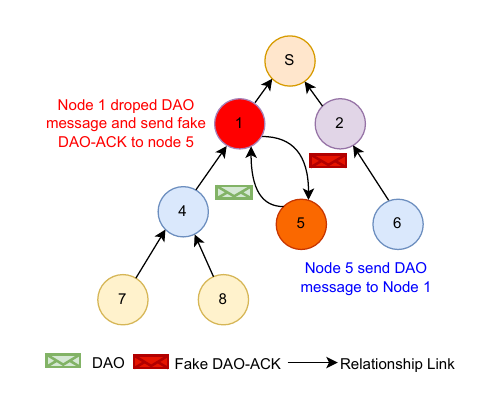}
			\caption{Dropped DAO Attack}
			\label{fig:DDAO}
		\end{subfigure}
		\begin{subfigure}[t]{0.45\linewidth}
			\centering
			\includegraphics[width=\linewidth]{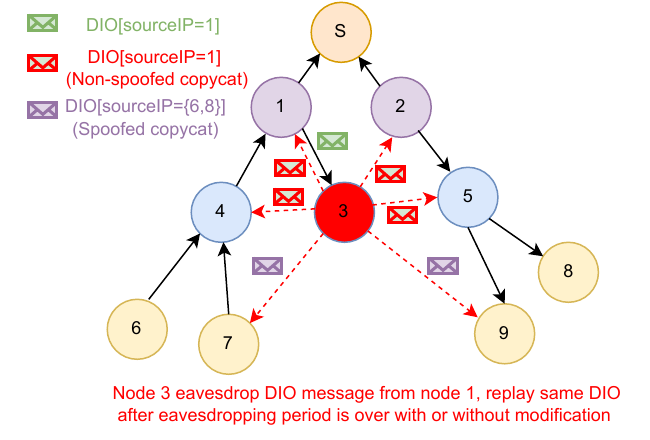}
			\caption{Copycat Attack}
			\label{fig:copycat}
		\end{subfigure}
		\caption{Illustration of various RPL-based Routing Attacks.}
		\label{fig:2X2_RTO_ED_DDAO_CC_grid}
	\end{figure}
	
	\subsection{Rank Attack} \label{subsec_rankattack}
	The rank feature or rules of RPL are used to exploit the rank attacks. The various rank attacks are summarized below. 
	\begin{itemize}
		\item Worst Parent Selection: Kiran \textit{et al.} \cite{kiran2022ids} first implemented the Worst Parent Selection attack where an attacker node modifies the Objective Function to identify the worst-ranked parent among all the neighbors even if it is advertising the actual rank. The objective of this attack is to create an un-optimized route, thereby increasing the end-to-end delay.
		\item Decrease Rank: Dvir \textit{et al.} \cite{dvir2011vera} discussed an attack where a malevolent node illegitimately publishes a lower rank to attract most of the network traffic of other nodes by posing as a preferred parent node closest to the root node. The primary goal is to attract the legitimate nodes present in a network and their traffic to reduce or degrade the network performance.
		\item Increase Rank: Le \textit{et al.} \cite{le2011specification} outlined an attack on the routing operation where a malicious node deliberately publishes a higher rank with the worst routing metrics, compelling the adjacent nodes to choose a different node as the parent. The intent is to disrupt the routing topology and cause a significant end-to-end delay.
		\item Rank Attack Using Objective Function: Rehman \textit{et al.} \cite{rehman2016rank} proposed a rank attack where a hostile node advertises a lower rank than its actual rank and the minimal routing metric based on ETX among observed routing metrics of neighbors. This attack is more severe as it reduces the packet delivery ratio.
		\item Divide and Conquer-based attack (DCB-Attack):\\ Boudouaia \textit{et al.} \cite{boudouaia2020divide} proposed an attack where a malicious node uses three different types of malicious behavior to launch an attack. The malicious node transmits the modified DIO control messages with a higher rank to the victim nodes. Then it sends another DIO message with a lower rank value to the victim nodes so that the victim node opts for it as the best parent. Further, a malicious node chooses a node having the highest rank as its best parent and forwards the complete traffic to the malevolent parent. Then the malicious node restarts the whole process again from the beginning. This type of attack increases global energy consumption as well as traffic delay and disrupts the topology. 
	\end{itemize}
	
	\subsection{Repair Attack} \label{subsec_repairattack}
	The repair mechanism of RPL is used to launch this type of attack. The various types of Repair attacks are summarized below 
	\begin{itemize}
		\item Local Repair: Le \textit{et al.} \cite{le2011specification} discussed a repair attack where a corrupt node exploits the attack in two ways: (1) broadcasts the infinite rank to all the neighbors. After receiving the rank update, the nodes search for the new parent in the direction of the root node. (2) illegitimately changes its DODAG ID to leave the DODAG and join another new DODAG. This causes every node to start its local repair mechanism and recalculate its routes, thereby leading to a degradation of resources and an increase in the overhead of control messages.
		\item Version Number: Dvir \textit{et al.} \cite{dvir2011vera} introduced a Version Number attack where a malicious node deliberately increases the version number in a DIO control message and broadcasts it to all of its adjacent nodes. This forces the DODAG to rebuild, leading to routing loops, topological inconsistency, and increased control packet overhead.
		\item DODAG Inconsistency: Sehgal \textit{et al.} \cite{sehgal2014addressing} addressed a DODAG inconsistency attack where a hostile node exploits the direction flag ``O" and topology repair flag "R" and sets them in opposite directions before forwarding the packet into the network. The consequences of this attack are the exhaustion of energy or resource, increased control overhead, and high end-to-end delay.
	\end{itemize}
	
	\subsection{Flooding Attack} \label{subsec_floodingattack}
	The Flooding attacks generate a huge number of packets and send them toward the legitimate node or network to exhaust their resources. The Flooding attacks are categorized in the section below.
	\begin{itemize}
		\item DIO Flooding: Wallgren \textit{et al.} \cite{wallgren2013routing} implemented the flooding attack where malicious nodes send the HELLO message (also known as DIO message in RPL) with the highest signal strength and better routing metrics to present them as a neighbor. Later, if any nodes want to join the malicious nodes, they reduce their signal strength to normal, becoming out of range. It is also known as the HELLO flood attack. The primary objective of such attacks is to exhaust the resource of legitimate nodes and increase the control message overhead.
		\item DIS Flooding: DIS flooding attack is launched by flooding the DIS control messages (unicast or multicast) within the network to reset the respective trickle timer of nodes and reply with multiple DIO messages. This attack leads to routing disruption, node energy exhaustion, and higher control message overhead. Verma \textit{et al.} \cite{verma2019addressing} addressed the DIS flooding attack and proposed a mitigation scheme based on safe DIS parameters. Medjek \textit{et al.} \cite{medjek2021multicast} addressed the effect of Multicast DIS (M-DIS) and proposed a mitigation scheme based on response delay and timer adjustment. Pu \textit{et al.} \cite{pu2019spam} introduced and evaluated the performance of spam DIS attacks where the hostile node sends multicast DIS control messages with bogus identities. 
	\end{itemize}
	
	\subsection{Storing Mode Attack} \label{subsec_storingattack}
	The storing operation mode stores routing information in the table for downward routes. RPL's storage mode lets hostile nodes perform routing attacks. The various types of storing mode attacks are summarized in the section below.
	\begin{itemize}
		\item Routing Table Falsification: Mayzaud \textit{et al.} \cite{mayzaud2016taxonomy} described the routing attack where a malicious node alters or forges the DAO control message to publish the fake route for other nodes. This will create a fake downward route that does not exist, resulting in sub-optimization of DODAG, packet drops, and increasing delay and network congestion.
		\item Routing Table Overload: Mayzaud \textit{et al.} \cite{mayzaud2016taxonomy} discussed the routing attack where a malicious node sends an ample amount of fraudulent DAO control messages (information on fake routes) to overload the victim nodes routing table, preventing them from building legitimate routes in the future. Duroyon \textit{et al.} \cite{duroyon2021stir} has proposed a novel method, STIR to mitigate this type of attack. The Fig. \ref{fig:routingtableoverlaod} illustrate the routing table overload attack in which Node 6 create the fake nodes and send fake DAO to node 2.
		\item DAO Inconsistency: Mayzaud \textit{et al.} \cite{mayzaud2016taxonomy} considered a routing attack where a malicious node places a flag for Forwarding-Error (F) in the routing packet and sends this information to its parent node that will later repudiate the licit downward route and create a sub-optimal DODAG. Pu \textit{et al.} \cite{pu2018mitigating} has proposed a dynamic threshold mechanism for mitigating this attack.
	\end{itemize}
	
	\subsection{Replay Attack}  \label{subsec_replayattack}
	The replay attack stores the past control message, i.e., DIS, DIO, DAO, and DAO-ACK, received from the neighboring nodes. These packets are later unicast or multicast into the network to make inconsistent paths. Below are The various types of replay attacks.   
	\begin{itemize}
		\item DIO Suppression: Perazzo \textit{et al.} \cite{perazzo2017dio} presented a novel DIO Suppression attack where a malicious node frequently sends a consistent DIO message to all the neighbors in the network. If the neighbors receive the same consistent DIO message many times, they are convinced that there is no change in network topology, and they suppress their DIO messages. The route quality will deteriorate, and in the worst situation, the network could become fractured.
		\item Routing Choice Intrusion Attack: Zhang \textit{et al.} \cite{zhang2015intrusion} introduced an attack where malicious nodes figure out the current routing choice rules, capture the ICMPv6 packets, and then broadcast the forged ICMPv6 packet with legitimate identity after some amount of time. The fundamental objective of this attack is to create routing loops, end-to-end delay, and increase energy consumption. 
		\item DAO Insider: Ghaleb \textit{et al.} \cite{ghaleb2018addressing} presented a new DAO insider attack where malicious nodes transmit forged DAO messages received from legitimate nodes to their parent, triggering DAO transmission from across all intermediate parents between the malicious node and the root. This attack increases power consumption and latency. 
	\end{itemize}
	
	\subsection{Denial of Service (DoS) Attack} \label{subsec_dosattack}
	DoS attack creates a situation where a legitimate node exhausts its resources or memory, responding to the excessive traffic or messages generated from the network. The various DoS attacks are summarized in the section below. 
	\begin{itemize}
		\item Hatchetman: Pu \textit{et al.} \cite{pu2018Hatchetman} has proposed an attack named Hatchetman, where malevolent nodes forge the received packet source-route header and broadcast the invalid packet containing the error route to genuine nodes. The genuine nodes drop all the error packets and reply with a volume of error messages to DODAG. In doing this, the nodes exhaust their energy and communication bandwidth, leading to the network denial of service attack. The Fig. \ref{fig:Hatchetman} depict the hatchetman attack in which node N3 manipulated source route header to launch attack.
		\begin{figure}[h]
			\centering 
			\includegraphics[width=.6\columnwidth]{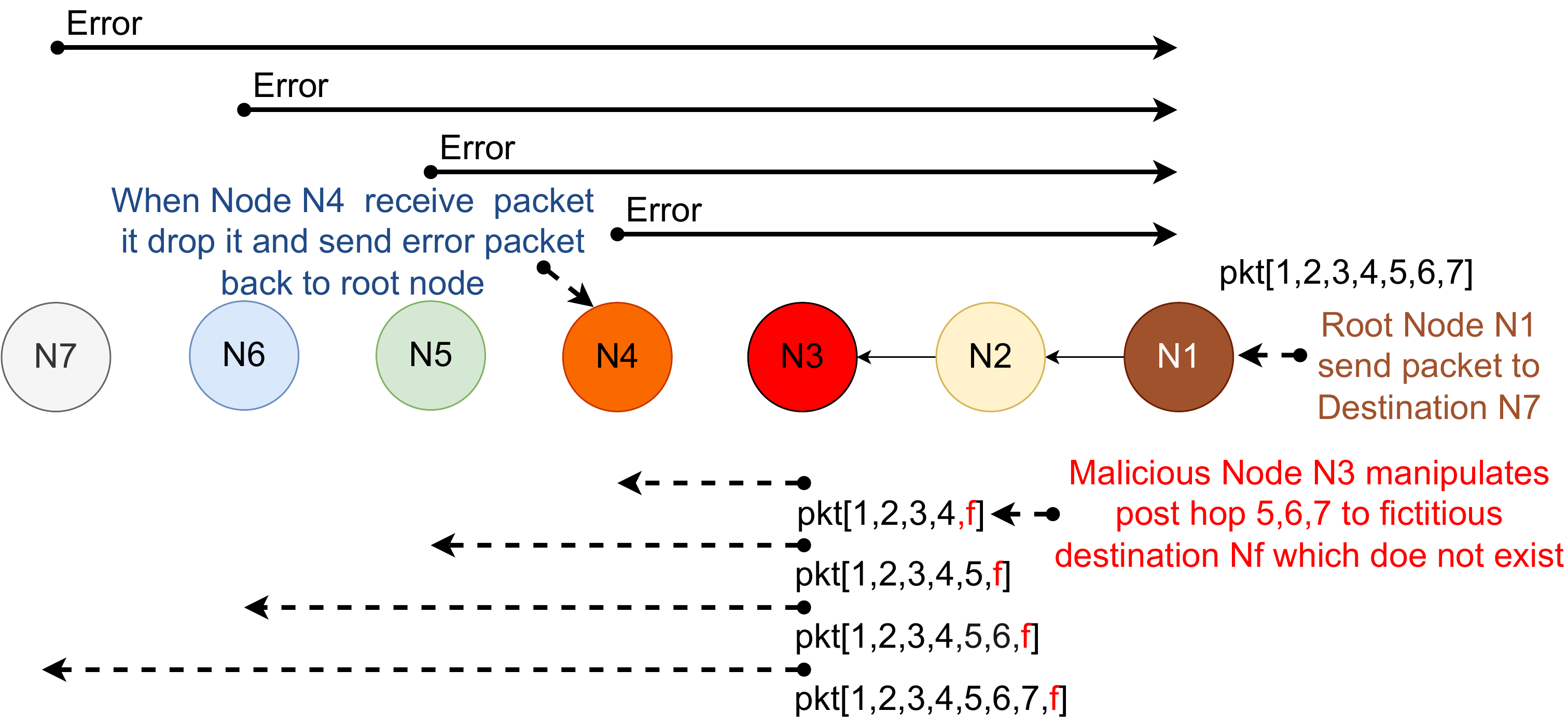}
			\caption{Illustration of Hatchetman Attack}
			\label{fig:Hatchetman}
		\end{figure}
		\item Energy Depletion: Pu \textit{et al.} \cite{pu2019energy} presented an Energy Depletion attack where several malicious nodes generate packets and broadcast them in the forwarding path toward the legitimate node. This will increase legitimate nodes' energy consumption and result in service denial. The author has proposed a misbehavior detection technique to detect this attack.  Fig. \ref{fig:energydepletion} explains the working of energy depletion attack where node 5 flood the packet to deplete the energy of node 1.
	\end{itemize}
	
	\subsection{Packet Dropping Attack} \label{subsec_pktdroppingattack}
	This attack is launched to drop or alter specific network traffic by exploiting legitimate nodes in the network to create un-optimized paths, thereby reducing the packet delivery ratio. The various packet-dropping attacks are summarized in the section below.  
	\begin{itemize}
		\item Blackhole: Raza \textit{et al.} \cite{raza2013svelte} introduced a blackhole attack where a hostile node relinquishes all the packets which are supposed to be forwarded in the network. This isolates a node or set of nodes from the network and decreases the packet delivery ratio.
		\item Induced Blackhole: Chen \textit{et al.} \cite{chen2018analysis} proposed an induced blackhole attack where malicious nodes continuously drop all the targeted router packets without modifying the legitimate router's internal logic. The malicious nodes use a jammer to launch or stop this attack which blocks outgoing routing updates of neighbors without modifying the incoming packet or MAC-layer ACK frames of a targeted router.    
		\item Coordinated Blackhole: Essaadi \textit{et al.} \cite{essaadi2021detection} proposed a coordinated blackhole attack, a type of binary blackhole attack \cite{zhang2019cuckoo} where a network of bad actors collaborates to carry out a single attack. They also proposed a Hop-Count Reachability (HCR) based mitigation mechanism against the attack.  
		\item Dropped DAO: Sheibani \textit{et al.} \cite{sheibani2022lightweight} proposed an attack where a malicious node relinquishes the DAO message received from a legitimate node and reverts with a counterfeit DAO Acknowledgment (DAO-ACK) to prevent the creation of new downward routes. This attack drops all the downward packets, reducing the packet delivery ratio. The Fig. \ref{fig:DDAO} show the Dropped DAO attack where node 1 drop the DAO message received from node 5 and send fake DAO-ACK in response.
		\item Sinkhole: Wallgren \textit{et al.} \cite{wallgren2013routing} described the Sinkhole attack where malicious nodes decrease their rank to attract network traffic. The network traffic is altered or dropped after a malicious node becomes the network's preferred parent.
		\item Grayhole/Selective Forwarding (SF): Wallgren \textit{et al.} \cite{wallgren2013routing} discussed a grayhole attack, also known as the selective forwarding attack, where a malicious node sends filtered RPL control packets and relinquishes the remaining packets in a network to distort the routing path.    
	\end{itemize}
	
	\subsection{Identity Attack} \label{sub_B_idnetity_sba}
	This attack exploits the identity of a legitimate node to acquire access control to a wider network area. The various identity attacks are summarized in the section below.
	\begin{itemize}
		\item Clone ID Attack: Wallgren \textit{et al.} \cite{wallgren2013routing} proposed this attack where a malicious node replicates the identity of a legitimate node to multiple physical nodes to control the larger part of a network.
		\item Sybil attack: Zhang \textit{et al.} \cite{zhang2014sybil} proposed a sybil attack where malicious nodes use multiple logical identities for the same network physical node. The authors also discussed three variations of sybil attack: SA-1 (limited to a specific network range), SA-2 (distributed to network range), and SA-3 (distributed to network range with mobile malicious node).  
	\end{itemize}
	
	\subsection{Hybrid Attacks} \label{sub_C_hybrid_sba}
	Hybrid attacks are launched when two or more separate routing attacks are combined to cause disruption in the network. Various types of hybrid attacks are summarized below.
	\begin{itemize}
		\item Sink-Clone attack: Mirshahjafari \textit{et al.} \cite{mirshahjafari2019sinkhole+} proposed a hybrid sink-clone attack, an amalgamation of a sinkhole and clone ID attacks. A collection of malicious nodes, all of which share the same identity, advertise themselves for nearby nodes to route their traffic through them. This hybrid attack is more destructive, affects the network performance, and increases power consumption.   
		\item Copycat attack: Verma \textit{et al.} \cite{verma2020addressing} addressed the copycat attack, where a malicious node intercepts DODAG Information Object (DIO) messages from its neighbors and then replays the intercepted DIO messages with a fixed interval (with or without modification). The main goal is to introduce congestion and inference into the network, resulting in a degraded quality of service. The Fig. \ref{fig:copycat} illustrate the working of copycat attack in which node 3 eavesdrop DIO message received from node 1 and later on it replay it to perform attack.
	\end{itemize}
	
	\subsection{Cross-layer Attacks} \label{sub_D_crosslayer_sba}
\begin{figure}[h]
		\centering 
		\includegraphics[width=.8\columnwidth]{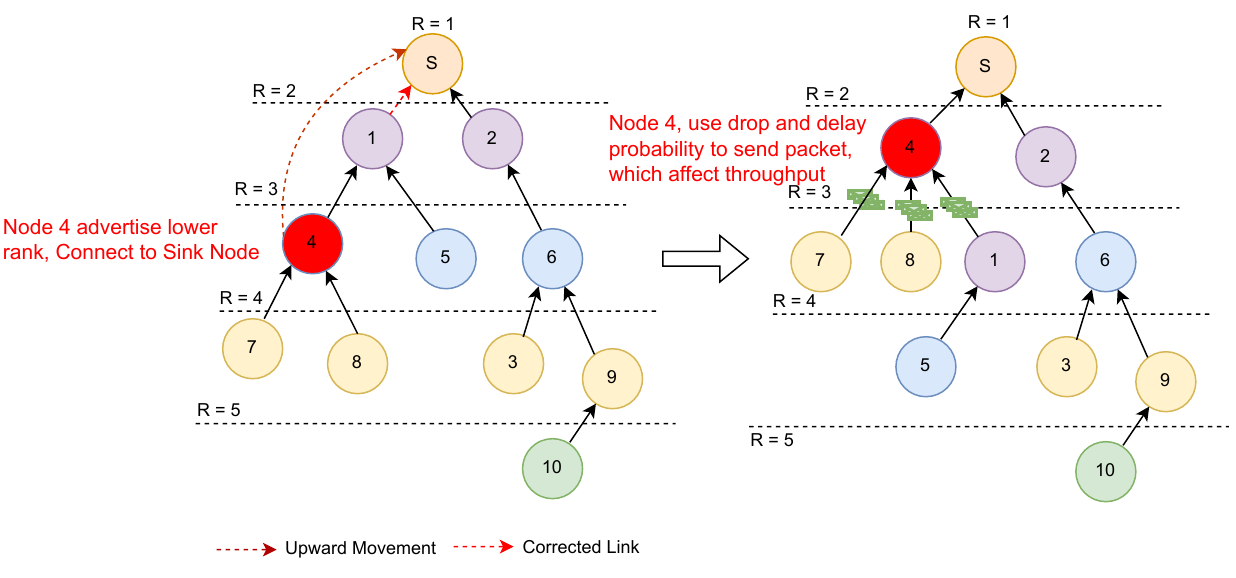}
		\caption{Illustration of Cross-Layer Attack}
		\label{fig:crosslayer}
	\end{figure}

	Cross-layer attacks are attacks where a malicious node exploits the abundance of communication among the MAC layer, network layer, and application layer to launch an attack that degrades the performance of multiple layers. \\
	\indent Asati \textit{et al.} \cite{asati2018rmdd} first proposed an IoT cross-layer attack named Rank Manipulation and Drop Delay (RMDD). A malicious node first lowers its rank among all neighboring nodes to become a preferred parent, resulting in the re-routing of network traffic through this malicious node. Once the traffic is diverted, it will either add delay or drop the packet, reducing throughput at the application layer. The Fig. \ref{fig:crosslayer} explain the cross layer attack where rank manipulation and dropping of messages concept used to launched attack.
	

	\subsection{Eavesdropping Attacks} \label{sub_E_eves_sba}
	Eavesdropping attacks are launched to steal data or information sent or received between nodes by exploiting unsecured communication links. Various types of eavesdropping attacks are summarized below.   
	\begin{itemize}
		\item Sniffing: Mayzaud \textit{et al.} \cite{mayzaud2016taxonomy} discussed a sniffing attack where a malicious node listens or sniffs packets sent across the network to obtain routing information such as DODAG ID, version number and deployment of nodes' ranks in network topology. Due to this attack's passive nature, detection is challenging.
		\item Traffic Analysis: Mayzaud \textit{et al.} \cite{mayzaud2016taxonomy} discussed traffic analysis, where a malicious node analyzes traffic patterns and characteristics to obtain routing information and identify parent/child relationships to provide a view of network topology. The impact of this type of attack is determined by the node's location. More traffic is inspected, and more sensitive routing information is exposed the closer it is to the root node. 
	\end{itemize}
	\vspace{-1.5em}
	
	\begin{table}[h!]
		\caption{Impact Assessment of state-of-the-art Routing Attacks}
		\label{tbl_impactassementofattacks}
		\renewcommand{\arraystretch}{1.2}
		\resizebox{\linewidth}{!}{
			\begin{tabular}{c p{2.7cm} p{2.0cm} p{2.0cm} p{1.7cm} c c c c c c c c c c }
				\toprule
				\textbf{} &
				\textbf{} &
				\textbf{} &
				\textbf{} &
				\textbf{} &
				\textbf{} &
				\multicolumn{9}{c}{\textbf{Impact Assessment}}\\ 
				\cmidrule(rl){7-15}
				\textbf{\textbf{Ref}} &
				\textbf{\begin{tabular}{l}Attack\\Name\end{tabular}} & 
				\textbf{\begin{tabular}{l}Attack\\Nature\end{tabular}} & 
				\textbf{\begin{tabular}{l}Attack\\Type\end{tabular}} & 
				\textbf{\begin{tabular}{l}Mode\\ Operation\end{tabular}} & 
				\textbf{\begin{tabular}{c}CIA\\Principle\end{tabular}} &
				\textbf{\begin{tabular}{c}Resource\\Consumption\end{tabular}} &
				\multicolumn{4}{c}{\textbf{\begin{tabular}{c}Routing\\Decision\end{tabular}}} &
				\multicolumn{3}{c}{\textbf{\begin{tabular}{c}Performance\\Metrics\end{tabular}}} &
				\textbf{Isolation}\\
				\cmidrule(rl){7-7}
				\cmidrule(rl){8-11}
				\cmidrule(rl){12-14}
				\cmidrule(rl){15-15}
				\textbf{} &
				\textbf{} &
				\textbf{} &
				\textbf{} &
				\textbf{} &
				\textbf{} &
				\textbf{EC} &
				\textbf{FR} &
				\textbf{RL} &
				\textbf{UR} &
				\textbf{TF} &
				\textbf{PDR} &
				\textbf{CMO} &
				\textbf{E2ED} &
				\textbf{SD} \\
				\midrule
				\cite{kiran2022ids}	&	Worst Parent Selection	&	Insider	&	Active/Passive	&	Non-Storing	&	I/A	&	L	&	Y	&	Y	&	Y	&	L	&	L	&	L	&	M	&	Y	\\
				\cite{dvir2011vera}	&	Decrease Rank	&	Insider	&	Active	&	Non-Storing	&	A	&	M	&	Y	&	Y	&	Y	&	M	&	L	&	L	&	L	&	N	\\
				\cite{le2011specification}	&	Increase Rank	&	Insider	&	Active	&	Non-Storing	&	A	&	M	&	Y	&	Y	&	Y	&	M	&	L	&	L	&	M	&	N	\\
				\cite{rehman2016rank}	&	Rank Attack using OF &	Insider	&	Active	&	Non-Storing	&	I/A	&	L	&	Y	&	N	&	Y	&	L	&	H	&	H	&	H	&	N	\\
				\cite{boudouaia2020divide}	&	Divide and Conquer &	Insider/Outsider	&	Active	&	Non-Storing	&	I/A	&	M	&	Y	&	N	&	Y	&	M	&	L	&	L	&	M	&	N	\\
				\cite{le2011specification}	&	Local Repair	&	Insider/Outsider	&	Active	&	Non-Storing	&	I/A	&	M	&	N	&	N	&	Y	&	L	&	M	&	M	&	L	&	Y	\\
				\cite{sehgal2014addressing}	&	DODAG Inconsistency	&	Insider	&	Active	&	Non-Storing	&	I/A	&	H	&	N	&	N	&	Y	&	H	&	L	&	H	&	H	&	Y	\\
				\cite{dvir2011vera}	&	Version Number	&	Insider	&	Active	&	Non-Storing	&	I/A	&	H	&	Y	&	Y	&	Y	&	H	&	L	&	M	&	L	&	N	\\
				\cite{wallgren2013routing}	&	DIO Flooding	&	Insider/Outsider	&	Active	&	Non-Storing	&	A	&	H	&	N	&	N	&	N	&	M	&	H	&	H	&	L	&	N	\\
				\cite{verma2019addressing}	&	DIS Flooding	&	Insider/Outsider	&	Active	&	Non-Storing	&	A	&	H	&	N	&	N	&	Y	&	M	&	M	&	H	&	M	&	N	\\
				\cite{mayzaud2016taxonomy}	&	Routing Table Falsification	&	Insider	&	Active	&	Storing	&	I/A	&	M	&	Y	&	N	&	Y	&	M	&	H	&	L	&	M	&	N	\\
				\cite{mayzaud2016taxonomy}	&	Routing Table Overload	&	Insider	&	Active	&	Storing	&	I/A	&	L	&	Y	&	N	&	Y	&	L	&	L	&	L	&	L	&	N	\\
				\cite{mayzaud2016taxonomy}	&	DAO Inconsistency	&	Insider	&	Active	&	Non-Storing	&	I/A	&	H	&	Y	&	N	&	Y	&	M	&	H	&	L	&	L	&	N	\\
				\cite{perazzo2017dio}	&	DIO Suppression	&	Insider/Outsider	&	Active	&	Non-Storing	&	A	&	H	&	Y	&	N	&	Y	&	L	&	H	&	L	&	L	&	Y	\\
				\cite{zhang2015intrusion}	&	Routing Choice Intrusion	&	Insider	&	Active/Passive	&	Non-Storing	&	I/A	&	M	&	Y	&	Y	&	Y	&	L	&	L	&	L	&	M	&	N	\\
				\cite{ghaleb2018addressing}	&	DAO Insider	&	Insider	&	Active	&	Non-Storing	&	I/A	&	M	&	N	&	N	&	Y	&	H	&	M	&	H	&	M	&	N	\\
				\cite{pu2018Hatchetman}	&	Hatchetman	&	Insider	&	Active/Passive	&	Non-Storing	&	I/A	&	H	&	N	&	N	&	N	&	H	&	H	&	L	&	M	&	N	\\
				\cite{pu2019energy}	&	Energy Depletion	&	Insider	&	Active	&	Non-Storing	&	A	&	H	&	N	&	N	&	Y	&	H	&	L	&	L	&	L	&	Y	\\
				\cite{raza2013svelte}	&	Blackhole	&	Insider	&	Active/Passive	&	Non-Storing	&	A	&	L	&	N	&	N	&	N	&	L	&	H	&	L	&	L	&	Y	\\
				\cite{chen2018analysis}	&	Induced Blackhole	&	Insider	&	Active/Passive	&	Non-Storing	&	A	&	L	&	N	&	N	&	Y	&	M	&	H	&	L	&	L	&	Y	\\
				\cite{essaadi2021detection}	&	Coordinated Blackhole	&	Insider	&	Active/Passive	&	Non-Storing	&	A	&	L	&	N	&	N	&	Y	&	L	&	H	&	L	&	L	&	Y	\\
				\cite{sheibani2022lightweight}	&	Dropped DAO	&	Insider	&	Active/Passive	&	Non-Storing	&	I/A	&	L	&	N	&	N	&	Y	&	M	&	L	&	M	&	L	&	N	\\
				\cite{wallgren2013routing}	&	Sinkhole	&	Insider	&	Active	&	Non-Storing	&	I/A	&	L	&	N	&	N	&	Y	&	M	&	L	&	L	&	L	&	N	\\
				\cite{wallgren2013routing}	&	Grayhole/SF &	Insider	&	Active/Passive	&	Non-Storing	&	A	&	L	&	N	&	N	&	Y	&	M	&	M	&	L	&	L	&	N	\\
				\cite{wallgren2013routing}	&	Clone ID 	&	Insider	&	Active	&	Non-Storing	&	I/A	&	L	&	Y	&	N	&	Y	&	L	&	L	&	L	&	L	&	N	\\
				\cite{zhang2014sybil}	&	Sybil 	&	Insider	&	Active	&	Non-Storing	&	I	&	L	&	Y	&	N	&	Y	&	M	&	M	&	L	&	L	&	N	\\
				\cite{mirshahjafari2019sinkhole+}	&	Sink-Clone 	&	Insider	&	Active	&	Non-Storing	&	I/A	&	M	&	Y	&	N	&	Y	&	M	&	M	&	L	&	L	&	N	\\
				\cite{verma2020addressing}	&	Copycat 	&	Insider/Outsider	&	Active	&	Non-Storing	&	I/A	&	M	&	N	&	N	&	N	&	H	&	M	&	M	&	L	&	N	\\
				\cite{asati2018rmdd}	&	Rank Manipulation and Drop Delay &	Insider	&	Active/Passive	&	Non-Storing	&	I/A	&	L	&	Y	&	Y	&	Y	&	M	&	H	&	L	&	H	&	N	\\
				\cite{mayzaud2016taxonomy}	&	Sniffing	&	Insider/Outsider	&	Active	&	Non-Storing	&	C	&	L	&	N	&	N	&	N	&	L	&	L	&	L	&	L	&	N	\\
				\cite{mayzaud2016taxonomy}	&	Traffic Analysis	&	Insider	&	Active	&	Non-Storing	&	C	&	L	&	N	&	N	&	N	&	L	&	L	&	L	&	L	&	N	\\
				\cite{le2013impacts}	&	Neighbor Attack	&	Insider/Outsider	&	Active	&	Non-Storing	&	I/A	&	L	&	Y	&	N	&	Y	&	L	&	L	&	L	&	M	&	N	\\
				\cite{sahay2021novel}	&	Novel Partitioning 	&	Insider	&	Active/Passive	&	Non-Storing	&	I/A	&	L	&	Y	&	N	&	Y	&	M	&	H	&	M	&	L	&	Y	\\
				\cite{baghani2021dao}	&	DAO Induction	&	Insider	&	Active	&	Non-Storing	&	I/A	&	L	&	N	&	N	&	N	&	M	&	M	&	L	&	M	&	N	\\
				\cite{wallgren2013routing}	&	Wormhole	&	Insider	&	Active/Passive	&	Non-Storing	&	I/A	&	L	&	Y	&	N	&	Y	&	M	&	L	&	L	&	L	&	N	\\
				\bottomrule
		\end{tabular}}
		\begin{tablenotes}
			\scriptsize
			\item C: Confidentiality
			\item I: Integrity 
			\item A: Availability 
			\item EC: Energy Consumption
			\item FR: Fake Routes/Falsification of Routes
			\item RL: Routing Loop
			\item UR: Un-optimized Route
			\item TF: Traffic Flow/Congestion
			\item PDR: Packet Delivery Ratio
			\item CMO: Control Message Overhead
			\item E2ED : End-to-End Delay
			\item SD : Sub-optimal DODAG
			\item L: Low
			\item M: Mid
			\item H: High
			\item Y: Yes
			\item N: No
		\end{tablenotes}
	\end{table}
	\subsection{Miscellaneous} \label{subsec_miscattack}
	The miscellaneous attacks presented below have not been categorized, but they also exploit the RPL features to launch routing attacks. 
	\begin{itemize}
		\item Neighbor Attack: Le at al. \cite{le2013impacts} proposed a neighbor attack where a malicious node forwards any unmodified DIO message in the network. When a valid node receives a DIO message, it assumes it may have a new neighbor out of range. Although it may be beyond range, if the new node broadcasts a high rank, the genuine node chooses to make it a preferred parent and changes the route. The network experiences a slight end-to-end latency due to this kind of attack.
		\item Novel Partitioning Attack: Sahay \textit{et al.} \cite{sahay2021novel} proposed a novel partitioning attack where a hostile node skips the route registration process. This can be done by cutting the parent selection step at the node joining or blocking the DAO output whenever a new DIO message is received. This will create segregation of legitimate nodes from DODAG and reduce the packet delivery rate of the sink node.
		\item DAO Induction: Baghani \textit{et al.} \cite{baghani2021dao} proposed a DAO induction attack where a malicious node exploits the DIO message and regularly increments the DAO Trigger Sequence Number (DTSN) to create the crafted control message, causing nodes in the malicious sub-DODAG to send redundant control messages. Eventually, it will increase the packet loss rate and end-to-end latency.
		\item Wormhole: Wallgren \textit{et al.} \cite{wallgren2013routing} discussed a wormhole attack where two or more malicious nodes forge an out-of-bound connection called a tunnel over a wired or wireless medium to forward legitimate network traffic without the involvement of a border router. This will create an optimized path in the network.
	\end{itemize}
	\indent The discussion above has provided an overview of the routing attacks described in the attack taxonomy. When an attacker or malicious node launches an attack on a network, it affects various parameters, degrading the performance. The impact on various parameters has been investigated as presented in Table \ref{tbl_impactassementofattacks} and real-life scenarios and impact of some prominent RPL attacks are explained in Table \ref{tab:rel_world_scanerio}. 
	\begin{table}[h!]
		\color{black}
		\centering
		\scriptsize
		\caption{\textcolor{black}{Real-life scenarios and impact of some prominent RPL attacks}}
		\label{tab:rel_world_scanerio}
		\begin{tabular}{p{2cm} p{10cm} p{3cm}}
			\toprule
			\textbf{Attack Name}	&	\textbf{Real-life Scenario}	&	\textbf{Impact}	\\ \midrule
			Blackhole Attack	&	In a healthcare IoT system, an attacking node (device) may advertise an optimum routing of the patient monitoring data that is to be forwarded via the malicious node. It will receive all the critical patient information without forwarding them further. Due to this, there could be loss of critical information for diagnosis, leading to incomplete information when any emergency case arises.  	&	Time sensitive patients' data loss and fatal delay in emergency.  	\\ \midrule
			Sybil Attack	&	In automated IoT enabled factory, a malicious node can acquire different fake identities which overwhelms the other nodes. This can lead to incorrect routing decision and impede real-time surveillance of machinery.   	&	Resource exhaustion and delay in data transmission.	\\ \midrule
			Rank Attack	&	In a post disaster communication network, a malicious node can advertise its rank to be lower than the actual. This can result in routing of critical data through it, eventually delaying the alerts for emergency events.	&	Inefficiency in response to emergencies.	\\ \midrule
			Version Number Attack
			&	In a RPL-based energy monitoring system of a smart grid, an attacker alters the version number of DODAG.  This will rebuild the DODAG routing tables, resulting in network instability and delays in the reporting of energy consumption data. 	&	Higher control message overhead due to topology changes, resulting in downtime in service.	\\ \midrule
			DAO Suppression Attack	&	In a smart robotic logistic IoT network, a malicious node can delay forwarding of DAO messages. This hinders the timely routing updates of the sensor nodes data causing delays in real-time inventory tracking for logistic operations.	&	Inconsistency in routing updates and performance lag in IIoT.	\\ \midrule
			DIS Flood Attack
			&	In a wildlife monitoring IoT network, a malicious node can continuously send the DIS messages, overwhelming the tracking devices. This causes energy depletion of devices used for tracking wildlife.	&	Energy depletion and loss of crucial tracking data.	\\ \midrule
			Dropped DAO Attack & A Dropped DAO attack might prevent the node from forwarding the messages of the DAO, which are very crucial to be sure of bidirectional routes from the root to the nodes. This could put people in unsafe situations and even cause loses in maintenance and money.  & Data loss, command and control failure, and energy wastage. \\ \midrule
			Routing Table Falsification Attack & In a smart agriculture setup, nodes monitor soil moisture, temperature, and humidity throughout a vast farm. In this setup, an attacker may advertise false routing information in the form of a fake shorter path to the root, pretend to be a root, and attract traffic and divert traffic to non-existent nodes or loops. Consequently, legitimate nodes update the routing tables based on such false information thinking that the compromised node is a shortest path to the root, which may be a health hazard to the crops, bring about economic losses, and face maintenance challenge.  & Data disruption, network instability, energy drain, DoS.  \\ \bottomrule
		\end{tabular}
	\end{table}
	\begin{figure}[h!]
		\centering
		\includegraphics[width=.7\columnwidth]{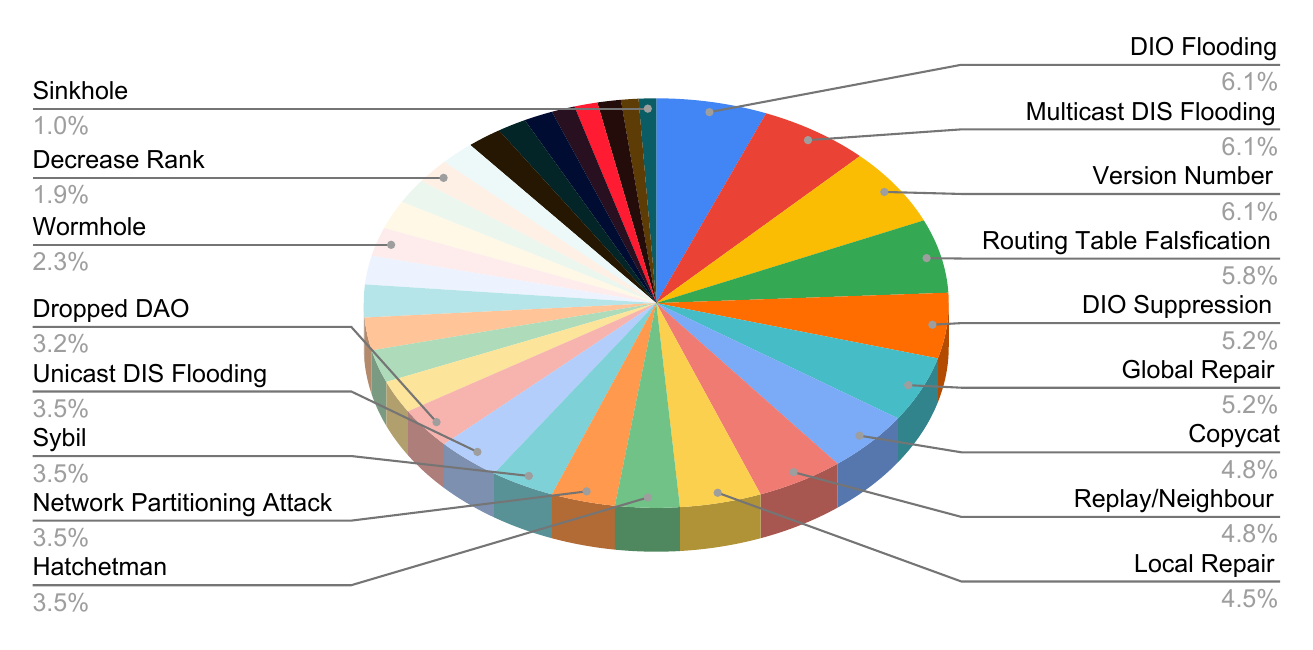}
		\caption{Adverse Impact of Routing Attacks}
		\label{fig_impactofattack}
	\end{figure}
	They include the nature of the attack (e.g., an insider or outsider node launching the attack), the type of attack (active or passive), the attacking node (e.g., storing or non-storing node), and the effect on confidentiality, integrity, and availability and the impact on various parameters resource consumption, routing decision, performance metrics and isolation of nodes. The overall impact has been classified as Low, Mid, and High.
	We have also perform impact analysis of individual attacks which helps the researchers while designing accurate and efficient defense solutions for identifying attacks. The cumulative impact of each attack is presented in Fig. \ref{fig_impactofattack} by analyzing the level of each of the parameters affected. The observation drawn from the figure that the most disruptive attacks on RPL are Version Number and the least disruptive is Increase Rank.    
	
	\section{Classification of Defense Mechanisms for RPL Attacks}  \label{sec_defensclassfication}
\begin{figure*}[h!]
		\centering 
		\includegraphics[width=1\linewidth]{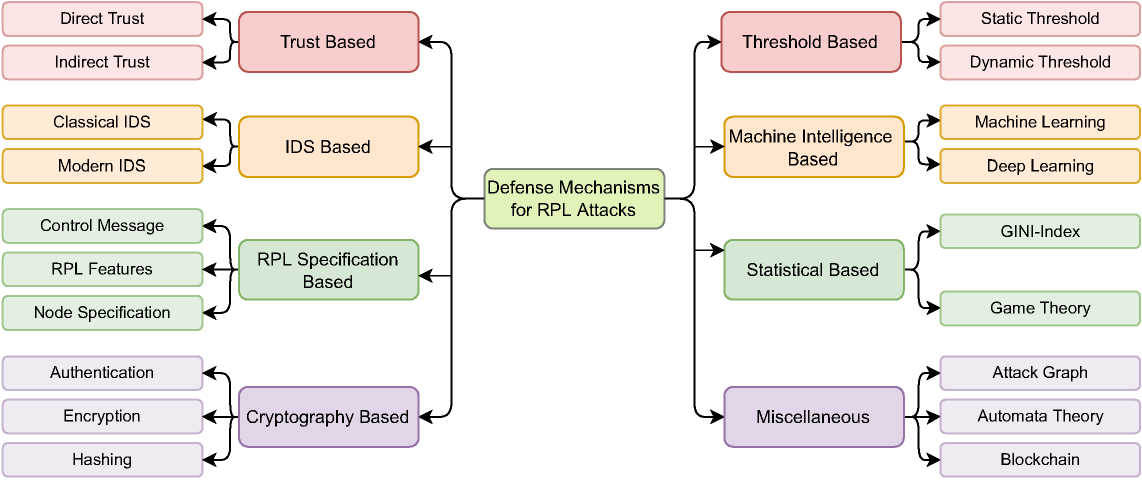}
		\caption{Taxonomy for Classification of Defense Mechanisms for RPL Attacks}
		\label{tx_defense}
	\end{figure*}

	The primary goal of this survey is to provide a comprehensive picture of RPL-based routing attacks and the associated defense mechanisms for the IoT.
	Only a few studies provide a classification or taxonomy for defense mechanisms against RPL-based routing attacks. \newline
	\indent Zarpelao \textit{et al.} \cite{zarpelao2017survey} proposed a taxonomy for IDS for the IoT that includes placement strategies (distributed, centralized, and hybrid) and detection methods (e.g., signature, anomaly, and specification). Verma \textit{et al.} \cite{verma2020security} presented a taxonomy for RPL-based routing attack defense mechanisms on secure protocols (using cryptographic, threshold, and trust mechanisms) and intrusion detection systems.
	Almusaylsim \textit{et al.} \cite{almusaylim2020proposing} presented a taxonomy of IDS based on detection strategy (processing strategy, monitoring technique, and detection methodology). Seyfollahi \textit{et al.} \cite{seyfollahi2021review} presented a taxonomy of IDS methods based on placement strategy and detection methods in their review.
	Raoof \textit{et al.} \cite{raoof2018routing} discussed classification based on mitigation mechanisms and IDS. Similarly, in their studies, Bang \textit{et al.} \cite{bang2022assessment} presented a classification based on mitigation techniques.
	\newline \indent \textcolor{black}{The existing studies, such as Almusaylim \textit{et al.} \cite{almusaylim2020proposing} and Zarpelao \textit{et al.} \cite{zarpelao2017survey}, have focused on IDS-based defense mechanisms. On the other hand, Verma \textit{et al.} \cite{verma2020security}, Raoof \textit{et al.} \cite{raoof2018routing}, and Bang \textit{et al.} \cite{bang2022assessment} have emphasized attack mitigation techniques rather than focusing solely on IDS. As far as RPL security research is concerned, multiple mitigation techniques and strategies beyond IDS have been proposed. This motivated us to develop a novel taxonomy that classifies defense solutions for RPL based on the type of defense strategy, i.e., trust, specification, cryptography, threshold, machine intelligence, statistical, IDS, and miscellaneous. These defense strategies are further sub-classified into specific key mechanisms emphasized by the authors of the defense mechanisms. We believe our proposed taxonomy, shown in Fig. \ref{tx_defense}, will help researchers better understand the depth of RPL security.}
	The subsection, \ref{subsec_rplspec_based} to \ref{subsec_misc_based}, provides a category-by-category overview of proposed defense mechanisms for RPL-based routing attacks presented in the literature. The performance of each proposed defense mechanism is evaluated using parameters such as the overhead in terms of control messages or memory consumption, energy consumption in terms of the amount of energy consumed, the detection rate in terms of attacker identification using \textit{Low-Mid-High}. Similarly the other performance parameters in proposed defense solution, whether they are lightweight in terms of computation involved in attack detection, and mobility in terms of mobile nodes present in the network topology as \textit{present-Yes and absent-No}. The criteria for accessing the performance of defence solutions in terms of detection rate are \textit{Low $<$50}, \textit{Mid 50-80}, and \textit{High $>$80}.
	\subsection{RPL Specification based Defense Mechanism} \label{subsec_rplspec_based}
	The RPL specification-based approaches designed the defense approaches using RPL features such as rank, version number, and control messages. This section provides an overview of defense mechanisms based on RPL specifications. 
	\subsubsection{Control Message}
	The control messages (DIS, DIO, DAO) are used to design the defense solution. \newline
	\indent Ghaleb \textit{et al.} \cite{ghaleb2018addressing} addressed a new type of insider attack called DAO insider attack, where a bogus DAO control message is sent to the parent node regularly by the malicious node. SecRPL, a proposed mechanism for mitigating DAO insider attacks, restricts the DAO message for every destination by a parent node. Every parent node keeps a counter for received DAO messages from the child node; if the child node's DAO sending rate exceeds a predefined threshold, the parent node discards all received DAO from the malicious child and blocks that child node. In terms of forwarded DAO, the proposed technique has a higher overhead. \newline
	\indent Shafique \textit{et al.} \cite{shafique2018detection} proposed an intrusion detection system (IDS) for detecting Rank attacks implemented at the root/sink node. A node IP address, preferred parent IP address, and rank are encrypted with the shared key in the DAO message at the node side, and the DAO message is decrypted with the same key at the sink node. Node current rank and node parent rank are compared; if the rank rule is broken, the node is classified as malicious. The accuracy and confidence interval are used to assess performance. The proposed method detects attacks with 100\% accuracy in normal conditions and varies depending on mobility conditions. Because the detection module is located at the sink node, this approach claims to have a very low overhead. To improve attack detection, the author will include more routing metrics such as throughput, hop count, delay, and bandwidth.\newline
	\indent Wadhaj \textit{et al.} \cite{wadhaj2020mitigation} first investigated the impact of DAO attacks on latency, energy consumption, and dependability. The author also proposed two mitigation schemes for DAO insider attacks, SecRPL1 and SecRPL2. The concept is to limit the number of DAO messages sent to each child node and specific node. In SecRPL1, each parent counts the DAO control messages received from the child nodes and discards DAO control messages that exceed the maximum defined limit. This work is an extension of \cite{ghaleb2018addressing}. The number of DAO forwarded by a specific node in SecRPL2 is limited by a predefined limit defined by \textit{DAO\_FORWARD\_MAX}. The proposed method performed well in terms of power consumption, packet delivery ratio, overhead, and latency. \newline
	\indent Verma \textit{et al.} \cite{verma2020mitigation} proposed the Secure-RPL mitigation scheme for DIS flooding attacks in RPL-based 6LoWPAN networks. The effects of the DIS flooding attack on network performance are measured in this technique. When a DIS attack occurs, SecRPL prevents the unnecessary reset of the trickle timer and reduces control message transmission. Power consumption and control packet overhead are used to evaluate performance. The authors utilized Cooja, a network simulator designed for Contiki-based devices. Cooja's power tracker is a tool for obtaining radio event information for each node, such as total time radio on (ON), transmitting radio (TX), receiving radio (RX), and interfered radio (INT). The authors proposed developing a defense mechanism to mitigate future DIO suppression attacks. \newline
	\indent Abhinaya \textit{et al.} \cite{abhinaya2021secure} proposed a secure RPL (SRPL) protocol for detecting and mitigating DIS flooding attacks. The concept of load balancing and route discovery is used in this approach, which distributes loads across different nodes in DODAG to avoid DIS flooding attacks and improve network life span. The main idea is to send DIS messages within the safe interval with the maximum allowed request while taking the remaining energy level of sensor nodes into account. The performance is measured in terms of end-to-end delay, packet loss ratio, control message overhead, and overall packet delivery ratio and network lifespan.  \newline
	\indent Sahay \textit{et al.} \cite{sahay2021novel} proposed a novel partitioning attack (NPA) where a malicious node isolates the other nodes in the IoT-LLN from the root node. This attack can be carried out by omitting the route registration process during node joining and DAG maintenance. The author also proposed a defense mechanism against NPA, by making DAO acknowledgment mandatory when a node joins the RPL network. To prove the proposed solution, an analytical model is proposed, and validation is performed using a simulation process. The proposed method may impact network convergence and is best suited for RPL secure mode operation.  \newline
	\indent Baghani \textit{et al.} \cite{baghani2021dao} proposed a novel attack called DAO induction attack where a malicious node provokes other nodes to send special control messages on a regular basis in the network, affecting the packet delivery ratio and end-to-end delay. To avoid this situation, a lightweight, reactive defense mechanism has been proposed where nodes can only accept DAO Trigger Sequence Number (DTSN) updates from neighboring nodes except the DAO root node. The proposed solution adds no overhead to the network and achieves a high detection rate. The author will investigate a new attack by combining DAO induction and Sybil attacks and evaluate the effect of DAO induction on a real-world multi-hop network.  \newline
	\indent Medjek \textit{et al.} \cite{medjek2021multicast} proposed a novel Multicast DIS (M-DIS) attack and examined its impact on control message overhead and power consumption in static and dynamic networks. RPL-MRC is a proposed defense mechanism that avoids multicast DIS messages by introducing the maximum response code (MRC) field in RPL DIO messages. A response delay mechanism is added to the dis-input function, and a timer readjustment mechanism is added to the new DIO interval function in the proposed mitigation scheme. The proposed defense mechanism reduces energy consumption and control message overheads significantly. \newline
	\indent Alsukayti \textit{et al.} \cite{alsukayti2022lightweight} proposed CDRPL, a collaborative and distributed security mechanism for the detection and mitigation of version number attacks in RPL. The proposed scheme ensures quick and accurate attack detection by maintaining performance parameters such as network stability, overhead, and energy consumption. The CDRPL has a lightweight collaborative verification mechanism that allows nodes to collaborate to maintain the legitimacy of received DIO messages and ensure the RPL network global repair mechanism accepts only legitimate VN updates. The packet delivery ratio, network stability, response latency, and accuracy are used to assess performance. In terms of network overhead and power consumption, CDRPL outperforms the existing SRPL-RP \cite{a2020detection}, RPL + Shield \cite{arics2019new}, but it suffers from control message overhead. \newline
	\indent Nandhini \textit{et al.} \cite{nandhini2022lightweight} proposed a lightweight rank attack detection and isolation mechanism known as RAD against internal rank attacks. The RAD (Rank attack detection) mechanism used non-cryptographic hash values to prevent rank attacks while maintaining control message integrity. The proposed mechanism modifies and integrates rank, parent rank, and calculated hash values in DAO messages. These values are compared to the values stored in the information table upon receiving DAO messages by the root node; if a mismatch occurs, an alarm is generated. The average packet delivery ratio, control packet overhead, and delay are used to assess performance. When compared to SRPL-RP \cite{a2020detection}, LEADER \cite{karmakar2021leader}, and SBIDS \cite{shafique2018detection}, this approach achieves 96\% accuracy, and DAO sampling technique is used to improve the accuracy as well as reduce the energy consumption by 40\%. 
	\begin{table}[ht]
		\caption{RPL Specification Based Defense Mechanism for RPL-based Routing Attacks}
		\scriptsize
		\resizebox{\linewidth}{!}{
			\begin{tabular}{l p{.5cm} c c c p{3cm} p{7cm} c c c c c}
				\toprule
				\textbf{Ref} & 
				\textbf{Year} & 
				\multicolumn{3}{c}{\textbf{Scope}} & 
				\textbf{Attack Focused} & 
				\textbf{Methods/Algorithm} & 
				\multicolumn{5}{c}{\textbf{Performance Assessment}} \\
				\cmidrule(rl){3-5} 
				\cmidrule(rl){8-12} 
				\textbf{} &
				\textbf{} &
				\textbf{D\tnote{1}} &
				\textbf{C\tnote{2}} &
				\textbf{M\tnote{3}} &
				\textbf{} &
				\textbf{} &
				\textbf{P1\tnote{4}} &
				\textbf{P2\tnote{5}} &
				\textbf{P3\tnote{6}} &
				\textbf{P4\tnote{7}} &
				\textbf{P5\tnote{8}}
				\\
				\midrule
				\cite{le2016specification} & 2016 & \ding{51} & \ding{53} & \ding{51} &	Rank, Sinkhole, Local Repair, DIS Flooding  &	Semi-auto Profiling Technique &	L & - &	H &	N &	L \\ 
				\cite{ghaleb2018addressing}  & 2018	& \ding{53} & \ding{53} & \ding{51} & DAO Insider & Restrict the number of Forwarded DAO by parent and destination &	H &	- &	- &	N &	H \\
				\cite{shafique2018detection}  & 2018	& \ding{51} & \ding{53} & \ding{53} &	Rank Attack & Encryption of DAO Massage and Rank Rule &	L &	- &	H &	Y &	H \\ 
				\cite{arics2019new} & 2019	& \ding{53} & \ding{53} & \ding{51} &	Version Number	& Elimination of VN updates and Shield to Analyzing DIO Message	& L &	Y &	H &	N &	L \\ 
				\cite{kaliyar2020lidl} & 2020 & \ding{51} & \ding{51} & \ding{51} &	Sybil, Wormhole & Highest Rank Common Ancestor & L & Y & H & N & L \\
				\cite{zaminkar2020sos} & 2020	& \ding{51} & \ding{53} & \ding{53} &	Sinkhole & Node Rating and Ranking Mechanism and Average Packet Transmission RREQ & - &	- &	H &	N &	-  \\ 
				\cite{wadhaj2020mitigation} & 2020 & \ding{53} & \ding{53} & \ding{51} & DAO Insider  &	SecRPL1 - Restrict the number of Forwarded DAO per Child and SecRPL2 by specific node	& H & - & - & N & H \\ 
				\cite{verma2020mitigation} & 2020 & \ding{51} & \ding{53} & \ding{51} &	DIS Flooding  & RPL Constant - Safe DIS Transmission Interval and Max Permitted DIS Request & L & - & - & Y & L \\ 
				\cite{medjek2021multicast} & 2021	& \ding{53} & \ding{53} & \ding{51} &	Multicast DIS Flooding & Response Delay and Timer Readjustment, Maximum Response Code inspired by Multi-cast Listener Queries &	L &	- &	- &	Y &	L \\ 
				\cite{boudouaia2021rpl} & 2021	& \ding{53} & \ding{53} & \ding{51} &	Divide and Conquer & Threshold of Minimum Rank and Maximum Rank &	L &	- &	H &	N &	L \\
				\cite{abhinaya2021secure}  & 2021 & \ding{51} & \ding{53} & \ding{51} &	DIS Flooding  & Safe DIS Communication Interval, Max Permitted DIS Request, Remaining Energy Level of Sensors & L & - & - &	Y &	L \\
				\cite{sahay2021novel} & 2021 & \ding{53} & \ding{53} & \ding{51} & Novel Partitioning Attack & Route Registration Process and DAO Acknowledgment &	L &	- &	- &	- &	- \\ 
				\cite{baghani2021dao} & 2021 & \ding{51} & \ding{53} & \ding{51} &	DAO Induction &	Monitoring DTSN update at Root & N &	Y & H &	N &	- \\ 
				\cite{sahay2022mitigating}  & 2022	& \ding{53} & \ding{53} & \ding{51} &	Worst Parent Selection & Optimal Parent Set during Topological Construction Phase &	L &	- &	- &	N &	L\\ 
				\cite{kiran2022ids} & 2022 & \ding{51} & \ding{53} & \ding{53} & Worst Parent Selection  &	NMapper compare the rank of parent with neighbors & M & - &	H &	N &	-\\
				\cite{bang2022embof}  & 2022	& \ding{51} & \ding{53} & \ding{51} &	Rank Attack	&	Echelon Metric Based Objective Function & L	& Y & H & N & L\\
				\cite{alsukayti2022lightweight}  & 2022 & \ding{51} & \ding{53} & \ding{51} & Version Number & Collaborative Verification of DIO & L & Y & H & N & L\\ 
				\cite{nandhini2022lightweight} & 2022 & \ding{51} & \ding{53} & \ding{51} &	Rank attack & Non cryptographic hash & L & Y & H & N & L \\ 
				\cite{nandhini2022enhanced} & 2022 & \ding{51} & \ding{53} & \ding{51} &	Rank attack & Rate limiting of DIO & L & Y & H & N & L \\
				\cite{alsirhani2022securing} & 2022 & \ding{51} & \ding{53} & \ding{51} &	Spam DIS & Number of incoming DIS & - & Y & H & N & L\\
				\cite{goel2023cra} & 2023 & \ding{51} & \ding{53} & \ding{51} &	Dropped DAO & Challenge Response Pair based Authentication of DAO-ACK  & L & Y & H & Y & M\\
				\bottomrule
			\end{tabular}
		}
		\label{tbl_rplspecificarion}
		\begin{tablenotes}
			\item D: Detection
			\item C: Characterization
			\item M: Mitigation
			\item P1 : Overhead
			\item P2 : Lightweight
			\item P3 : Detection Rate
			\item P4 : Mobility
			\item P5 : Energy Consumption
			\item Y: Yes and N: NO
			\item L: Low, M: Medium, and H: High
			\item - : Not Defined 
		\end{tablenotes}
	\end{table}
	\newline
	\indent Nandhini \textit{et al.} \cite{nandhini2022enhanced} proposed enhanced rank attack detection (E-RAD) mechanism for early detection and isolation of rank in an RPL network. The proposed E-RAD is based on limiting the rate of DIO control message generation and solicitation of DIS control messages to isolate rank attacker nodes. The rate-limiting trickle algorithm concept, where each node maintains the DIO counter and its trickle timer, is used to measure the consistency of DIO messages. This method also used a non-cryptographic hash value and rank to check for inconsistencies in DAO messages to identify malicious nodes. The performance of the system is measured in terms of packet delivery ratio, energy consumption, and control message overhead. E-RAD has higher accuracy than SRPL-RP \cite{a2020detection} and SBIDS \cite{shafique2018detection}, both of which are 97.23\% in grid-centered topology. \newline
	\indent Alsirhani \textit{et al.} \cite{alsirhani2022securing} proposed a DIS Spam Attack Mitigation (DISAM) approach against the Spam DIS attack in the RPL-Blockchain based Internet of things. The proposed DISAM approach runs on each node in the RPL network distributively. The idea behind this approach is to keep track of the number of incoming DIS messages with no DAO received from neighbor nodes wishing to join the DODAG, and if the count exceeds the defined threshold, it indicates that there is a spam DIS request in the network. As a mitigation scheme, the victim nodes reject or discard all incoming DIS messages for a set period of time. The performance is measured in energy consumption, and a high detection rate is obtained in case of low DIS packet injection. \\ \indent 
	Goel \textit{et al.} \cite{goel2023cra} proposed a challenge-response authentication based detection mechanism against the Dropped Destination Advertisement attack (DDAO) in RPL-based IoT known as CRA-RPL. They modified the control message and used the reserved field to implement a challenge-response pair mechanism to authenticate the DAO-ACK received in response to the DAO message in the RPL non-storing mode. The Prime Sequence Codes generate the challenge-response pairs based on the prime number used by nodes to identify the false DAO-ACK. The proposed approach is lightweight in nature and achieves a high detection rate while effectively restoring network performance.
	\subsubsection{RPL Features}
	This section will review defense mechanisms designed using RPL features such as version, rank, and objective function.  \newline
	\indent Kaliyar \textit{et al.} \cite{kaliyar2020lidl} proposed a LiDL defense mechanism against Sybil and wormhole attacks based on the highest-rank common ancestor (HRCA). The main idea of detecting the attack is to finding the ancestor with the highest rank in the network tree. The proposed approach detects and localizes the attack early, speeding up the mitigation process. The three functions used in proposed approach:initial joining, periodic timer verification, applying the detection algorithm at the time of periodic timer expiry and sending the alarm message, and performing the mitigation process based on that. The proposed approach provides rapid mitigation with low overhead. \newline
	\indent Zaminkar \textit{et al.} \cite{zaminkar2020sos} proposed a new protocol called SoS-RPL, built on top of RPL and is used to detect sinkhole attacks. SoS-RPL is divided into two sections: the first performs node ranking and rating based on distance measurement, and the second identifies misbehaving nodes using average packet transmission RREQ (APT-RREQ). To detect the sinkhole attack, the main focus is on the rank present in the DIO message. The DIO message is malicious if the rank difference between the node and the source node is greater than the node and its parent node. Similarly, APT-RREQ detects fake RREQ packets generated by malicious nodes. The proposed detection method has a detection rate of greater than 96.19\%. \newline
	\indent Boudounia \textit{et al.} \cite{boudouaia2021rpl} proposed a defense mechanism in the RPL for Divide and Conquer-based (DCB) attacks. The DCB attack occurs when a malicious node simultaneously increases and decreases its rank in order to attack traffic and then chooses the worst parent to forward the entire traffic, resulting in an un-optimized path in the network. The proposed defense mechanism makes use of the concept of rank thresholds with minimum (\textit{ThRankmin}) and maximum (\textit{ThRankmax}) values. If the node's rank is greater than the maximum rank threshold but less than the minimum rank threshold, it is malicious. This method's performance is measured in average network hops, global energy consumption, and detection rate. \newline
	\indent Kiran \textit{et al.} \cite{kiran2022ids} proposed a defense mechanism to detect worst parent selection (WPS) attacks. This is the first paper to implement the worst parent selection attack in RPL-based IoT networks by modifying the objective function of the malicious node to select the worst-rank node as the parent. DWA-IDS, an extension of \cite{raza2013svelte}, compares the rank of parent nodes with neighbor nodes and generates an alarm message if it is greater than \textit{MinHopRankIncrease} to mitigate the worst parent selection attack. The proposed approach had the lowest overhead while achieving a true positive rate of up to 95\% and a detection rate of 100\%. 
	\newline
	\indent Aris \textit{et al.} \cite{arics2019new} proposed two lightweight mitigation mechanisms, one based on elimination and the other on shield against version number attack. To mitigate the majority of version number attacks, the elimination mechanism restricts the VN updates from the leaf node. The shielding mechanism, VN update, is only acceptable if most neighbors claim the network's better position in terms of rank. These two defense mechanisms have been tested in four different network topologies, and both reduce the effect of a version number attack. The proposed approach significantly improves performance parameters such as delay by up to 87\%, average power consumption by up to 63\%, control message overhead by up to 71\%, and packet delivery ratio by up to 86\%. \newline
	\indent Bang \textit{et al.} \cite{bang2022embof} proposed a lightweight detection and mitigation scheme against rank attacks based on echelon metric-based objective function RPL (EMBOF-RPL). The EMBOF-RPL improved the functionality of the DIO message by including the Echelon metric within the DIO packet; additionally, it will aid in the routing decision. The parent selection is based on the echelon metric, with a ten-second interval for each node that wishes to join the DODAG. Isolation latency, detection accuracy, packet delivery ratio, end-to-end delay, memory overhead, and power consumption are all used to assess performance. The proposed approach outperformed SVELTE \cite{raza2013svelte}, SecTrust \cite{airehrour2019sectrust}, and SBIDS \cite{shafique2018detection}.
	\subsubsection{Node Specification}
	\indent Le \textit{et al.} \cite{le2016specification}  proposed a specification-based intrusion detection system (IDS) against network topology-based attacks in RPL, such as rank attacks, sinkhole attacks, local repair attacks, neighbor attacks, and DIS attacks. To build the rules for an intrusion detection agent, a semi-auto profiling technique is used to construct high-level specifications that include legitimate protocol states and transition statistics (IDA). There are two phases: the first creates an extended finite state machine (EFSM) from RPL traces using inductive logic programming (ILP) techniques, and the second translates RPL profile knowledge for the detection of attacks implemented on IDS agents. The performance is evaluated using a true positive rate of 100\% reported with a 6.3\% overhead. \newline
	\indent Sahay \textit{et al.} \cite{sahay2022mitigating} proposed an Enhanced RPL (ERPL) mitigation scheme for the worst parent selection attack, based on the concept of selecting only those parents in the candidate set that have the best path towards the sink node during the topological construction phase. The novel function generates the optimal parent set by ignoring DIO messages with inferior sink paths and decision variables that include the neighboring node in the candidate parent set. In terms of energy consumption, network convergence, packet delivery ratio, and network overhead, the proposed scheme outperforms. The proposed mechanism is best suited for packet delivery ratios, a primary focus in IoT applications.
	\subsection{Intrusion Detection System (IDS) Based Defense Mechanism} \label{subsec_IDS_based}
	An intrusion detection system (IDS) can send an alert whenever malicious activity is detected in the network traffic. This section provides an overview of intrusion detection-based defense mechanisms, divided into two major categories: classical and modern. \subsubsection{Classical IDS}
	The classical IDS relies on traditional methods to detect routing attacks, classified as an anomaly, signature, specification, and hybrid. Anomaly-based IDS first creates a normal behavior profile based on network traffic and then compares network activities to the created profile to identify anomalies in terms of known routing attacks. Signature-based IDS uses a previously stored signature to identify a routing attack. Specification-based IDS is similar to anomaly-based IDS, but it creates a network profile rather than a behavior profile to detect routing attacks. Hybrid IDS is created by combining two or more traditional IDS. \newline
	\indent Shin \textit{et al.} \cite{shin2019detection} proposed a novel anomaly-based intrusion detection system to detect packet-dropping attacks. The proposed approach used a distributed attack detection and a centralized network decision-making mechanism for alarm generation. The malicious packet-dropping attack detection process computes packet-dropping probabilities for each node, which are calculated as the ratio of the number of packets sent and received by a node. If a node's packet-dropping probability exceeds the sum of the average packet-drop probability and the threshold specified in the time window, the node is classified as malicious. The central mode sends an instruction to remove a malicious node from the network. The proposed method had a high detection rate of 94\%. \newline
	\indent Nikravan \textit{et al.} \cite{nikravan2018lightweight} proposed a decentralized lightweight online/offline identity-based signature for detecting rank spoofing and version number attacks. To detect attacks, five algorithms are used: setup, extract, OffSign, OnSign, and UnSign. Setup and extraction create public parameters and private keys for nodes. UnSign verifies signatures, while OnSign and OffSign generate online and offline signatures. This approach is compared to VeRA \cite{dvir2011vera}, and TRAIL \cite{perrey2016trail}, and the proposed signature is designed to remain unaffected by variations in the network size and the individual nodes' fixed algorithms which ensures the scalability of the solution.
	The proposed approach's performance is measured in energy consumption and computational cost, which are extremely low. \newline
	\indent Althubaity \textit{et al.} \cite{althubaity2020specification} proposed FORCE (Forged Rank and Routing Metric Detector), a distributed specification-based intrusion detection system for Rank-related attacks such as Decrease Rank, Worst Parent Selection, and Rank attack using Objective Function. The key idea behind this approach is that each node analyses the control messages (DIO and DAO) received during routing from neighbors locally and generates an alert if one of the IDS-defined rules is violated. The detection rate, true positive rate, and false positive rate are used to assess performance. The authors claimed that the proposed approach has a 100\% detection rate and a low overhead when compared to SVELTE \cite{raza2013svelte}. Overall, the proposed method is lightweight, has a high detection rate, and is appropriate for resource-constrained networks. A new algorithm will be combined with FORCE in the future to detect more attacks.
	\begin{table}[!]
		\caption{Intrusion Detection System Based defense Mechanism for RPL-based Routing Attacks}
		\scriptsize
		\resizebox{\linewidth}{!}{
			\begin{tabular}{l p{.5cm} c c c p{3cm}  p{7cm} c c c c c}
				\toprule
				\textbf{Ref} & 
				\textbf{Year} & 
				\multicolumn{3}{c}{\textbf{Scope}} & 
				\textbf{Attack Focused} & 
				\textbf{Methods/Algorithm} & 
				\multicolumn{5}{c}{\textbf{Performance Assessment}}\\
				\cmidrule(rl){3-5} 
				\cmidrule(rl){8-12} 
				\textbf{} &
				\textbf{} &
				\textbf{D\tnote{1}} &
				\textbf{C\tnote{2}} &
				\textbf{M\tnote{3}} &
				\textbf{} &
				\textbf{} &
				\textbf{P1\tnote{4}} &
				\textbf{P2\tnote{5}} &
				\textbf{P3\tnote{6}} &
				\textbf{P4\tnote{7}} &
				\textbf{P5\tnote{8}} \\
				\midrule
				\cite{raza2013svelte} & 2013 & \ding{51} & \ding{53} & \ding{53 } & Spoofed, Selective Forwarding, Sinkhole & 6Mapper-Network Graph Inconsistency Detection, Detect Filtered Node, Finding Rank Inconsistency, and Mini-firewall &	L &	- &	M &	N &	L \\ 
				\cite{surendar2016indres}  & 2016 & \ding{51} & \ding{53} & \ding{51} & Sinkhole & Evidence Theory, and Ranking & M & - & - & N & M \\ 
				\cite{khan2017trust} & 2017 &	\ding{51} & \ding{53} & \ding{51} & Selective Forwarding, Sinkhole, Version Number	& Trust Evaluator (Direct Trust Values), Neighbor, Clustered or Tree based Trust dissemination &	- &	- &	H &	N & -\\
				\cite{bostani2017hybrid} & 2017 & \ding{51} & \ding{53} & \ding{53} & Sinkhole, Selective Forwarding & Specification and Anomaly Agents Based IDS, Voting Mechanism and Optimal Path Forest  &	H &	N &	- &	N &	-\\
				\cite{arshad2018colide} & 2018 & \ding{51} & \ding{53} & \ding{53} &	Multi-Stage Attack & Node Level Module, Edge Router Module & L & Y & - & N & L \\
				\cite{nikravan2018lightweight}  & 2018 & \ding{51} & \ding{53} & \ding{53} & Version Number, Rank 	& Identity Based Offline and Online Signature &	L	& Y & - & N &	L \\
				\cite{mirshahjafari2019sinkhole+} & 2019	& \ding{51} & \ding{53} & \ding{53} &  Sink-Clone & 6LoWPAN Mapper and IDS Module-rankID & M & - & H & N & M \\
				\cite{shin2019detection} & 2019 & \ding{51} & \ding{53} & \ding{53} &	Packet Dropping Attack & Packet Drop Probability & - &	- &	H &	N &	- \\
				\cite{althubaity2020specification} & 2020	& \ding{51} & \ding{53} & \ding{51} & Worst Parent Selection, Decrease/Increase Rank, Rank Attack using OF & Forged Rank and Routing Metric Detector, Parent Child Relation Ship & M & Y &	H & N & L \\
				\cite{verma2020cosec}  & 2020 & \ding{51} & \ding{53} & \ding{51} & Copycat & Outlier Detection-Interquartile Range (IQR) & M & - & H & Y & - \\ 
				\cite{gothawal2020anomaly} & 2020 & \ding{51} & \ding{53} & \ding{53} &	Selective Forwarding, Rank Attack, Local Repair, Neighbor Attack & Stochastic and Evolutionary Game Models & H & - & H &	N &	H \\
				\cite{violettas2021softwarized} & 2021 & \ding{51} & \ding{51} & \ding{51} & 13 Attacks* &	 Software-Defined Networking & L & Y & H & N & M \\
				\cite{agiollo2021detonar} & 2021 &	\ding{51} & \ding{53} & \ding{53 } & RADAR Dataset & Packet Sniffing Approach, Auto Regressive Integrated Moving Average &	L &	- &	H &	N &-\\
				\cite{ray2023novel} & 2023 & \ding{51} & \ding{51} & \ding{53 } & Rank, Version Number  & Active Probing and Discrete Event Modeling  & L & Y &	H &	N & L \\
				\cite{deveci2023evolving} & 2023 &	\ding{51} & \ding{53} & \ding{53} & Worst Parent, Increased Version, Hello Flood, Decreased Rank & Genetic Programming Tree based IDS & L & Y & H & N & L\\
				\cite{bhale2024hybrid} & 2024 &	\ding{51} & \ding{53} & \ding{51} & Sinkhole & SHAP and Hidden Markov Model & - & Y & H & - & L\\ 
				\bottomrule
			\end{tabular}
		}
		\label{tbl_IDS}
		\begin{tablenotes}
			\item D: Detection
			\item C: Characterization
			\item M: Mitigation
			\item P1 : Overhead
			\item P2 : Lightweight
			\item P3 : Detection Rate
			\item P4 : Mobility
			\item P5 : Energy Consumption
			\item Y: Yes and N: NO
			\item L: Low, M: Medium, and H: High
			\item - : Not Defined 
		\end{tablenotes}
	\end{table}
	\newline
	\indent Raza \textit{et al.} \cite{raza2013svelte} proposed a real-time intrusion detection mechanism named SVELTE to detect spoofed or altered, selective-forwarding, and sinkhole types of routing attacks. SVELTE is made up of two major components: a 6LoWPAN mapper that is used to create RPL routing states and an IDS module that uses a hybrid approach and is integrated with a mini-firewall to protect against global attacks. Network graph inconsistency, checking node availability, and routing validity components are present in the IDS modules for detection purposes. The detection rate and true positives were evaluated as performance parameters. The detection rate is 100\% for some settings of experiments and the reported true positive rate is 90\% for some setting indicating the 10\% false alarms in some cases. SVELTE can be extended to detect wormhole attacks by extending the 6LoWPAN mapper with signal strength information for each node, as well as Clone ID and Sybil attacks by adding location information that can be used to build the network's physical map.\newline
	\indent Bostani \textit{et al.}\cite{bostani2017hybrid} proposed a real-time anomaly and specification agent-based IDS mechanism to detect Sinkhole and selective forwarding attacks. The specification agent IDS (SA-IDS) sends information about potential malicious nodes by analyzing input and output traffic to 6BR. An anomaly agent-based IDS (AA-IDS) uses a MapReduce-based approach, runs on 6BR, and performs clustering based on an optimal path forest algorithm to detect anomalies. Based on the local and global decisions of SA-IDS and AA-IDS, the root node makes the final decision about anomalies based on the voting mechanism. The performance was evaluated in terms of true positive rate, false positive rate, and accuracy. The proposed hybrid approach can be deployed in the smart city environment. \newline
	\indent Arshad \textit{et al.} \cite{arshad2018colide} proposed a Collaborative Intrusion Detection (COLIDE). In this effective signature-based IDS, intruders are detected by collaborating with sensor node devices and border node routers. The IDS has two major components: node-level detection and edge router detection. A detection engine at the node level detects routing attack attempts based on existing signatures. An edge router detection system consists of three components: an alert collector, a correlation agent, and a detection agent. An alert collector gathers alerts from node-level monitoring components by communicating with IoT devices. Correlation agents provide countermeasures by correlating malicious events at the network and system levels as monitored by node-level monitors. The detection agent used an anomaly-based approach to detect attacks based on alerts collected and correlated in the IDS. The performance metrics are energy and processing overhead. \newline
	\indent Mirshahjafri \textit{et al.} \cite{mirshahjafari2019sinkhole+} proposed a Sink-Clone hybrid attack, a combination of the Sinkhole and Clone ID attacks, and a hybrid IDS to detect the attack. In the hybrid Sink-Clone attack, a group of malicious nodes with the same identity displays the lowest routing cost and entices all adjacent nodes to forward their packets through them. To detect a Sink-Clone attack, the proposed hybrid approach combines the SVELTE \cite{raza2013svelte} and sinkhole detection approaches. IDS consists of two components: 6LowPAN Mapper and IDS modules, which contain an additional rankID module to detect Clone ID attacks. The proposed method reduces false positives while increasing the detection rate. The detection rate is reported to be 100\%, decreasing as the number of nodes increases. \newline
	\indent Agiollo \textit{et al.} \cite{agiollo2021detonar} proposed a signature and anomaly-based intrusion detection system with a high attack detection rate to detect 14 well-known RPL attacks using a packet sniffing approach. This work used the RADAR - Routing attack dataset for RPL, which contains traces of 14 RPL attacks on 16 static scenarios. In this method, 11 features are chosen for attack detection, an auto-regressive moving average model is used for anomaly detection, and a set of rules is used to classify the attacks into one of 14 categories. \newline
	\indent Violettas \textit{et al.} \cite{violettas2021softwarized}  proposed an ASSET, a software-defined network-based intrusion detection system, to detect, characterize, and mitigate 13 types of RPL attacks. ASSET comprises four specifications and three anomaly-based mechanisms for detecting attacks and implementing various mitigation strategies. ASSET is a three-tiered architecture that includes a data communication plane, a control plane, and an application plane. The data communication plane includes an RPL protocol stack, cross-level configuration hooks, control packet statics, and node-level anomaly detection. The control plane is in charge of communication statics. The application plane includes a graphical user interface (GUI) for IDS visualization and configuration. ASSET's performance demonstrates an adaptable intrusion detection system with minimal communication overhead. 
	\subsubsection{Modern IDS}
	The modern IDS builds on non-traditional methods such as probability, evidence theory, self-organizing maps, and software-defined networks. \newline
	\indent Surendar \textit{et al.}  \cite{surendar2016indres}  proposed a Sinkhole detection and response system (InDRes). In this method, a malicious node is first detected, then isolated from the network, and finally, the entire topology is rebuilt. To compute evidence value, the Dempster-Shaffer evidence theory and probability distribution functions are used. InDRes architecture includes: leader node selection, calculating the packet drop count, detecting malicious nodes based on evidence, detecting malicious nodes based on ranking, and steps for detecting malicious nodes based on evidence value. The packet drop ratio, packet delivery ratio, normalized overhead, average energy consumption, and throughput are all used to assess performance. Based on the outcome, this approach improves the disadvantage of INTI \cite{cervantes2015detection} and SVELTE \cite{raza2013svelte}. The author suggests developing behavioral rules based on numerical analysis and employing an optimization techniques approach in the future. \newline
	\indent Khan \textit{et al.} \cite{khan2017trust} proposed a trust-based IDS to detect denial of service attacks such as selective forwarding, sinkholes, and version number attacks. There are three algorithms in the proposed IDS: trust evaluator, direct trust of neighboring member nodes, and trust value combination. The trust evaluator calculates trust values based on opinion triangles (belief, disbelief, and uncertainty). The direct trust of neighboring member nodes is defined by the forwarding check, ranking check, and version number check. The border router or cluster head aggregates all the trust values and creates the reputation of nodes; if any node has a higher value of distrust from its neighbors, the node is classified as malicious. The performance is measured in terms of intruders detected false positives, false negatives, undetected positives, and undetected negatives. \newline
	\indent Verma \textit{et al.}\cite{verma2020cosec} proposed CoSec-RPL as an Intrusion Detection System for non-spoofed copycat attacks. The outlier detection technique is used to demonstrate significant node behavior. When a DIO message is received, CoSec-RPL is embedded in the DIO processing method. It measures two threshold values to ensure a safe DIO interval and a block threshold, as well as one additional parameter for fine-tuning the CoSec-RPL. Copycat attacks can be detected by monitoring the difference between two DIO messages. When the DIO message time difference is less than or equal to the DIO interval, the neighbor is labeled malicious and added to the blacklist. As performance indicators for non-spoofed copycat attacks, the packet delivery ratio, average end-to-end delay, and average power consumption were used. \newline
	\indent Gothawal \textit{et al.} \cite{gothawal2020anomaly} proposed a game model-based Intrusion Detection System for detecting rank attacks, local repair attacks, neighbor attacks, and DIS attacks. The proposed approach employs two types of games: stochastic for detecting attacks and evolutionary for confirming attacks. The attack was verified using a clustered network topology. The proposed method operates in two stages: first, observing network topology construction and packet forwarding behavior to derive states, and then, based on the first phase, developing rules to detect attacks. The detection accuracy, throughput, delay, normalized overhead, energy consumption, and packet loss are the performance metrics. The deviation is used for compassion and produces the best results for less than 50 nodes network. \\ 
	\indent Ray \textit{et al.} \cite{ray2023novel} proposed a novel attacker identification mechanism for version and rank attacks that employ the concept of intelligent, active probing and detection via an IDS based on a Discrete Event System (DES). The proposed IDS operates in a centralized manner, receiving input from leaf nodes and classifying traffic behavior as normal or attack using active probing. The IDS is composed of three components: Packet Sniffer, which captures packets (both control and data), RQST\_RSP\_HANDLER(), which extracts meaningful information about network events from packets; and DES Diagnoser, which performs attacker identification based on event information. The proposed IDS performance is evaluated using simulation and testbed experiments on the FIT-IoT LAB platform. The average energy consumption is 14872\textit{mJ}, while the accuracy of attacker detection is very high (99.1\%). \\ \indent 
	Deveci \textit{et al.} \cite{deveci2023evolving} proposed a lightweight IDS against the worst parent attack, hello flooding, decrease rank and increase version number attack based on Genetic Programming (GP). The solution runs by central ID node which collect and aggregate the network traffic periodically received from monitoring nodes in order to detect the attacks which generate the malicious traffic. The main objective of this approach to optimize the detection accuracy and reduce the communication overhead (energy and  memory consumption ) of Intrusion Detection. The high detection accuracy around 92.2\% is achieved after 30\% of generations. \\ \indent
	Bhale \textit{et al.} \cite{bhale2024hybrid} proposed a hybrid IDS for detecting and mitigating sink hole attacks in RPL-based IoT networks. The proposed EaHIDS approach consists of four phases:data collection, feature extraction, application of a lightweight ML model to edge nodes, and blacklisting of malicious nodes. The SHAP approach is used to extract features, while the Hidden Markov Model is used to detect anomalies. The proposed approach was evaluated on the Cooja simulation environment, machine learning datasets such as IRAD, NLA, and FIT-IoT Testbed, and it showed high accuracy and low energy consumption. 
	\subsection{Trust Based Defense Mechanisms} \label{subsec_trust_based}
	Trust-based defense mechanisms establish trust relationships between neighbors based on trust value to mitigate routing attacks. This section gives an overview of trust-based defense mechanisms. 
	\subsubsection{Direct Trust}
	The following section discusses the behavioral trust-based defense mechanisms. \newline
	\indent Cervantes \textit{et al.}\cite{cervantes2015detection} proposed an intrusion detection of sinkhole attack on 6LoWPAN enabled Internet of Things named INTI for sinkhole attack detection. INTI performs four major operations: cluster configuration, routing monitoring, attack detection, and attack isolation. The cluster configuration uses a leader-based clustering of nodes and a beta distribution to predict future behavior based on past activities. Routing performance is monitored by counting the number of incoming and outgoing streams. The attacks were detected using a trust and reputation-based evaluation. Finally, once the malicious nodes have been identified, an alarm is broadcast to notify nearby nodes. This method detected 92\% of static scenarios and 75\% of mobile devices. \newline
	\indent Airehrour \textit{et al.} \cite{airehrour2016securing} proposed a lightweight trust-based routing protocol for RPL to overcome the Blackhole attack. The proposed approach first computes trust for each node, which is the ratio of number of packets delivered and the number of packets received. The computed trust is sorted in decreasing order of values, node with the highest trust value making the routing decision for the optimal path. The throughput and packet loss rate are used to evaluate the performance of the proposed approach. When these two parameters are compared to RPL-MRHOF, the packet loss rate in trust-based RPL is 40\%-60\% versus 60\%-100\% in MRHOF-RPL. To build the defense solution in the future, the author would use energy metrics to monitor the energy level of nodes. \newline
	\indent Pishdar \textit{et al.} \cite{pishdar2021pcc} proposed a trust-based Parent Change Control (PCC)-RPL defense solution against the worst parent attack. The method is divided into two stages: monitoring and detection. In monitoring, all parents continuously observe their children's behavior in terms of preferred parent change. During the detection phase, if any node engages in malicious activity, the parent reduces the trust level of the child nodes and notifies the root node to generate an alert for the suspicious node. Trust levels are classified into five categories ranging from very low to very high. The performance is measured in control overhead, attack detection delay, precision, and energy consumption. In comparison to SVELTE \cite{raza2013svelte}, the PCC-RPL has a lower computational overhead and power consumption. \newline
	\indent Prathapchandran \textit{et al.} \cite{prathapchandran2021trust} proposed a lightweight trust-based model named RFTrust that uses the concepts of Random Forest (RF) and Subjective Logic (SL) for sinkhole attack detection and isolation. Trust metrics, calculated based on delay, packet delivery ratio, energy consumption, and honesty, have been used to detect attacks. Similarly, the isolation of malicious nodes was performed using the proposed RFTrust model, with the help of RF in the case of direct trust and SL in the case of indirect trust. The RF algorithm classified the nodes as either trusted or malicious. In comparison to existing approaches SoS-RPL \cite{zaminkar2020sos}, INTI  \cite{cervantes2015detection}, InDRes \cite{surendar2016indres}, the proposed model is more efficient in terms of false positive rate, false negative rate, and accuracy.
	\newline
	\indent Bang \textit{et al.} \cite{bang2021novel} proposed a novel decentralized architecture for mitigating Sybil attacks (SA-1, SA, and SA-3) in the context of a smart home. The hybrid approach employed the concept of parent selection based on each node's trust value and geographical location. Malicious nodes are identified by the use of specially configured nodes known as monitoring nodes, which monitor network traffic and store the data in trace tables. This approach outperforms the existing approaches SecTrust \cite{airehrour2019sectrust} and LiDL \cite{kaliyar2020lidl} in terms of control message overhead, average power consumption, packet delivery ratio, and accuracy. The proposed method achieved up to 100\% accuracy. In the future, the functionality of monitoring nodes will be expanded to mitigate multiple or hybrid attacks.  \newline
	\indent Patel \textit{et al.} \cite{patel2021trust} proposed and implemented a lightweight trust-integrated RPL protocol (TRPL) against the Blackhole attack. This approach calculates trust value based on successful interaction between two nodes and is used in the RPL protocol parent selection process. The trust value is calculated on a regular basis for each node and is embedded in the routing decision. TRPL also builds feedback chains between nodes to monitor node behavior. This method detects and isolates the Blackhole attack while consuming no power or increasing network overhead. TRPL met two objectives: first, high detection accuracy, and second, improved data delivery ratio. In the future, the TRPL will be enhanced to detect selective forwarding attacks.  \newline
	\indent Kim \textit{et al.} \cite{kim2022physical} proposed PITrust. This trust path routing mechanism uses a physical identification mechanism based on RSSI and a centralized trust scheme to detect Sybil attacker nodes in an RPL network. The PITrust procedure is divided into three steps: RSSI variance observation, RSSI pairwise distance computation (RPD), and path trust based on a modified objective function. When the RSSI variance exceeds the predefined RSSI threshold, an alarm is generated via DIO message for potential sybil attacker nodes. When nodes receive an alarm signal, they compute the RPD, and, based on positive and negative values, they decide whether the node is honest or malicious. Finally, trust is measured using the PITrust algorithm based on the trust parameter in the objective function. The performance metrics are packet delivery ratio, detection latency, energy consumption, and communication overhead. In the future, a new security scheme for coordinated cyber attacks on IoT networks will be developed. 
	\begin{table}[!]
		\caption{Trust Based Defense Mechanism for RPL-based Routing Attacks}
		\scriptsize
		\resizebox{\linewidth}{!}{
			\begin{tabular}{l p{.5cm} c c c p{3cm}  p{6.5cm} c c c c c}
				\toprule
				\textbf{Ref} & 
				\textbf{Year} & 
				\multicolumn{3}{c}{\textbf{Scope}} & 
				\textbf{Attack Focused} & 
				\textbf{Methods/Algorithm} & 
				\multicolumn{5}{c}{\textbf{Performance Assessment}}\\
				\cmidrule(rl){3-5} 
				\cmidrule(rl){8-12} 
				\textbf{} &
				\textbf{} &
				\textbf{D\tnote{1}} &
				\textbf{C\tnote{2}} &
				\textbf{M\tnote{3}} &
				\textbf{} &
				\textbf{} &
				\textbf{P1\tnote{4}} &
				\textbf{P2\tnote{5}} &
				\textbf{P3\tnote{6}} &
				\textbf{P4\tnote{7}} &
				\textbf{P5\tnote{8}} \\
				\midrule
				\cite{cervantes2015detection} & 2015 &	\ding{51} & \ding{53} & \ding{51} & Sinkhole & Watchdog, Reputation and Trust-Beta Distribution & - & - & H & Y & - \\ 
				\cite{airehrour2016securing} & 2016 & \ding{53} & \ding{53} & \ding{51} &	Blackhole & Compute Trust for Optimal Routing Decision, Effective Feedback & 	 L &	- &	- &	N &	-\\
				\cite{airehrour2019sectrust}  & 2019 & \ding{51} & \ding{53} & \ding{51} &	Rank Attack,  Sybil & Direct Trust and Recommended Trust & L & - &	H &	N &	- \\ 
				\cite{thulasiraman2019lightweight} & 2019 & \ding{51} & \ding{53} & \ding{53} &	Sybil, DoS & Trust Value and Average Received Signal Strength Indicator &	L &	Y &	- &	Y &	- \\ 
				\cite{ul2021ctrust} & 2021 & \ding{51} & \ding{53} & \ding{51} & Blackhole & Trust on Forwarding behavior of Node and Trust Calculation on Controller &	L &	- &	H &	- &	L \\
				\cite{pishdar2021pcc} & 2021 & \ding{51} & \ding{53} & \ding{53} & Sinkhole,  Blackhole, Rank Reduction/Promotion, Wormhole  & Parent Change Control &	L &	- &	H &	N &	L \\
				\cite{bang2021novel} & 2021	& \ding{51} & \ding{53} & \ding{51} & Sybil & Geographical Location of Node and Trust based Parent Selection & L & Y & H & Y & M \\ 
				\cite{patel2021trust} & 2021 & \ding{51} & \ding{53} & \ding{51} &	Blackhole &	Trust Value based on successful interaction between two nodes &	N	& -	& - & N	& L \\ 
				\cite{prathapchandran2021trust}  & 2021 & \ding{51} & \ding{53} & \ding{51} & Sinkhole &	Random Forest and Subjective Logic &L &	Y &	H &	N &	L \\ 
				\cite{patel2022reputation}  & 2022	& \ding{51} & \ding{53} & \ding{51} &	Selective Forwarding & Reputation based in Forwarding behavior of Node &	N & Y & - &	N &	L\\
				\cite{jiang2022secure}  & 2022 & \ding{51} & \ding{53} & \ding{51} &	Selective Forwarding & Self Trust on the basis of forwarding behavior of packets & - & Y & H & N & L\\
				\cite{kim2022physical} & 2022 & \ding{51} & \ding{53} & \ding{53} & Sybil attack  & Trust Path Routing based in RSSI & L & - & H & N & L\\
				\bottomrule
			\end{tabular}
		}
		\label{tbl_trust}
		\begin{tablenotes}
			\item D: Detection
			\item C: Characterization
			\item M: Mitigation
			\item P1 : Overhead
			\item P2 : Lightweight
			\item P3 : Detection Rate
			\item P4 : Mobility
			\item P5 : Energy Consumption
			\item Y: Yes and N: NO
			\item L: Low, M: Medium, and H: High
			\item - : Not Defined 
		\end{tablenotes}
	\end{table}
	\subsubsection{Indirect Trust}
	The following section discusses the trust metric-based defense mechanisms. \newline
	\indent Airehrour \textit{et al.} \cite{airehrour2019sectrust} proposed and implemented SecTrust-RPL, a time-based trust-aware RPL routing protocol for detecting and isolating rank and sybil attacks. The proposed approach is an extension of \cite{airehrour2016securing} work, and each node in the network calculates time-based trustworthiness based on successful packet exchange between nodes and positive feedback of packet exchange. The decision for attack detection and isolation is based on the direct and recommended trust of neighbor nodes. Compared to MRHOF-RPL, the proposed routing protocol has a lower packet loss rate and a higher detection rate. The simulation results are validated by real-world testbed experimentation in the context of the smart home. SecTrust-RPL will be expanded in the future to mitigate colluding attacks. \\
	\indent Thulasiraman \textit{et al.} \cite{thulasiraman2019lightweight} developed a mobile trust-based security architecture to choose a better routing path against Sybil identity and denial of service attacks. The modified rank computation for parent selection based on \textit{bestneighborValue} is used in this approach. To measure node behavior, the \textit{bestneighborValue} is calculated from the normalized objective function, trust value defined as 1 for trusted node and 0 for untrusted node, and average receive signal strength indicator. In an IoT network, the random waypoint model is used to create a mobile environment. The proposed approach outperforms all others in terms of PDR and control overhead. In the future, the architecture will be evaluated for energy consumption and expanded to accommodate a greater number of mobile nodes.  \newline
	\indent Hassan \textit{et al.} \cite{ul2021ctrust} proposed CTrustRPL, a trust-based defense mechanism against Blackhole attacks. The proposed method computed the trust value based on node forwarding behavior at the controller node in order to save node storage and energy. The controller layer is also in charge of trust aggregation, including rating and updating. The concept of subjective logic was used in the trust value, which is an opinion about a node in terms of belief, disbelief, and uncertainty. Packet loss rate and forwarding delay are two quality parameters used in trust composition and calculation. This approach outperforms Sec-Trust \cite{airehrour2019sectrust} in terms of energy efficiency, detection time, packet loss rate, and storage overhead.  \newline
	\indent Patel \textit{et al.} \cite{patel2022reputation} integrate a trust framework with RPL to detect and isolate a selective forwarding attack. The trust framework is based on node reputation, which describes each node forwarding behavior. Essentially, the reputation is calculated in terms of packet loss, taking into account both actual and normal packet loss. The RPL parent selection process is altered based on node reputation. Throughput, energy consumption, and packet loss rate are used to evaluate performance. Lower packet loss and energy consumption were achieved compared to the standard RPL protocol, which uses MRHOF as the objective function. The proposed reputation system performance will be tested in the future for larger networks with lower memory overhead.  \newline
	\indent Jiang \textit{et al.} \cite{jiang2022secure} implements a variety of advanced selective forwarding attacks, including protocol-based attacks, packet forwarding attacks, and bad-mouthing attacks. Furthermore, a lightweight centralized trust model combines the self-trust value to reflect the trustworthiness of the node in terms of packet forwarding behavior. This approach consists of three modules: detection, notification, and isolation. The detection module is responsible for analyzing trust values. The notification module encapsulates information and DIO messages and sends notifications to all nodes. The isolation module allows child nodes to re-select their parents in order to exclude malicious ones based on DIO message responses. Using ICMPv6 control packets, the author proposed a novel anomaly report mechanism for malicious node information. The proposed method achieved the highest detection accuracy while consuming the least amount of energy.
	\subsection{Machine Intelligence Based Defense Mechanism} \label{subsec_machine_based}
	The defense mechanism was designed using machine-based computational techniques.
	The benefit of fast computation speed for faster attack detection. This section provides an overview of defense mechanisms based on machine intelligence. 
	\subsubsection{Machine Learning (ML)}
	The defense solution was designed using machine learning algorithms.
	There are three techniques being developed to counter-routing attacks: supervised, unsupervised, and hybrid learning models. \newline
	\indent Foley \textit{et al.} \cite{foley2020employing} proposed a method for detecting combined attacks against two popular objective functions (OF0 and MRHOF) using machine learning techniques. To detect the attack, the first novel IoT dataset was created by cleaning, transformation and feature reduction, normalization, and sampling techniques based on power and network metrics of combined malicious attacks: rank and version, rank and blackhole, rank and sybil, and last decrease path metric against OF0 and MRHOF. The top-performing classifiers that detect attacks with high accuracy are the multi-level perceptron (MLP) and random forest (RF).  \newline
	\indent Osman \textit{et al.} \cite{osman2021ml} proposed a supervised machine learning-based light gradient boosting machine (ML-LGBM) model for version number attack detection. The ML-LGBM model is divided into four stages: the development of a version number attack dataset, feature extraction using step-forward feature selection, the LGBM algorithm, which uses a histogram-based approach to reduce memory storage and speed up training time, and parameter optimization using logistic regression. Precision, recall, accuracy, true negative rate, and false positive rate were the performance measures.  \newline
	\indent Mehbodniya \textit{et al.} \cite{mehbodniya2021machine} used supervised machine learning approaches such as Random Forest (RF), Naive Bayes (NB), and Logistic Regression (LR) to detect fake identity attacks, also known as sybil, in an IoT sensor network. The detection techniques consider the node's packet delivery rates at the time of attack detection. When an attack is detected, the radio transmission (TX) and radio reception (RX) for packet sending and receiving for each node are used as features for the intrusion detection process, and an alarm alert is sent to the user. The NB algorithm has the highest accuracy of 92.14\%, while LR has the lowest accuracy of 89.12\%.\newline 
	\indent Prakash \textit{et al.} \cite{prakash2022optimized} proposed a novel optimized voting ensemble classifier for improving ANIDS performance on the RPL-NIDDS17 dataset of an RPL-based IoT network. The SMOTE technique is used to balance the dataset. The hybrid approach is used for feature selection based on simulated annealing and Salp Swarm Optimization to select the best feature in the dataset. Ensemble machine learning techniques such as Decision tree (DT), K-nearest neighbor (KNN), Logistic Regression (LR), Support Vector Machine (SVM), and Deep Learning classifiers such as Bi-directional Long Shot Term Memory (Bi-LSTM) based voting classifier are designed for attack detection. The accuracy (94.4\%), average false negative rate (3.6\%), and attack detection rate (97.7\%) of the ensemble classifier are the performance parameters that are used to evaluate its performance. \newline
	\indent Osman \textit{et al.} \cite{osman2024ensemble}  proposed an Ensemble Learning-based intrusion detection system (IDS) for detecting DIS flooding, decreased rank, and version number in an RPL network. In his approach, the author used the stacking of different machine learning with optimal feature selection based on genetic algorithms (GA) to achieve remarkable accuracy in attack detection.  The authors created the RPL-ELIDS dataset, which consists of samples of RPL internal attacks such as decrease rank, version number, and DIS flooding compiled on cooja under the contiki operating system. The proposed ELG-IDS achieves a high detection rate and accuracy for single-model classification and an average for multi-model classification. In the future, dynamic IoT network scenarios will be used to evaluate the ELG-IDS model. \\ \indent 
	Paganraj \textit{et al.} \cite{paganraj2024dair} proposed DAIR-MLT approach uses machine learning algorithms to detect and avoid RPL-based routing attacks on IoT. The author generated the normal and attack samples of rank, version, and flooding attacks, and created the dataset named DA\_IoT\_Routing\_Normal and DA\_IoT\_Routing\_Attack. The network traces were processed using LR, RF, DT, NB, and KNN, with RF outperforming the other methods in terms of accuracy. Once the attacker node is identified, it is moved to the trash list, and any future messages from that node are ignored. The proposed method greatly improves packet delivery ratio, throughput, network lifetime, and energy consumption.
	\begin{table}[h]
		\caption{Machine Intelligence Based Defense Mechanism for RPL-based Routing Attacks}
		\scriptsize
		\resizebox{\linewidth}{!}{
			\begin{tabular}{l p{.6cm} c c c p{2.6cm} p{5.2cm} c c c c c}
				\toprule
				\textbf{Ref} & 
				\textbf{Year} & 
				\multicolumn{3}{c}{\textbf{Scope}} & 
				\textbf{Attack Focused} & 
				\textbf{ML/DL Model} & 
				\multicolumn{5}{c}{\textbf{Evalution Metrics}} \\
				\cmidrule(rl){3-5} 
				\cmidrule(rl){8-12} 
				\textbf{} &
				\textbf{} &
				\textbf{D\tnote{1}} &
				\textbf{C\tnote{2}} &
				\textbf{M\tnote{3}} &
				\textbf{} &
				\textbf{} &
				\textbf{M1\tnote{4}} &
				\textbf{M2\tnote{5}} &
				\textbf{M3\tnote{6}} &
				\textbf{M4\tnote{7}} &
				\textbf{M5\tnote{8}} \\
				\midrule
				\cite{yavuz2018deep} & 2018 & \ding{51} & \ding{53} & \ding{53} & IRAD Dataset & Multi Layer Perceptron, Naive Bayes and Sequential model & 96.30 & 95.00 & 96.70 & 94.30 & - \\
				\cite{foley2020employing} & 2020 & \ding{51} & \ding{53} & \ding{53} & Sybil, Blackhole, Rank, Version Number & Random Forest and Multi-Layer Perceptron & 87.08 & - & - & - &	-\\ 
				\cite{morales2021dense} & 2021 & \ding{51} & \ding{53} & \ding{53} &	Clone ID & Sparse Auto Encoding and Deep Neural Network & 99.65 & - & - & 99.65 & 2.65 \\
				\cite{osman2021ml}  & 2021 & \ding{51} & \ding{53} & \ding{53} & Version Number	&  Light Gradient Boosting ML & 99.60 & 99.0 & 99.60 & 99.30 & 140.20 \\ 
				\cite{shirafkan2021autonomous} & 2021 & \ding{51} & \ding{53} & \ding{53} & Version Number, Hello Flood & Hierarchical Semantic and Group Method of Data Handling Neural Network & 99.50 & 100 & 100 & 99.20 & 1.52 \\ 
				\cite{osman2021artificial} & 2021 & \ding{51} & \ding{53} & \ding{53} &	Decrease Rank & Random Forest Classifier and Artificial Neural Network & 97.14 & 97.03 &	97.01 &	97.00 &	- \\
				\cite{sharma2021aiemla} & 2021 & \ding{51} & \ding{53} & \ding{53} &	Decrease/Increase Rank, Hello Flood  & Artificial Neural Network with Cross Validation & 100 & 100 &	100 & 100 &	- \\
				\cite{sahay2022holistic} & 2021 & \ding{51} & \ding{53} & \ding{53} & Worst Parent Selection, Increase Rank, Decrease Rank & GAN, LSTM and Feed Forward Neural Network & 91.88 & 99.00 & 81.00 & 85.00  & - \\ 
				\cite{nayak2021deep} & 2021 & \ding{51} & \ding{51} & \ding{53} & IRAD Dataset &   Generative Adversarial Network-Classifier (GAN-C) and Support Vector Machine & 91.00 & 93.00 & 92.00 & 92.00 & - \\ 
				\cite{mehbodniya2021machine} & 2021 & \ding{51} & \ding{53} & \ding{53} & Sybil Attack &  Naive-Bayes, Random Forest, Logistic Regression & 92.14 & - & - & - & - \\ 
				\cite{prakash2022optimized} & 2022 & \ding{51} & \ding{53} & \ding{53} & RPL-NIDDS17 Dataset & Ensemble Voting Classifier & 96.40 & 95.26 & - & 96.45 & - \\
				\cite{albishari2022deep} & 2022 & \ding{51} & \ding{53} & \ding{53} & IRAD Dataset &   Deep Learning based Rarly Stage Detection & 98.85 & 97.50 & 98.33 & 97.01 & - \\
				\cite{alghamdi2023cascaded} & 2023 & \ding{51} & \ding{53} & \ding{53} & Sybil attack &   Federated Learning and Dynamic Trust Factor & 96.00 & - & 100 & 97.00 & -\\
				\cite{al2023hybrid} & 2023 & \ding{51} & \ding{53} & \ding{53} & Rank, DIS, Wormhole & Supervised Deep Artificial Neural Network and Semi-supervised Deep Autoencoder & 95.00 & 95.00 & 82.00 & 87.00 & -\\
				\cite{osman2024ensemble} & 2024 & \ding{51} & \ding{53} & \ding{53} & DIS, Decrease Rank and Version Number & Ensemble Learning with Genetic Algorithm & 99.03 & 99.50 & 99.50 & 99.50 & -\\
				\cite{budania2024oead}  & 2024 & \ding{51} & \ding{53} & \ding{53} & RADAR Dataset & Online Ensemble Learning with Drift Detection & 97.67 & 98.55 & 96.67 & 97.93 & -\\
				\cite{paganraj2024dair} & 2024 & \ding{51} & \ding{53} & \ding{51} & Decrease Rank, Version Number, Flooding  & Random Forest Classifier & - & - & - & - & -\\
				\bottomrule
			\end{tabular}
		}
		\label{tbl_ml}
		\begin{tablenotes}
			\item D: Detection
			\item C: Characterization
			\item M: Mitigation
			\item M1 : Accuracy (\%)
			\item M2 : Precision (\%)
			\item M3 : Recall (\%)
			\item M4 : F1-Score (\%)
			\item M5 : Training Time (Sec)
			\item - : Not Defined 
		\end{tablenotes}
	\end{table}
	\subsubsection{Deep Learning (DL)}
	\indent Yavuz \textit{et al.} \cite{yavuz2018deep} proposed a seven-layer sequential deep learning model to detect routing attack decrease rank, hello flood, and version number.
	The author proposed IRAD, a new dataset for decreasing rank, hello flood, and version number attack traces, which includes 64.2 million values and eighteen features. The dataset is normalized using the min-max transformation, and the feature importance rates are computed using the randomized decision tree method. The sequential model is used to build a deep neural network with seven layers, five of which are hidden, and a loss function of mean squared error (MSE). The proposed method has a high recall and precision, as well as an AUC and F1 score. \newline
	\indent Osman \textit{et al.}  \cite{osman2021artificial} proposed an artificial neural network-based Multi-Layer RPL attack detection (MLRPL) model for decreasing rank attack detection.
	The proposed model is divided into three stages: data pre-processing, feature extraction, and artificial neural network training. The model is tested using the IRAD DR attack \cite{yavuz2018deep} dataset. The random forest (RF) classifier is used to select features. The proposed MLRPL model consists of three hidden layers, a rectified linear function (ReLU) as the activation function and MSE as the error function.
	The proposed method achieved up to 97\% detection accuracy while improving recall and F1 score. \newline
	\indent Morales \textit{et al.} \cite{morales2021dense} proposed a dense neural network-based model for detecting identity attacks, also known as clone ID attacks. The simulation logs are used to generate the three datasets cloneid\_20n, cloneid\_50n, and cloneid\_100n, which are then pre-processed by applying dataset balancing, value transformation, and scaling. The most important features are identified using unsupervised training with autoencoders. The dense neural network-based supervised classification with binary cross entropy loss (BCEL) as an activation function was used for classification. The proposed model achieved a detection accuracy of 99.65\%. \newline
	\indent Shirafkan \textit{et al.} \cite{shirafkan2021autonomous} proposed a hybrid approach against the hello flood and version number attack based on a hierarchical semantic and group method of data handling (GMDH) neural network. The first phase is monitoring, which calculates the trust parameter supplied as an input to the analysis phase. It generates the hello flood and version number datasets and sends them to the central router. The planning phase is in charge of detecting attacks using the GMDH algorithm and node information. In the knowledge phase, the dataset and training are processed in order to build an IDS. Finally, during the execution phase, a GMDH-based neural network is designed to avoid the attack. The proposed model was 99.9\% accurate. In the future, the author will create a deep long short-term memory-based model to improve detection. \newline
	\indent Sharma \textit{et al.} \cite{sharma2021aiemla} proposed a supervised learning model based on an artificial neural network (ANN) for detecting decrease rank, hello flood, and increase version number attacks. Pre-processing, feature extraction, and normalization have been performed on the first routing attack dataset of three attacks. The proposed ANN model divides the dataset into benign and malicious classes, and the hidden layer in the neural network is identified by averaging the number of units in the input and output layers. Parameter tuning is done to optimize performance. Finally, ten cross-validation is performed to ensure accuracy. The recall, precision, and F1-score are used to assess performance. \newline
	\indent Sahay \textit{et al.} \cite{sahay2022holistic} proposed a holistic framework for predicting multiple types of RPL attacks in IoT LLNs using Blockchain and deep learning technology. To secure the IoT LLN data, the packet captured (pacp) file is generated from the IoT-LLN and stored in a private Blockchain-based smart contract. Two popular deep learning tools were used to extract spatial and temporal features from captured data: Graph Convolution Neural Network (GCN) and Long Short Term Memory (LSTM).
	The feedforward neural network divides the captured data into four categories based on the inputs from GCN, LSTM, and the warning pulse generated by the smart contract: resource attack (p1), topological attack (p2), traffic attack (p3), and normal state (pn). The proposed method for detecting specific attacks and determining how to store the large amount of data generated by IoT LLN in the Blockchain.  \newline
	\indent Nayak \textit{et al.} \cite{nayak2021deep} proposed a deep learning-based Generative Adversarial Network-Classifier (GAN-C) to detect routing attacks in the Industrial Internet of Things (IIoT). The GAN-C is a two-stage GAN and support vector machine combination (SVM). The parallel learning methodology is used in this model to reduce the training time significantly. The first stage GAN classifier, is a combination of two ANN models used to generate and detect adversaries, while the second stage classifier is used to identify and classify attacks. The proposed model's performance is evaluated using the IRAD dataset. \newline
	\indent Albishari \textit{et al.} \cite{albishari2022deep} proposed a novel scheme known as deep learning-based early-stage detection (DL-ESD) on the IRAD dataset of IoT. The model performance is improved by selecting distinct features using Linear Discriminant Analysis (LDA) and min-max scaling for feature normalization, and reduces training time with various deep learning techniques (LR, KNN, SVM, NB, MLP). The proposed DL-ESD used the idea of binary classification with lightweight deep learning techniques to classify the behavior of the IRAD dataset and MLP and achieved high accuracy during the testing and training phase with low running time. The performance parameters are accuracy (98.85\%), recall (98.33\%), precision (97.50\%), and F1-score (97.01\%). \newline
	\indent Alghamdi \textit{et al.} \cite{alghamdi2023cascaded} proposed a framework for real-time wormhole detection in an IoT network using Dynamic Trust Factor (DTF) and cascaded federated deep learning techniques (CNN and LSTM). The proposed approach collects critical information concerning security and privacy from nodes through decentralized training. Once a local model has been created, it is submitted to a federated learning aggregation agent. The global model is created while considering the node's identified security and privacy concerns. The global model assists individual nodes in analyzing and detecting wormhole attacks with high accuracy. The DTF values are determined using the QoS and social trust metrics, and they have complete knowledge of the trust associated with other nodes and are less than the defined threshold than malicious nodes. The proposed framework achieves high detection accuracy while reducing resource requirements and latency. \\ \indent
	AL Sawafi \textit{et al}.\cite{al2023hybrid} proposed a hybrid approach based on supervised and semi-supervised deep learning concepts for the classification of network traffic behavior in RPL-based IoT networks. The author created the IoTR-DS dataset, which contains traces of three popular RPL-based attacks: Rank, DIS Flooding, and Wormhole. The proposed DL-IDS (DAE-DANN) employs a hybrid model that includes the Deep Artificial Neural Network (DANN) as a supervised model and the Deep Auto-encoder (DAE) as a semi-supervised model for attack classification on the proposed IoTR-DS. The proposed model achieves high accuracy on IoTR-DS for both known attacks (pre-trained) and unknown attacks (untrained).\\ \indent
	Budania \textit{et al}.\cite{budania2024oead} a deep learning-based anomaly detection mechanism for RPL attacks was proposed, utilizing unsupervised ensemble learning. The proposed approach, OEAD (Online Ensemble-based Anomaly Detection), creates an ensemble model using Autoencoder (AE), iForest (ASD, and NDKSWIN) techniques. The approach employs initial model training, followed by fine tuning with a drift detector to identify significant changes in network traffic, and model updating for real-time anomaly detection. The performance of the proposed approach is evaluated on the RADAR dataset, which contains nine RPL attack samples, and the highest accuracy for Sybil attack is obtained. 
	\subsection*{\textcolor{black}{Role of Artificial Intelligence in Attack and Defense}}
	\textcolor{black}{Artificial Intelligence (AI) leverages enormous power to learn and analyze massive amounts of data and make predictions to improve the network's reliability \cite{hussain2020machine,waqas2022role}. The two popular techniques: Machine Learning and Deep Learning, are essential in developing security solutions that protect from abnormal traffic behavior and anomalies. Moreover, AI is a powerful tool having exceptional capabilities of data processing to be used offensively and defensively from a security perspective . The offensive use of AI will explore the security threats or weaknesses existing in the network to launch the attacks. The defensive use of AI will protect the network from internal and external attacks exploited due to security vulnerabilities \cite{comiter2019AIattack}.
		\begin{itemize}
			\item How AI attack can use to launch attack?: The attackers can use any of the below techniques to cause damage, evade detection, and degrade the trust of the defense mechanism.
			\begin{itemize}
				\item Exploit Learning Capability:The machine learning model learns from the statistical associations between data and patterns to achieve excellence. An attacker can disrupt the learning capability to craft the attack and undermine the excellence of the model.
				\item Exploit Dataset:The machine learning algorithms build their knowledge base from a set of patterns stored in the dataset while designing a model. This model had no prior base knowledge like humans and depends absolutely on the dataset. The AI system utilizes the fact to poison the dataset to launch the intelligent attack, which can be activated promptly. 
				\item Exploit Black-box Nature of ML Algorithm: The AI system is built with complex algorithms consisting of several inputs processed to make output. However, the internal working, black boxes, is challenging to comprehend. This nature makes it impossible to tell whether the existing machine-learning model is being compromised or working benignly. The AI-based attack is brutal to identify if they used the black box characteristics to launch it. 
			\end{itemize}
			\item How AI can use to protect from attack?: The AI can be used to protect attacks using robust algorithms, developing new dimension AI techniques \cite{ahanger2022state}. 
			\begin{itemize}
				\item Robust AI Algorithms: The protection against the AI attack can be obtained by designing the new robust AI algorithms which use hardening techniques to deceive themselves \cite{nguyen2021federated}. The hardening techniques can be borrowed from other domains which prove great success against the attack, such as Address Space Layout Randomization (ASLR). The popular organization DARPA works hard in this area and provides Guaranteeing AI Robustness Against Deception (GARD) as a good example.
				\item Self-Supervised Learning: The future research will be conducted to build the self-supervised mechanism to protect against AI attacks \cite{wu2023self}. The self-supervised learning algorithms combine the power of supervised and unsupervised learning to automatically build the label from the raw dataset, which can detect unknown attacks efficiently. The self-supervised learning uses Generative Adversarial Network (GAN) to build defensive AI approaches.
				\item Explainable AI: The Explainable AI is the key research area in the development solution for IoT security \cite{kalutharage2023explainable}. The explainable AI helps to understand the complex process involved in machine learning algorithms for human beings. The crucial aspects related to the trustworthiness of the dataset, machine learning models, and the applied algorithm is answerable by the explainable AI. The explainable AI may be explored to build the defense against the AI attack by understanding the black box nature of compromised machine learning models.
			\end{itemize}  
		\end{itemize}
	}
	\subsection{Threshold Based Defense Mechanisms} \label{subsec_thresold_based}
	The threshold-based defense mechanisms used static or dynamic threshold values. This subsection provides an overview of threshold-based defense mechanisms.
	\subsubsection{Static Threshold}
	The following section discusses static threshold-based defense solutions. \newline
	\indent Ahmed \textit{et al.} \cite{ahmed2016mitigation} proposed a blackhole attack mitigation technique based on local decisions and global verification processes. The proposed approach is divided into two phases: the first uses a local decision to observe the misbehaving activity of its neighbor's data packet, and if it exceeds a threshold, it is classified as a suspicious node; the second uses a global verification process to determine whether a suspicious node is a blackhole node based on verification messages received request (RREQ) and received result (RRES). End-to-end delay, true positive rate, false positive rate, and PDR were used to evaluate performance. The proposed approach effectively detects blackhole attacks in larger networks and improves the data delivery rate. (comes from miscellaneous) \newline
	\indent Patel \textit{et al.} \cite{patel2019blackhole} proposed a Strainer-based Blackhole Intrusion Detection in 6LoWPAN for the Internet of Things (SIEWE). The SIEWE architecture consists of two modules: one at the node level, known as the local module, and one at the border level, known as the global module. The method first filters out suspicious nodes based on high routing metrics such as RSSI and LQI values and adds those nodes to the suspect list. After creating the suspect list, the detection and verification process is applied to nodes in the suspect list based on the difference between incoming and outgoing packet counts that exceed the border router's defined threshold. The performance is measured in terms of packet delivery ratio for network topologies of 20, 30, 40, and 50 nodes.  \newline
	\indent Ioulianou \textit{et al.} \cite{ioulianou2019denial} proposed a lightweight intrusion detection system (IDS) based on the threshold to detect and mitigate DIS flooding attacks. IDS includes two modules: a centralized detection module implemented at the border router that identifies nodes with a high DIS message sending rate and classifies them as malicious nodes, and a distributed detection module implemented in each node that identifies normal behavior of nodes by using a packet sending interval threshold to limit the average packet interval of each node. True positive rate, false positive rate, IDS warning, and a number of messages sent to root nodes are used to evaluate performance. This method achieved a high detection rate while also having high overhead. \newline 
	\indent Seth \textit{et al.} \cite{seth2020detection} proposed a detection, verification, and isolation approach based on round trip time (RTT) against decreased rank attacks. The server node, also known as the root node, maintains the blacklist and whitelist tables to store information such as Node ID, rank, and average RTT. The server node collects all of the data and stores it in WT. The server node monitors the behavior of all nodes in the network. If the new RTT of a node exceeds the stored RTT value, the node is assigned to BT, and an alarm is generated. The proposed approach's accuracy rate is proportional to the percentage of malicious nodes present in the network, with high accuracy in the case of a few malicious nodes. \newline
	\indent Almusaylim \textit{et al.} \cite{a2020detection} proposed a Secure Routing Protocol (SRPL-RP) for the detection, mitigation, and isolation of rank and version number attacks. SRPL-RP is divided into five phases: the first monitors the freshness of DIO messages using the timestamp value; the second validates the legitimacy of the sending node by ID in the case of a lower threshold value; and the third detects, mitigates, and isolates attacks using a blacklist table, a monitoring table, and an alert function, the fourth and fifth define the conditions for defense against rank and version number attacks. The performance is measured in terms of packet delivery ratio, energy consumption, control packet overhead, and detection accuracy. SRPL-RP achieved a 98.48\% packet delivery ratio and 98.30\% detection accuracy. \\ \indent
	Aljufair \textit{et al.} \cite{aljufair2023mitigating} perform the impact assessment of DIS attack and proposed a mitigation mechanism based on DIO response named as \textit{DIO\_{resp}} against the attack. The proposed solution based on the work of \cite{verma2020mitigation} and it restrict the total number of send DIO from neighbours in the next trickle interval in response to DIS message send by a node. Whenever this the transmission DIO exceeds the predefined threshold then mitigation scheme suppress the DIO transmission. The performance of proposed mitigation technique is measured by energy consumption, traffic overhead and packet delivery ratio. \\ \indent  
	Sharma \textit{et al.} \cite{sharma2023qsec} proposed a Q-learning based detection mechanism known as QSec-RPL against the version number attack in RPL-based static and mobile IoT network. The main idea was to used Q-Learning approach based in discounting factor to detect the attack. The proposed approach run in every T second time, and it calculate the Q-value of each node based on number of DIO received. Whenever the version number changes in DIO message received by node, a patently is assigned as per Q-Learning strategy, and upper limit of Inter Quartile Range to detect the attacker nodes. The proposed approach achieve the high detection accuracy while minimizing the overhead.
	\begin{table}[h!]
		\caption{Threshold Based Defense Mechanism for RPL-based Routing Attacks}
		\scriptsize
		\resizebox{\linewidth}{!}{
			\begin{tabular}{l p{.5cm} c c c p{3cm}  p{6.6cm} c c c c c}
				\toprule
				\textbf{Ref} & 
				\textbf{Year} & 
				\multicolumn{3}{c}{\textbf{Scope}} & 
				\textbf{Attack Focused} & 
				\textbf{Methods/Algorithm} & 
				\multicolumn{5}{c}{\textbf{Performance Assessment}} \\
				\cmidrule(rl){3-5} 
				\cmidrule(rl){8-12} 
				\textbf{} &
				\textbf{} &
				\textbf{D\tnote{1}} &
				\textbf{C\tnote{2}} &
				\textbf{M\tnote{3}} &
				\textbf{} &
				\textbf{} &
				\textbf{P1\tnote{4}} &
				\textbf{P2\tnote{5}} &
				\textbf{P3\tnote{6}} &
				\textbf{P4\tnote{7}} &
				\textbf{P5\tnote{8}}\\
				\midrule
				\cite{sehgal2014addressing} & 
				2014 & \ding{53} & \ding{53} & \ding{51} &	DAG Inconsistency & Adaptive Threshold on Number of Trickle Timer Reset & L & - & - & N & L \\
				\cite{mayzaud2015mitigation}  & 2015 & \ding{53} & \ding{53} & \ding{51} & DAG Inconsistency & Dynamic Threshold on Network Characteristics for Trickle Timer Reset & L & - & - & N & L \\  
				\cite{ahmed2016mitigation} & 2016	& \ding{53} & \ding{53} & \ding{51} &	Blackhole & Local Decision and Global Verification Process	& - &	- &	H &	N &	- \\
				\cite{pu2018mitigating} & 2018	& \ding{53} & \ding{53} & \ding{51} &	DAO Inconsistency & Dynamic Threshold Mechanism on Forwarding Error Packets	& L &	- &	- &	N &	L \\
				\cite{pu2019energy}  & 2019	& \ding{51} & \ding{53} & \ding{51} &	Energy Depletion Attack &	Misbehavior Aware Detection Scheme	& - &	- &	H &	N &	H \\ 
				\cite{patel2019blackhole} & 2019 & \ding{51} & \ding{53} & \ding{51} &	Blackhole &	  Suspect List and Threshold of Packet Count &	- &	- &	- &	N &	L \\ 
				\cite{ioulianou2019denial} & 2019 & \ding{51} & \ding{53} & \ding{51} &	Selective Forwarding &	DIS Message Sending Rate and Packet Send Interval & L &	Y &	H &	N &	- \\ 
				\cite{seth2020detection} & 2020 & \ding{51} & \ding{51} & \ding{51} & Decrease Rank & Black List, White List, Round Trip Time & H & - & H & N &	- \\ 
				\cite{a2020detection} & 2020 & \ding{51} & \ding{53} & \ding{51} & Version Number, Rank Attack & Rank Threshold and Attack Status Table, Blacklist &	L &	- &	H &	N &	L\\ 
				\cite{sheibani2022lightweight} & 2022 & \ding{51} & \ding{53} & \ding{51} &	Dropped DAO & Packet eavesdropping and Threshold Behaviors (Positive/Negative) of Nodes &	L &	Y &	H &	N &	L \\
				\cite{aljufair2023mitigating}  & 2023 & \ding{51} & \ding{53} & \ding{53} &	DIS Flooding & DIO Response based Mitigation & L & - & H & N & L \\
				\cite{sharma2023qsec}  & 2023 & \ding{51} & \ding{53} & \ding{51} &	Version Number & Q-Learning Principle based on Q-Value of DIO Messages & L & Y & H & Y & L\\
				\bottomrule
			\end{tabular}
		}
		\label{tbl_threshold}
		\begin{tablenotes}
			\item D: Detection
			\item C: Characterization
			\item M: Mitigation
			\item P1 : Overhead
			\item P2 : Lightweight
			\item P3 : Detection Rate
			\item P4 : Mobility
			\item P5 : Energy Consumption
			\item Y: Yes and N: NO
			\item L: Low, M: Medium, and H: High
			\item - : Not Defined 
		\end{tablenotes}
	\end{table}
	\subsubsection{Dynamic Threshold}
	The following section discusses dynamic threshold-based defense solutions. \newline
	\indent Sehgal \textit{et al.} \cite{sehgal2014addressing} the DODAG Inconsistency attack, where a malicious node manipulates the IPv6 header option of an RPL packet using the "O" and "R" flags to launch attacks. The proposed defense mechanism used the concept of a dynamic threshold ranging from 0 to 20 to limit the number of trickle timer resets on the nature of the attacker and network conditions. The effect of a defined threshold value, computational overhead, and energy consumption are all used to evaluate performance. The proposed approach has low computational complexity and energy consumption, making it a good solution for DODAG inconsistency attacks.  \newline
	\indent Mayzuad \textit{et al.} \cite{mayzaud2015mitigation} proposed a dynamic threshold-based defense mechanism for the topological inconsistency attack. This work is an extension of \cite{sehgal2014addressing}, where a static parameter was used for an adaptive threshold to prevent trickle timer reset. Still, this approach used a dynamic threshold based on network condition and neighborhood size to improve attack detection and determine the actual trickle time reset. Control overhead and energy consumption are used to evaluate performance. The proposed approach can reduce overhead by up to 50\% while increasing the packet delivery ratio by up to 99\%. This approach is suitable for mitigating blackhole situations that cannot be mitigated by an adaptive threshold.  \newline
	\indent Pu \textit{et al.} \cite{pu2018mitigating} proposed dynamic threshold mitigation for DAO inconsistency attack. A parent dynamically adjusts the threshold based on the number of forwarding error packets received and the normal forwarding error rate. Each node maintains an FR (forwarding record) of its child node, then determines the threshold limit of accepting forwarding error packets using a dynamic threshold based on the attack pattern. Suppose the calculated dynamic threshold is less than the number of received forwarding error packets. In that case, the child node is treated as malicious and detected forwarding misbehavior increments by one at the record window end. The parent node isolates the malicious node when forwarding misbehavior reaches a threshold. Performance evaluation is done using energy consumption, and packet delivery ratio.  \newline
	\indent Pu \textit{et al.} \cite{pu2019energy} proposed a misbehavior aware detection scheme (MAD) to protect resource-constrained networks from a type of denial of service attack known as an energy depletion attack. To counter the energy depletion attack, each node keeps a record of received packets from its children within a specified time window in the observation table (OT), as well as the number of detected forwarding misbehavior of each neighbor node in the detection table (DT). When the number of misbehaved packets exceeds a dynamically calculated threshold, the node is classified as malicious and is isolated from the network. The packet delivery ratio, energy consumption, and detection rate are used to assess performance. The proposed approach achieved an approximate packet delivery rate of 90\% and a detection rate of 85\%.  \newline
	\indent Sheibani \textit{et al.} \cite{sheibani2022lightweight} proposed a new attack known as the Dropped Destination Advertisement Object (DDAO). A lightweight defense mechanism for attacks is also proposed, which monitors the DAO forwarding behavior of parent nodes. The watchdog mechanism is used to monitor the behavior of the parent and, based on the negative packet forwarding behavior of nodes while taking into account the attack detection threshold, the punishment isolates the malicious nodes, and the forgiveness avoids false detection of malicious nodes. The evaluation metrics are true positive rate, false positive rate, precision, accuracy, power consumption, and PDR. The proposed mechanism achieved a high detection rate and increased the packet delivery ratio by up to 158\% with little overhead. 
	
	
	\subsection{Cryptography Based Defense Mechanisms}
	Various cryptography-based techniques are used to design defense solutions. This subsection provides an overview of cryptography-based defense mechanisms. 
	\subsubsection{Authentication}
	The following sections discuss authentication-based defense solutions.  \newline 
	\indent Luangoudom \textit{et al.} \cite{luangoudom2020svblock} proposed a novel intrusion detection scheme svBLOCK based on SVELTE \cite{raza2013svelte} to detect and mitigate blackhole attacks. The main idea was to reconstruct DODAG and validate nodes from malicious behavior, as well as to provide authentication of control messages using encryption and isolate blackhole nodes from DODAG. The proposed approach uses fewer resources to detect malicious nodes while maintaining a low false-positive rate and a high detection rate. The true positive rate achieved is 98.5\%, while the false negative rate is 3.7\%. svBLOCK outperforms SVELTE \cite{raza2013svelte} in terms of packet delivery rate and power consumption. The proposed approach will be extended in the future to detect wormhole attacks. \newline
	\indent Karmakar \textit{et al.} \cite{karmakar2021leader} proposed LEADER, a low overhead-based increase and decrease rank attack detection scheme that uses a modified DODAG formation algorithm. The goal of this approach is to modify the DODAG construction process to detect rank-based routing attacks launched during topology formation and maintenance. The proposed approach modifies the DAO message and further ensures the integrity and authenticity of the exchanged messages during RPL DODAG formation using a lightweight message authentication code known as HMAC-LOCHA. The efficiency of the system is measured by energy consumption, false positive/negative rate, and detection accuracy. In comparison to SBIDS \cite{shafique2018detection}, LEADER performed better. 
	
	\subsubsection{Encryption}
	The following sections discuss encryption-based defense solutions.  \newline
	\indent Perry \textit{et al.} \cite{perrey2016trail} made two contributions. The first was an extension of VeRA \cite{dvir2011vera} to mitigate a new rank attack based on forgery and reply. The second was a topology authentication mechanism to mitigate routing topology attacks such as version number, rank spoofing, and rank replay. The Trust Anchor Interconnection Loop (TRAIL) mechanism was developed to inquire about actual path properties through attestation and message announcement. The main idea here is to validate the upward path up to the root node using a round trip message or rank integrity recursively from the validations of the upward path. TRAIL is an RPL protocol implementation on the RIOT platform that provides RPL security and is deployed on the DES mesh testbed. The scheme's performance is measured regarding network size and message overhead. \newline
	\indent Pu \textit{et al.} \cite{pu2022lightweight} proposed \textit{liteSAD}, a lightweight detection mechanism for sybil attacks that uses a Bloom filter (BF) and a physical unclonable function (PUF). A DODAG root sends a BF-DAO packet to each node in the \textit{liteSAD} approach. Bloom filter array contains each node's hash identifier and PUF response; when a node receives a DIS packet, it verifies the identifier and PUF from the local copy, and mismatches are used to identify the malicious node. To avoid DIS attacks, the author proposed a probabilistic approach to reduce DIO message broadcasting using textitproDIO. An extensive simulation was used to evaluate the performance of \textit{liteSAD} and \textit{proDIO}, which provide better performance in terms of detection latency, detection rate, and energy consumption. 
	\begin{table}[ht!]
		\caption{Cryptography Based Defense Mechanism for RPL-based Routing Attacks}
		\scriptsize
		\resizebox{\linewidth}{!}{
			\begin{tabular}{l p{.5cm} c c c p{3cm} p{6.5cm} c c c c c}
				\toprule
				\textbf{Ref} & 
				\textbf{Year} & 
				\multicolumn{3}{c}{\textbf{Scope}} & 
				\textbf{Attack Focused} & 
				\textbf{Methods/Algorithm} & 
				\multicolumn{5}{c}{\textbf{Performance Assessment}}\\
				\cmidrule(rl){3-5} 
				\cmidrule(rl){8-12} 
				\textbf{} &
				\textbf{} &
				\textbf{D\tnote{1}} &
				\textbf{C\tnote{2}} &
				\textbf{M\tnote{3}} &
				\textbf{} &
				\textbf{} &
				\textbf{P1\tnote{4}} &
				\textbf{P2\tnote{5}} &
				\textbf{P3\tnote{6}} &
				\textbf{P4\tnote{7}} &
				\textbf{P5\tnote{8}} \\
				\midrule
				\cite{dvir2011vera}  & 2011 & \ding{51} & \ding{53} & \ding{53} &	Version Number, Decrease Rank & Hash Chains of Rank and Version Number & L &	- &	- &	- &	-\\ 
				\cite{perrey2016trail} & 2016 &	\ding{51} & \ding{53} & \ding{53} &	Rank Replay/Spoofing & Path Validation using Encryption Chain and Rank Attestation using Bloom filter &	L &	- &	- &	N &	L\\ 
				\cite{conti2018split} & 2018 & \ding{51} & \ding{53} & \ding{53} &	Rank, Sybil	& Remote Attestation and Piggybacks & L &	Y &	- &	Y &	L \\
				\cite{luangoudom2020svblock}& 2020	& \ding{51} & \ding{53} & \ding{51} & Blackhole & Authentication of Control Message and Isolation of Malicious node &	L &	- &	H & N & H \\
				\cite{karmakar2021leader} & 2020 & \ding{51} & \ding{53} & \ding{53} & Decrease/Increase Rank &	Message Authentication Code (HMAC-LOCHA) & L & Y & M & N & L \\ 
				\cite{pu2022lightweight} & 2022 & \ding{51} & \ding{53} & \ding{53} &	Sybil	& Bloom Filter and Physical Unclonable Function & 	L &	Y &	H &	N &	L\\
				\bottomrule
			\end{tabular}
		}
		\label{tbl_mathamatical}
		\begin{tablenotes}
			\item D: Detection
			\item C: Characterization
			\item M: Mitigation
			\item P1 : Overhead
			\item P2 : Lightweight
			\item P3 : Detection Rate
			\item P4 : Mobility
			\item P5 : Energy Consumption
			\item Y: Yes and N: NO
			\item L: Low, M: Medium, and H: High
			\item - : Not Defined 
		\end{tablenotes}
	\end{table}
	\subsubsection{Hashing}
	The following sections discuss hashing-based defense solutions. \newline
	\indent Dvir \textit{et al.}  \cite{dvir2011vera} proposed VeRA, a cryptographic mechanism for version number and rank authentication against version number and rank attacks. With the help of a message authentication code, the VeRA security scheme computes a version number hash chain and a rank hash chain. When an attacker node wants to increase the version number, it must compute a pre-image of the version number's last hash chain element, which is impossible because the cryptographic hash function is one-way only. When a new node is created, all information is received via a DIO message. The performance is measured in terms of overhead, calculated as the estimated time required to build the authentication chain. \newline
	\indent Conti \textit{et al.} \cite{conti2018split} proposed a secure and scalable version of the RPL routing protocol called SPLIT to overcome the situation of rank and sybil attack. The proposed protocol is lightweight and employs remote attestation and piggyback techniques for RPL control messages. The DAO control messages are used for remote attestation with additional header field modifications, and additional overhead is avoided by piggybacking the messages. The attestation process includes two main components verifier and prover. The verifier performs the following functions: initial joining, verify trickle time, attestation, and report sending, whereas the prover performs the following functions: DODAG creation, verification trickle time, attestation, report gathering, and verification. The proposed protocol consumes little energy, making it suitable for resource-constrained networks such as IoT.
	\subsection{Statistics Based Defense Mechanisms}
	The statistical-based defense mechanism used various statistical approaches, i.e., GINI index or the Game Theory Model. 
	\subsubsection{Gini Index}
	The following sections discuss GINI-index-based defense solutions. \newline
	\indent Groves \textit{et al.} \cite{groves2019gini} proposed a detection and mitigation strategy for sybil attacks based on GINI index that required three steps: First, the traces of DIS messages received from newly joined nodes are stored in the trace table for an observation window period to measure forwarding behavior. Second, measure the dispersion of new node identities at the end of the observation window using GINI index, which measures the divergence between the probability distributions of the value of the target attributes. Whenever the GINI impurity exceeds the normal range, a sybil attack occurs. Third, initiate the attack mitigation procedure by restricting the DIO message rate. The effectiveness is measured by energy consumption, detection rate, and the number of sybil attacks detected. \newline
	\indent Pu \textit{et al.} \cite{pu2020sybil} introduced a GINI-based detection and mitigation scheme for the sybil attack. In the GINI-based mitigation technique, each node preserves a new node trace table to store the information of the received new DIS message for the newly joined node for each observation window, and the dispersity of each node is measured when the observation window ends. Based on GINI, impurity detection, and attack mitigation have been performed. Compared to \cite{ghaleb2018addressing,ahmed2016mitigation}, the performance improves in several ways: detection rate, isolation delay, and power consumption. The performance evaluation parameters are the rate of change of the DIO message and the energy consumption. A testbed consisting of TelosB nodes will be designed and deployed in real networks in the future. 
	\subsubsection{Game Theory}
	The following sections discuss game theory-based defense solutions. \newline
	\indent Kiran \textit{et al.} \cite{kiran2020towards} proposed S-MODEST, protecting the RPL protocol against packet-dropping attacks by employing a hybrid of the DODAG contextual trust model and the RPL rank variance factor, known as the non-cooperative game model and the Dumpster Shaffer theory. This approach is divided into trust-based routing behavior, lightweight defense mechanisms, and coalition formation based on the level of certainty in the context. In non-cooperative game theory, a player is a node with a tuple combination of high and low trust and energy weights. Similarly, as a utility function, strategy is decided by the parent and child, along with energy and weighted trust. The findings of the simulation demonstrate that S-MODEST outperforms SecTrust \cite{airehrour2019sectrust} in terms of detection accuracy. \\ \indent
	Sharma \textit{et al.} \cite{sharma2024lightweight} investigated the Hatchetman attack on the RPL network and contributed a lightweight security mechanism based on Game Theory. The game is played by the network's various sensor nodes, using strategies and actions. The players used the payoff metric to play the game, which determines the packet forwarding behaviour of the node and its parent node. During the attacker detection process, each node maintains the payoff metric; whenever an attacker node modifies the Source Routing Header (SRH), a checksum is computed. If the checksum differs from the previous one, the payoff matrix is updated, and the node is moved to the blacklist. This approach demonstrated superior performance in terms of packet delivery ratio, average end-to-end delay, and overhead.
	\subsection{Miscellaneous} \label{subsec_misc_based}
	This subsection provides an overview of defense mechanisms that are not classified by the above categories.
	\subsubsection{Attack Graph}
	Sahay \textit{et al.} \cite{sahay2018attack} investigated the vulnerability assessment on RPL rank property and performed an analysis on the possible threats that can be exploited with rank property by creating an attack graph. Exploiting the rank property has a negative impact on RPL, resulting in several RPL attacks such as decreased rank, increase rank, and worst parent selection. These attacks result in network isolation, excessive resource consumption, and under-optimization. The attack graph-based detection technique has been developed to detect rank attacks and attacks that occur as a result of rank attacks. Observations for various attacks were derived from simulations on parameters such as the number of average power consumption, control messages, average beacon interval, packet loss, and ETX. 
	\subsubsection{Automata Theory}
	Gothawal \textit{et al.} \cite{gothawal2021intelligent} proposed an  IDS for Efficient Routing (RAIDER) for RPL attacks, which is a lightweight, intelligent IDS that uses the concept of an automata model to identify the node's behavior and reduce the impact of routing attacks such as rank, DAO inconsistency, sinkhole, and denial of service. RAIDER used a context-aware decision model based on automata theory. Decisions are made from transition flows and compared to predefined thresholds to detect RPL attacks with improved detection accuracy and lower power consumption. Performance metrics, such as energy consumption, packet delivery ratio, overhead, delay, network lifetime, and attack detection accuracy, are used to evaluate performance. RAIDER reduces delay by 88\% and energy consumption by 25\% when compared to SecTrust-RPL \cite{airehrour2019sectrust}. 
	\subsubsection{Blockchain}
	Sahay \textit{et al.} \cite{sahay2020novel} designed a novel Blockchain-based framework for generating real-time alerts on RPL attacks such as decreased rank, increase rank, and worst parent selection. The proposed approach is divided into three major phases: DODAG advertisement, node joining (parent selection, route registration, and node advertisement), and DODAG maintenance. The Blockchain Network (BCN) records LLN activity and forwards it to security analysts (SA) for evaluating LLN behavior and detecting anomalies. The private Blockchain is built with the Ethereum client, and smart contracts are written in Solidity.
	\subsubsection{Fuzzy Logic}
\begin{figure}[h!]
		\centering
		\includegraphics[width=.85\columnwidth]{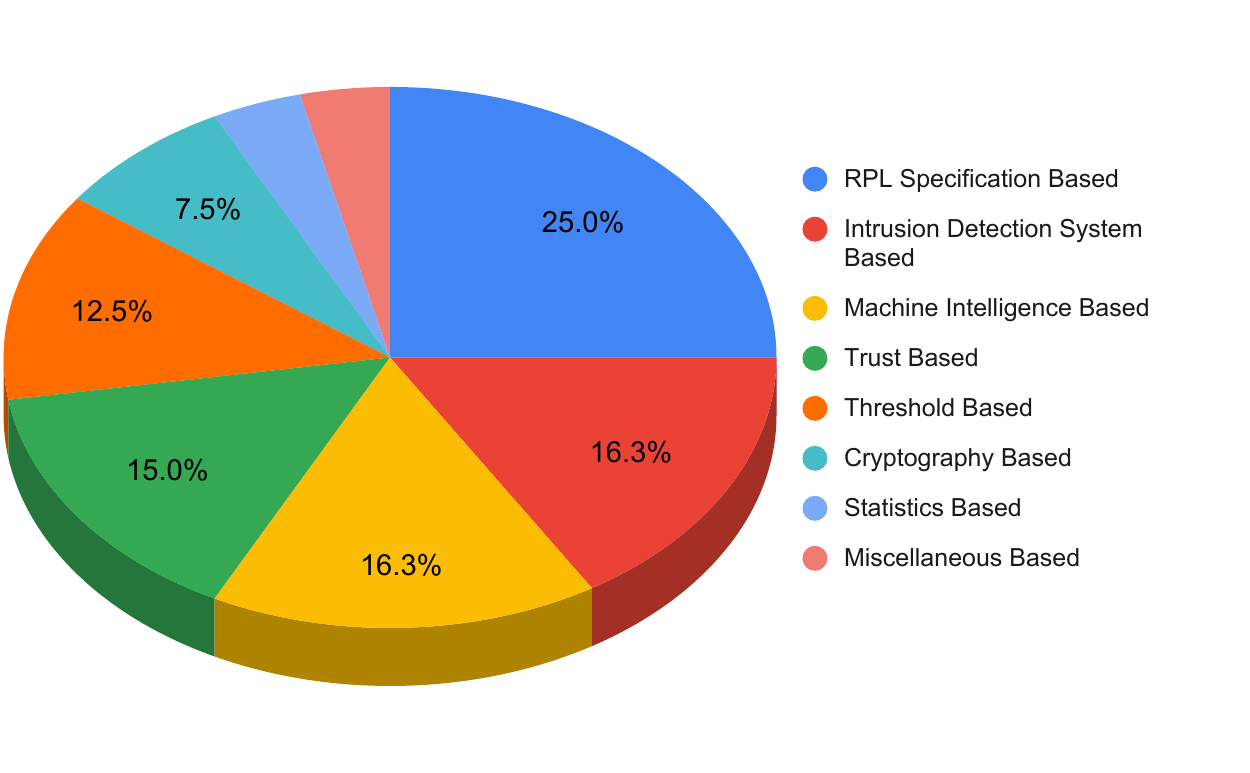}
		\caption{Distribution of RPL-based Defense Mechanisms}
		\label{fig_catewise_defensemech}
	\end{figure}

	\indent Farzaneh et al. \cite{farzaneh2020new} proposed a fuzzy logic-based local repair attack mitigation scheme. Fuzzy logic uses residual energy (RE), expected transmission count (ETX), and node distance to determine state. The graph membership function calculates fuzzy values with low, mid, and high levels for each input. The proposed approach has two phases: first phase checks the common nodes within same transmission range, if yes then fuzzy process based on three input is initiated otherwise discarding the local repair message request and second phase decide the attack situation based on fuzzy system, a node request for local repair message which is discarded by fuzzy system otherwise request for local repair message has been accepted. This method accurately detects local repair attacks. The same approach will be used to detect  RPL-specific attacks in the future. \newline 
	\indent A quantified category-wise distribution of defense mechanisms is given in Fig. \ref{fig_catewise_defensemech}. 
	Intrusion detection systems and RPL specification-based defense mechanisms are the most popular against the routing attacks that account for highest among all defense mechanisms. The next category of defense mechanism is Machine Intelligence and Trust based, which covers a significant role. However, it seems there is less interest in Cryptography and Statistics based defense mechanisms.
	Lastly, we analyzed the performance metrics used by the researchers in their security solutions against routing attacks and the performance accomplished by various defense mechanism is shown in Table \ref{tab:performace_metrics}. 
	The highest three performance metrics are packet delivery ratio, energy consumption, and accuracy. Every defense solution focuses on lowering energy consumption, increasing packet delivery ratio, and higher accuracy. Key parameters of interest are average residual energy, CPU usage, first response time, IDS warning message sent to IDS root, the number of child nodes connected to the attacking nodes, and percentage of nodes segregated.
	\begin{table}[h!]
		\centering
		\tiny
		\color{black}
		\caption{Performance Assessment of various performance metrics for RPL-based Defense Solutions}
		\label{tab:performace_metrics}
		\renewcommand{\arraystretch}{1.2}
		\begin{tabular}{cccccccccccccccccc}
			\rotatebox{90}{Reference}	&	\rotatebox{90}{Packet Delivery Ratio (\%)}	&	\rotatebox{90}{Packet Loss Ratio (\%)}	&	\rotatebox{90}{Average Power Consumption(mW)}	&	\rotatebox{90}{Control Message Overhead}	&	\rotatebox{90}{Network Overhead (\%)}	&	\rotatebox{90}{Memory Consumption (Bytes)}	&	\rotatebox{90}{Throughput (kbps)}	&	\rotatebox{90}{True Positive Rate (\%)}	&	\rotatebox{90}{False Positive Rate (\%)}	&	\rotatebox{90}{True Negative Rate (\%)}	&	\rotatebox{90}{False Negative Rate (\%)}	&	\rotatebox{90}{Detection Rate (\%)}	&	\rotatebox{90}{Accuracy (\%)}	&	\rotatebox{90}{Precision (\%) }	&	\rotatebox{90}{Recall (\%) }	&	\rotatebox{90}{F1-Score (\%)}	&	\rotatebox{90}{Delay (ms)}	\\ \midrule
			\midrule
			\cite{le2016specification}	&	-	&	-	&	-	&	-	&	6.3	&	-	&	-	&	100	&	-	&	-	&	-	&	-	&	-	&	-	&	-	&	-	&	-	\\
			\cite{ghaleb2018addressing}	&	100	&	-	&	3.5	&	-	&	-	&	-	&	-	&	-	&	-	&	-	&	-	&	-	&	-	&	-	&	-	&	-	&	-	\\
			\cite{shafique2018detection}	&	-	&	-	&	1.8	&	-	&	-	&	-	&	-	&	-	&	-	&	-	&	-	&	-	&	100	&	-	&	-	&	-	&	-	\\
			\cite{arics2019new}	&	-	&	-	&	-	&	-	&	-	&	-	&	-	&	100	&	100	&	98	&	-	&	-	&	-	&	-	&	-	&	-	&	-	\\
			\cite{kaliyar2020lidl}	&	99	&	-	&	68.29	&	-	&	-	&	8574	&	-	&	100	&	-	&	-	&	-	&	-	&	-	&	-	&	-	&	-	&	-	\\
			\cite{zaminkar2020sos}	&	94.17	&	-	&	-	&	-	&	-	&	-	&	2700	&	-	&	-	&	-	&	-	&	96.19	&	-	&	-	&	-	&	-	&	-	\\
			\cite{wadhaj2020mitigation}	&	98	&	-	&	5.7	&	-	&	-	&	-	&	-	&	-	&	-	&	-	&	-	&	-	&	-	&	-	&	-	&	-	&	-	\\
			\cite{verma2020mitigation}	&	-	&	-	&	36.2	&	750	&	-	&	5191	&	-	&	-	&	-	&	-	&	-	&	-	&	-	&	-	&	-	&	-	&	-	\\
			\cite{medjek2021multicast}	&	-	&	-	&	5.2	&	1000	&	-	&	-	&	-	&	-	&	-	&	-	&	-	&	-	&	-	&	-	&	-	&	-	&	-	\\
			\cite{boudouaia2021rpl}	&	81	&	-	&	2.69	&	-	&	-	&	-	&	-	&	-	&	-	&	-	&	-	&	-	&	-	&	-	&	-	&	-	&	-	\\
			\cite{abhinaya2021secure}	&	96	&	-	&	-	&	400	&	-	&	-	&	-	&	-	&	-	&	-	&	-	&	-	&	-	&	-	&	-	&	-	&	-	\\
			\cite{sahay2021novel}	&	80	&	-	&	-	&	520	&	-	&	-	&	-	&	-	&	-	&	-	&	-	&	-	&	-	&	-	&	-	&	-	&	-	\\
			\cite{baghani2021dao}	&	90	&	-	&	-	&	-	&	-	&	-	&	-	&	-	&	-	&	-	&	-	&	-	&	-	&	-	&	-	&	-	&	-	\\
			\cite{kiran2022ids}	&	96	&	-	&	-	&	-	&	-	&	-	&	-	&	95	&	-	&	-	&	-	&	-	&	-	&	-	&	-	&	-	&	-	\\
			\cite{bang2022embof}	&	-	&	-	&	1.14	&	-	&	-	&	55007	&	-	&	-	&	-	&	-	&	-	&	-	&	-	&	-	&	-	&	-	&	-	\\
			\cite{alsukayti2022lightweight}	&	97.98	&	-	&	1.66	&	950	&	-	&	-	&	512	&	-	&	-	&	99.32	&	1.48	&	-	&	99	&	-	&	-	&	-	&	-	\\
			\cite{nandhini2022lightweight}	&	94.35	&	-	&	-	&	1451	&	-	&	-	&	-	&	-	&	-	&	-	&	-	&	-	&	98.1	&	-	&	-	&	-	&	-	\\
			\cite{nandhini2022enhanced}	&	95.53	&	-	&	-	&	1023	&	-	&	-	&	-	&	-	&	-	&	-	&	-	&	-	&	97.23	&	-	&	-	&	-	&	-	\\
			\cite{goel2023cra}	&	100	&	-	&	4.76	&	2264	&	-	&	59603	&	-	&	-	&	-	&	-	&	-	&	-	&	-	&	-	&	-	&	-	&	-	\\
			\cite{raza2013svelte}	&	100	&	-	&	-	&	-	&	-	&	-	&	-	&	100	&	-	&	-	&	-	&	-	&	-	&	-	&	-	&	-	&	-	\\
			\cite{surendar2016indres}	&	98	&	-	&	-	&	-	&	-	&	-	&	4800	&	-	&	-	&	-	&	-	&	-	&	-	&	-	&	-	&	-	&	-	\\
			\cite{bostani2017hybrid}	&	96.02	&	-	&	-	&	-	&	-	&	-	&	-	&	76.19	&	5.92	&	-	&	-	&	-	&	-	&	-	&	-	&	-	&	-	\\
			\cite{arshad2018colide}	&	-	&	-	&	1.98	&	-	&	-	&	-	&	-	&	-	&	-	&	-	&	-	&	-	&	-	&	-	&	-	&	-	&	-	\\
			\cite{mirshahjafari2019sinkhole+}	&	100	&	-	&	6.25	&	-	&	-	&	-	&	-	&	-	&	-	&	-	&	-	&	-	&	-	&	-	&	-	&	-	&	-	\\
			\cite{shin2019detection}	&	-	&	-	&	-	&	-	&	-	&	-	&	-	&	95	&	3	&	-	&	-	&	94	&	-	&	-	&	-	&	-	&	-	\\
			\cite{althubaity2020specification}	&	-	&	-	&	-	&	912	&	-	&	-	&	-	&	-	&	-	&	-	&	-	&	100	&	-	&	-	&	-	&	-	&	-	\\
			\cite{verma2020cosec}	&	88	&	-	&	-	&	-	&	-	&	-	&	-	&	-	&	-	&	-	&	-	&	-	&	-	&	-	&	-	&	-	&	-	\\
			\cite{gothawal2020anomaly}	&	-	&	8.89	&	-	&	-	&	-	&	-	&	2508	&	-	&	-	&	-	&	-	&	-	&	90	&	-	&	-	&	-	&	-	\\
			\cite{violettas2021softwarized}	&	-	&	-	&	-	&	-	&	-	&	-	&	-	&	-	&	-	&	-	&	-	&	-	&	100	&	-	&	-	&	-	&	-	\\
			\cite{agiollo2021detonar}	&	-	&	-	&	-	&	-	&	-	&	-	&	-	&	-	&	-	&	-	&	-	&	80	&	100	&	-	&	-	&	-	&	-	\\
			\cite{ray2023novel}	&	98.76	&	-	&	1.36	&	-	&	-	&	-	&	6540	&	-	&	-	&	-	&	-	&	-	&	99	&	-	&	-	&	-	&	-	\\
			\cite{deveci2023evolving}	&	-	&	-	&	-	&	-	&	-	&	-	&	-	&	-	&	-	&	-	&	-	&	-	&	92.2	&	-	&	-	&	-	&	-	\\
			\cite{bhale2024hybrid}	&	98.2	&	-	&	2.78	&	-	&	-	&	-	&	-	&	-	&	-	&	-	&	-	&	-	&	95.49	&	94.93	&	93.49	&	93.34	&	-	\\
			\cite{cervantes2015detection}	&	-	&	-	&	-	&	-	&	-	&	-	&	-	&	-	&	-	&	-	&	-	&	92	&	-	&	-	&	-	&	-	&	-	\\
			\cite{airehrour2016securing}	&	-	&	40	&	-	&	-	&	-	&	-	&	-	&	-	&	-	&	-	&	-	&	-	&	-	&	-	&	-	&	-	&	-	\\
			\cite{airehrour2019sectrust}	&	-	&	28	&	-	&	-	&	-	&	-	&	-	&	-	&	-	&	-	&	-	&	-	&	-	&	-	&	-	&	-	&	-	\\
			\cite{thulasiraman2019lightweight}	&	100	&	-	&	-	&	-	&	-	&	-	&	-	&	-	&	-	&	-	&	-	&	-	&	-	&	-	&	-	&	-	&	-	\\
			\cite{ul2021ctrust}	&	-	&	0.38	&	-	&	-	&	-	&	-	&	-	&	-	&	-	&	-	&	-	&	-	&	-	&	-	&	-	&	-	&	-	\\
			\cite{pishdar2021pcc}	&	-	&	-	&	0.99	&	-	&	-	&	-	&	-	&	100	&	-	&	-	&	-	&	-	&	-	&	-	&	-	&	-	&	-	\\
			\cite{bang2021novel}	&	100	&	-	&	1.3	&	4000	&	-	&	-	&	-	&	-	&	-	&	-	&	-	&	-	&	100	&	-	&	-	&	-	&	-	\\
			\cite{patel2021trust}	&	-	&	28	&	1.2	&	-	&	-	&	-	&	3400	&	-	&	-	&	-	&	-	&	-	&	-	&	-	&	-	&	-	&	-	\\
			\cite{prathapchandran2021trust}	&	94	&	-	&	-	&	-	&	-	&	-	&	2900	&	-	&	1.4	&	-	&	1.8	&	-	&	85	&	-	&	-	&	-	&	3.2	\\
			\cite{patel2022reputation}	&	-	&	20	&	-	&	-	&	-	&	-	&	3800	&	-	&	-	&	-	&	-	&	-	&	-	&	-	&	-	&	-	&	-	\\
			\cite{kim2022physical}	&	90	&	-	&	-	&	-	&	-	&	-	&	-	&	-	&	-	&	-	&	-	&	-	&	-	&	-	&	-	&	-	&	-	\\
			\cite{yavuz2018deep}	&	-	&	-	&	-	&	-	&	-	&	-	&	-	&	-	&	-	&	-	&	-	&	-	&	96.3	&	95	&	96.7	&	94.3	&	-	\\
			\cite{foley2020employing}	&	-	&	-	&	-	&	-	&	-	&	-	&	-	&	-	&	-	&	-	&	-	&	-	&	87.08	&	-	&	-	&	-	&	-	\\
			\cite{morales2021dense}	&	-	&	-	&	-	&	-	&	-	&	-	&	-	&	-	&	-	&	-	&	-	&	-	&	99.65	&	-	&	-	&	99.65	&	-	\\
			\cite{osman2021ml}	&	-	&	-	&	-	&	-	&	-	&	-	&	-	&	-	&	-	&	-	&	-	&	-	&	99.6	&	99	&	99.6	&	99.3	&	-	\\
			\cite{shirafkan2021autonomous}	&	-	&	-	&	-	&	-	&	-	&	-	&	-	&	-	&	-	&	-	&	-	&	-	&	99.5	&	100	&	100	&	99.2	&	-	\\
			\cite{osman2021artificial}	&	-	&	-	&	-	&	-	&	-	&	-	&	-	&	-	&	-	&	-	&	-	&	-	&	97.14	&	97.03	&	97.01	&	97	&	-	\\
			\cite{sharma2021aiemla}	&	-	&	-	&	-	&	-	&	-	&	-	&	-	&	-	&	-	&	-	&	-	&	-	&	100	&	100	&	100	&	100	&	-	\\
			\cite{sahay2022holistic}	&	-	&	-	&	-	&	-	&	-	&	-	&	-	&	-	&	-	&	-	&	-	&	-	&	91.88	&	99	&	81	&	85	&	-	\\
			\cite{nayak2021deep}	&	-	&	-	&	-	&	-	&	-	&	-	&	-	&	-	&	-	&	-	&	-	&	-	&	91	&	93	&	92	&	92	&	-	\\
			\cite{mehbodniya2021machine}	&	-	&	-	&	-	&	-	&	-	&	-	&	-	&	-	&	-	&	-	&	-	&	-	&	92.14	&	-	&	-	&	-	&	-	\\
			\cite{prakash2022optimized}	&	-	&	-	&	-	&	-	&	-	&	-	&	-	&	-	&	-	&	-	&	-	&	-	&	96.4	&	95.26	&	-	&	96.45	&	-	\\
			\cite{albishari2022deep}	&	-	&	-	&	-	&	-	&	-	&	-	&	-	&	-	&	-	&	-	&	-	&	-	&	98.85	&	97.5	&	98.33	&	97.01	&	-	\\
			\cite{alghamdi2023cascaded}	&	-	&	-	&	-	&	-	&	-	&	-	&	-	&	-	&	-	&	-	&	-	&	-	&	96	&	-	&	100	&	97	&	-	\\
			\cite{al2023hybrid}	&	-	&	-	&	-	&	-	&	-	&	-	&	-	&	-	&	-	&	-	&	-	&	-	&	95	&	95	&	82	&	87	&	-	\\
			\cite{osman2024ensemble}	&	-	&	-	&	-	&	-	&	-	&	-	&	-	&	-	&	-	&	-	&	-	&	-	&	99.03	&	99.5	&	99.5	&	99.5	&	-	\\
			\cite{budania2024oead}	&	-	&	-	&	-	&	-	&	-	&	-	&	-	&	-	&	-	&	-	&	-	&	-	&	97.67	&	98.55	&	96.67	&	97.93	&	-	\\
			\cite{mayzaud2015mitigation}	&	99.9	&	-	&	-	&	790	&	-	&	54370	&	-	&	-	&	-	&	-	&	-	&	-	&	-	&	-	&	-	&	-	&	-	\\
			\cite{ahmed2016mitigation}	&	99	&	-	&	-	&	-	&	-	&	-	&	-	&	98.7	&	2.2	&	-	&	-	&	-	&	-	&	-	&	-	&	-	&	1.55	\\
			\cite{pu2019energy}	&	89	&	-	&	-	&	-	&	-	&	-	&	-	&	-	&	-	&	-	&	-	&	-	&	-	&	-	&	-	&	-	&	-	\\
			\cite{patel2019blackhole}	&	92	&	-	&	-	&	-	&	-	&	-	&	-	&	-	&	-	&	-	&	-	&	-	&	-	&	-	&	-	&	-	&	-	\\
			\cite{ioulianou2019denial}	&	-	&	-	&	-	&	-	&	-	&	-	&	-	&	100	&	7	&	-	&	-	&	-	&	-	&	-	&	-	&	-	&	-	\\
			\cite{seth2020detection}	&	-	&	-	&	-	&	-	&	-	&	-	&	-	&	-	&	-	&	-	&	-	&	-	&	94.5	&	-	&	-	&	-	&	-	\\
			\cite{a2020detection}	&	98.48	&	-	&	-	&	991	&	-	&	-	&	-	&	-	&	-	&	-	&	-	&	-	&	98.3	&	-	&	-	&	-	&	-	\\
			\cite{sheibani2022lightweight}	&	93	&	-	&	1.25	&	-	&	-	&	-	&	-	&	99	&	2	&	-	&	-	&	-	&	99.96	&	97.5	&	-	&	-	&	-	\\
			\cite{sharma2023qsec}	&	-	&	-	&	-	&	-	&	-	&	-	&	-	&	-	&	-	&	-	&	-	&	-	&	97.5	&	-	&	-	&	-	&	-	\\
			\cite{conti2018split}	&	99	&	-	&	0.09	&	-	&	-	&	-	&	-	&	-	&	-	&	-	&	-	&	-	&	-	&	-	&	-	&	-	&	-	\\
			\cite{luangoudom2020svblock}	&	80	&	-	&	1.1	&	-	&	-	&	-	&	-	&	98.5	&	3.7	&	-	&	-	&	-	&	-	&	-	&	-	&	-	&	-	\\
			\cite{karmakar2021leader}	&	-	&	-	&	-	&	-	&	-	&	-	&	-	&	-	&	17	&	-	&	-	&	-	&	100	&	-	&	-	&	-	&	1.8	\\
			\cite{pu2022lightweight}	&	-	&	-	&	-	&	-	&	-	&	-	&	-	&	-	&	-	&	-	&	-	&	96	&	-	&	-	&	-	&	-	&	3.1	\\
			\bottomrule
		\end{tabular}
	\end{table}
	\clearpage
	\subsection*{\textcolor{black}{Real World Applicability of Defense Mechanisms}}
	\textcolor{black}{The below discussion explains the RPL-based defense solution in real world scenarios.
		\begin{itemize}
			\item Smart Home: Machine learning-based IDS modules can be integrated to detect anomalies in the gateway devices. It can monitor any unexpected patterns in network traffic such as changes in DODAG Rank or excessive DIS messages. It can isolate any suspicious device from the home network and alert the owner. Trust-based mechanisms can ensure that malicious or malfunctioning devices cannot disrupt communication, maintaining the reliability of systems like security cameras and alarms. IDS solutions leverage early detection of threats like sinkhole or wormhole attacks, minimizing the impact on critical systems such as fire alarms and intrusion sensors. Similarly, threshold-based solutions can identify resource exhaustion attacks, ensuring devices like smart thermostats and speakers remain operational. 
			\item Industrial IoT: Cryptographic defense mechanisms can help in the early detection of threats like sinkhole or selective forwarding attacks and allows timely response, ensuring minimal downtime in industrial processes. Similarly, machine learning-based solutions adapt to evolving threats, offering dynamic protection for Industrial IoT systems in rapidly changing industrial environments. The location-based data can be used to implement geographical validation mechanisms in edge gateways. A potential malicious node can be identified by verifying the claimed positions of the nodes with its location. Trust-based mechanisms can prevent malicious nodes from participating in the routing process, ensuring consistent data flow and protecting critical operations like machine-to-machine communication. 
			\item Advanced Metering Infrastructure (AMI): AMI systems are vulnerable to various attacks, including sinkhole, selective forwarding, wormhole, dropped DAO, and routing table falsification attacks. Trust-based systems prevent compromised nodes from participating in routing, ensuring reliable and secure data delivery between smart meters and the control center. Likewise, threshold-based defense mechanisms can protect against resource exhaustion attacks, ensuring nodes remain functional and the AMI network maintains stability RPL trust-based, and threshold-based defense mechanisms can significantly enhance the security and reliability of AMI by mitigating these threats and help reduce downtime and maintenance costs. 
		\end{itemize}
	}
\section{Evaluation Tools \& Datasets}
\label{sec_evalution_dataset}
This section summarizes the evaluation or validation tools (such as simulators and testbeds) and datasets used for RPL-based security study.
\subsection{RPL Attack Datasets}
Different datasets have been created from RPL-based attacks and are available publicly or on a request basis. Table \ref{tbldataset} lists various datasets developed by researchers globally. In Table \ref{tbldataset},  various parameters to evaluate the datasets are included like "Dataset Name" and "Type" (Synthetic, i.e., those are not publicly available, whereas "Standard" which are available for research purposes), "Attacks Covered" tells the RPL attack samples used in the dataset, "Records" represent totals number of samples in a dataset (benign and malicious), "Platform" tells simulator used to develop the dataset. \\
\textcolor{black}
{
	\indent IRAD \cite{yavuz2018deep} dataset contains $64\times10^{6}$ instances of decrease rank, version number, and hello flood attack. This dataset has been developed  mainly for the evaluation of deep learning based intrusion detection. Assessment of the IRAD on variable number of nodes (10, 20, 100, 1000) shows that 10\% or 20\% of the nodes are malicious, while the rest are benign. Developers of IRAD also proposed a deep learning model for attack detection that provides high precision, good area under the curve (AUC) value, F1 score, recall for decrease rank, version number, and hello flood attacks. \newline
	\indent RPL-NIDDS17 \cite{verma2019elnids} dataset contains traces of seven different types of RPL-based routing attacks. RPL-NIDDS17 dataset classified the 20 features into basic, normal and timed categories with additional 2 label attributes for indicating attack and normal type of traffic. The dataset is evaluated using Boosted Trees, Bagged Trees, Subspace Discriminant and RUS Bootsed Trees ensemble classifier.  It is shown the the machine learning based classifiers perform well in terms of AUC and Accuracy. Boosted Trees achieved highest accuracy on RPL-NIDDS17.\\
	\indent IoT-RPL \cite{aydogan2019central} dataset is created by implementing two well known attacks, i.e., Hello Flood and Version Number Attack on  IIoT scenario. This dataset evolved from 26 different IIoT TelosB nodes running 20 minutes simulation time, and approximately 10\% nodes are selected as malicious nodes. Genetic programming-based Intrusion detection mechanism is  developed for addressing both attacks and their performance is measured in terms of accuracy, true positives and false positives. \\
	\indent RPL Attack Dataset \cite{medjek2021fault} is a publicly available dataset that contains multi class dataset containing traces of multiple attacks. It contains traces of six attacks. For machine learning algorithms, the dataset provides two class distribution, i.e., attack and non-attack. Three different types of topologies containing 25, 50 and 100 Tmote Sky motes with 2, 4 and 10 malicious nodes. The dataset is developed from one hour simulation in Cooja, and the traffic is captured by Radio logger plugin in the form of PCAP format. Finally, min-max scaling is used to scale the dataset in interval of 0 to 1, and then seven class dataset is generated.\\
	\indent Routing Attack Dataset for RPL-based  IoT \cite{canbalaban2020cross} dataset is generated from simulation of 50 nodes for one hour with variable attacker density (2, 6, 10, 20). It contains traffic Version Number, Worst Parent and Hello Flood attack. Authors also have developed a neural network based cross layer intrusion detection system for RPL based attack and a testbed on the dataset. The proposed model achieved the highest detection rate of 97.52\% \\
	\indent IoT-DDoS \cite{al2020real} dataset is generated from three different DoS attacks including Selective Forwarding, Blackhole and Flooding attack from the real-time dataset generation framework. 30 different types of Z1 motes were placed on network to produce the network data of attacks for 96 hours. The sniffer is used to aggregate the traffic and send it to the queue system which sends the traffic for feature selection on a defined time window. The feature selection unit selects the feature of data link, network and application Layers one by one. After data labeling the dataset is generated with 16 different features.\\
	\indent IOT Dataset\cite{foley2020employing} is a novel dataset generated from the power and network metrics of IoT attacks against the popular objective functions OF0 and MRHOF of RPL protocol. RF and MLP machine learning algorithm are evaluated on this dataset and showed the effective results.\\
	\indent RADAR\cite{agiollo2021detonar} known as Routing Attack Dataset for RPL contains 14 different RPL attacks and generated from network of 16 static IoT nodes. RADAR is produced from NetSim Simulator. 80 different simulation scenarios each of 1500 seconds simulation time were executed to extract traffic of 14 different attacks in the form of PACP files. A novel attack detection mechanism named DETONAR is proposed for detecting of routing attacks. DETONAR is based on packet sniffing approach which gives more than 80\% detection accuracy for 10 attacks. \\
	\begin{table}[h!]
		\caption{Dataset proposed for RPL-based Routing Attacks}
		\label{tbldataset}
		\scriptsize
		\renewcommand{\arraystretch}{1.5}
		\resizebox{\linewidth}{!}{
			\begin{tabular}{p{.6cm} c p{3cm} p{1cm} p{5.2cm} p{1cm} p{1.2cm} c c p{.8cm} }
				\toprule
				\textbf{Ref} & \textbf{Year} & \textbf{Dataset} & \textbf{Type} & \textbf{Attack Covered} & \textbf{Records} & \textbf{Attributes}&  \multicolumn{2}{c}{\textbf{Samples}} & \textbf{Platform} \\ \cmidrule(rl){8-9}
				& & \textbf{Name}& & & &  &\textbf{Benign} & \textbf{Malicious} &  \\\midrule
				\cite{yavuz2018deep} & 2018 & IRAD & Standard & Version Number, Decrease Rank and Hello Flood & 9550795 & 18(17F+1L) & 3395601 &	6155194 & Cooja \\
				\cite{verma2019elnids} & 2019 &	RPL-NIDDS17 & 	Standard &	Clone ID, Local repair, Hello flooding, Selective forwarding, Sinkhole, Blackhole, Sybil &	465318 & 22(20F+2L) & 431981	& 33337	& NetSim \\
				\cite{aydogan2019central} &	2019 &	IoT-RPL & Synthetic	& Hello Flood Attack and Version Number &	- &	50 & - & - & Cooja \\
				\cite{medjek2021fault} & 2019 &	2-Class / Multi-class & Synthetic & Decrease Rank, Sinkhole, Version Number, Hello Flood, Blackhole, Selective Forwarding, and Multiclass &	106192798  & 19(17F+2L) & - & - & Cooja	\\
				\cite{canbalaban2020cross} & 2020 & Routing Attack Dataset for RPL-based IoT	& Standard & Version Number, Worst Parent, Hello Flood &	-  & 27 & - & - & Cooja \\
				\cite{al2020real} &	2020 &	IoT-DDoS &	Synthetic &	Selective Forwarding, Blackhole and Flooding &	4195537 & 16  & - & - &	Cooja \\
				\cite{foley2020employing} &	2020 &	IoT Dataset	& Synthetic  & Rank and Blackhole, Rank and Version, Decreased Path Metric, and Rank and Sybil & - & 24(23F+1L) & - &	- &	Cooja \\
				\cite{agiollo2021detonar} &	2021 &	RADAR & Standard &	Blackhole, Sinkhole, Selective Forwarding, Continuous Sinkhole, Sybil, Clone ID, Version Number, Wormhole, Replay, Rank, Worst Parent Selection, DIS Flooding and Local Repair & - &	15\* &	- &	- &	NetSim \\
				\cite{said2021efficient} &	2021 &	IDC and EDC	& Synthetic	& Rank, Version Number and Flooding	& - & - &	- &	- &	Cooja \\
				\cite{morales2021dense}	& 2021 & cloneid\_20n, cloneid\_50n, cloneid\_100n & Synthetic & Clone ID & 1492579, 1576668, 1232862 & 20(19F+1L) & - &	- &	Cooja \\
				\cite{osman2021ml} &	2021 &	VNA Dataset	& Synthetic &	Version Number & 1376231 &	17 &	884860 &	491371 & Cooja \\
				\cite{bokkamachine} &	2021 &	RPL Attack Dataset & Synthetic & Sinkhole, Blackhole, Sybil, Selective Forwarding, DIS Flooding, DIO Suppression Attack & 117510	& 23(21F+2L) & 95342 & 22168 & Netsim \\
				\cite{al2023hybrid} & 2023 & IoTR-DS & Synthetic & DIS, Wormhole, Rank & 380772 & 16(15F+1L) & 337508 & 43224 & Cooja \\
				\cite{ROUT-4-2023} & 2023 &	ROUT-4-2023 & Standard & Blackhole, Decreased Rank, Flooding, Version Number & 1639989  & 18(16F+2L) & 579992 & 1059997 & Cooja\\
				\cite{osman2024ensemble} & 2024 & RPL-ELIDS & Synthetic & Version Number, Decrease Rank, DIS Flooding, Multiclass & 1566352  & 18(17F+1L) & 1219284 & 347068 & Cooja\\
				\cite{paganraj2024dair} & 2024 & DA\_IoT\_Routing & Synthetic & Hello Flood, Decrease Rank, Increase Version Number & 9561  & 18(17F+1L) & 4096 & 5465 & Cooja\\
				\textcolor{black}{\cite{omar2024uos_iotsh_2024}} & \textcolor{black}{2024} & \textcolor{black}{UOS\_IOTSH\_2024} & \textcolor{black}{Synthetic} & \textcolor{black}{Sinkhole} & \textcolor{black}{1,771,880}  & \textcolor{black}{14(13F+1L)} & \textcolor{black}{59381} & \textcolor{black}{1712499} & \textcolor{black}{Cooja}\\
				\bottomrule
			\end{tabular}
		}
	\end{table}
	\indent Intrusion Detection Component dataset (IDC), and Event Detection Component dataset (EDC) \cite{said2021efficient} are developed by SERCOM Lab, University of Carthage in the Smart Hospital based IoT System. This dataset was developed with the aim to design anomaly detection system. IDC dataset stores the network states in term of energy consumed by devices such as radio transmission or reception and radio interfered energy. Similarly, EDC dataset stored the environmental data such as light, temperature and humidity. Both the datasets contains the 1000 instance out of 200 instance used for training purposes.\\
	\indent Clone ID dataset \cite{morales2021dense} is an identity based attack dataset developed from the CloneID attacks. This dataset was developed with the aim for implementing an artificial intelligence framework for intrusion detection, and intrusion protection system. The dataset is created from the simulation of variety of node such as 20, 50 and 100 and respective dataset is labeled as cloneid\_20n, cloneid\_50n and  cloneid\_100n by considering the 10\% nodes are malicious and rest are benign nodes. There are two class of attack and normal in the dataset. SAE based encoders with Deep Neural Network for were implemented and tested for attacks detection which achieved 99.65\% accuracy on the cloneid\_50n dataset.\\
	\indent VNA Dataset\cite{osman2021ml} is developed for lightweight Version Number attack detection based on machine learning model which achieved 99.6\% accuracy. The raw dataset is created in the form of JavaScript Object Notation (JSON) format from PCAP using the Wireshark tool. Subsequently, python script is used to extract the features from the JSON, and after processing 17 features are chosen to be part of the dataset. The dataset contains the record from the malicious node scenario of 1, 2 and 3 nodes out of 10 and 20 nodes obtained from simulation 10 minutes.\\
	\indent RPL Attack Dataset \cite{bokkamachine} created to test seven different machine learning algorithms namely K-Nearest Neighbors (KNN), Random Forest (RF), Logistic Regression (LR), Decision Trees (DT), Gaussian Naive Bayes(GNB),  Multilayer Perceptron classifiers (MLP) and AdaBoost (AdB). The dataset has two label, i.e., attack and normal used to classify the traces. The range of 60-90\% for training and 40-10\% for testing the machine learning the algorithms. Decision Tree classifier achieved the highest precision and recall among other machine learning algorithms. \\
	\indent IoTR-DS \cite{al2023hybrid} was proposed for hybrid intrusion detection algorithms which uses the concept of supervised and semi-supervised learning to classify the RPl-based routing attacks. This dataset contain the attack sample of Rank, DIS and Wormhole attack and an Hybrid IDS based approach for detection of those attack with 95\% accuracy and 87\% F1-Score. There are total twelve features which contains the details of data packet which was sent towards root such has control message (count of DIS, DIO, DAO) , UDP messages and overall packet delivery ratio.\\
	\indent ROUT-4-2023  \cite{ROUT-4-2023} is RPL attack dataset consist of four different types of RPL attacks named Flooding Attack, Blackhole Attack, Decreased Rank Attack and DODAG Version Number. When the dataset is analyzed, total 18 features are there with normal sample around 67\% and attacks sample is 33\% which make it imbalanced. The primary focus of this dataset is to build the efficient intrusion detection mechanism for RPL-based IoT network using artificial intelligence techniques.\\	
	\indent RPL-ELIDS \cite{osman2024ensemble} proposed a dataset consisting the attack sample of DIS flooding, Version Number and Decrease Rank and proposed a ensemble learning based IDS which usage a genetic algorithm for feature selection. To develop the dataset a variety of network topology is created with 10, 20 nodes with 1,2,3 as malicious node. Initially the number of features in dataset was 113, after certain iterations for removing the redundant features, they are reduced to 17 with 1 class label. The proposed IDS system achieve 97.90\% accuracy for the developed dataset.\\
	\indent UOS\_IOTSH\_2024 \cite{omar2024uos_iotsh_2024} is a comprehensive dataset developed at University of Sharjah for real time RPL-based network on Cooja simulator focused on Sinkhole attack. This dataset consider the small and medium IoT network scenario where single and dual attacker were deployed at various location of topology to gathered the network traffic. There are total 13 features having information about source, destination and per second count of control message as important feature.  \\
}
\subsection{IoT Testbeds}
IoT testbeds provide a real-time environment for creating, developing, debugging, troubleshooting, and testing the proposed solution. IoT testbed facilitates the deployment and execution of the IoT application over real hardware, which is programmed to monitor the IoT devices and analyze the performance of applications. Table \ref{tbltestbed} shows the widely used testbeds for the Internet of Things to develop RPL-based solutions and compares the testbeds on the various parameter such as type of access, provides the detail about the registration process required to access the testbed, a user interface to operate the testbed, the number of nodes and hardware motes deployed in the testbeds, programming language to create our application in the testbed, operating system supported by the testbeds.\\ \textcolor{black}{ 
	\indent WISEBED \cite{chatzigiannakis2009wisebed} testbed project is supported by European Universities and Research Institutes. It is a large scale wireless sensor network which is deployed at University of Lübeck. The Open Federation Alliance (OFA) defines web services to connect with this federated testbed with the help of overlay network. The WISEBED can be viewed as as global network of intelligent agents and human users connected together with wireless networks. The main aim is to provide a wireless sensor network testbed which provides variety of interfaces, connect several hundred nodes, unified algorithmic and software environment togther with support for mobility.\\
	\indent LOG-a-TEC\cite{LOG-a-TEC-testbed} provides a diverse set of testbeds in various verticals such as air quality monitoring, photovoltaic energy production forecasting, smart motor-home and new vertical in the area of Internet of Things using LR-WPAN and LPWAN technologies such as 6LoWPAN and ultra wide band communication. The hardware platform used is VESNA. VERSNA is developed by Sensor Lab using ARM Cortex-M3 core micro controller support multiple frequency bands. Currently VESNA supports the Contki-NG of version 4.5 and accessed via LCSP protocol and 6LOWPAN using the login credentials. Various experiments that are available include 6LoWPAN statistics, game-theoretical interference mitigation and over-the-air programming and dataset for LPWAN spectrum trace, UWB localization and BLE fingerprints on receive signal strength on a web portal.\\
	\indent FIT IoT-LAB \cite{adjih2015fit} is a large scale IoT testbed deployed at six different cities in France connected with REST API and CLI support. The FIT IoT-LAB testbed has world wide user base of more than 5000 user and across countries and till date more than 200 thousand of experiments were conducted on FIT IoT-LAB. It has a wide range of features like open platform, open software and open tools and capabilities to support multi-radio, multi- platform, multi-OS and multi-topology. In FIT IoT-LAB, a user can choose a specific hardware board with its own operating system like CONTIKI-NG, RIOTS and Free RIOTS to develop firmware that can run on different physical topologies. The architecture has three main components namely Open Node (ON) (a low power device), Control Node (CN) (responsible for operation and choosing power source for Open Node) and Gateway (GW) connecting Open Node to Control Node using serial ports using Linux based computers. Some use cases include visualization of WiFi traffic and its impact assessment.\\
	\indent NetSecIoT \cite{NetsecIoT} is an IoT testbed which supports 6LoWPAN protocol implemented in RIOT Operating System. The main components of this testbed are Display Node (a http client), Gateway Node which scan the CoAP devices and translate http request into CoAP to access the testbed, RIOT IoT Node, a constrained device implementing Netsec IoT Protocol and a Non-Embedded IoT Node for running Raspberry Pi. To setup the testbed, three steps are needed: creation of Raspbian image with 6LoWPAN support, setting up RIOT OS on running on top of Ubuntu OS for compilation, and flashing the developed RIOT applications.}

\begin{table}[!h]
	\centering
	\scriptsize
	\caption{Testbed for RPL-based Internet of Things}
	\renewcommand{\arraystretch}{1.4}
	\resizebox{\linewidth}{!}{
		\begin{tabular}{p{.8cm} p{1.5cm} p{2cm} p{2cm} p{1cm} p{1.5cm} p{3.5cm} p{3cm}}
			\toprule
			\textbf{Ref} 	&	\textbf{Testbed Name} 	&	\textbf{Access Type} 	&	\textbf{User Interface}
			&	\textbf{Number of Node}	&	\textbf{Programming Language} &	\textbf{OS Supported}	&	\textbf{Hardware Nodes}	\\
			\midrule
			\cite{chatzigiannakis2009wisebed}	&	WISEBED	&	Close Access	&	GUI	&	550	&	-	&	Contiki OS and TinyOS	&	Mica2, MicaZ, SunSPOT and TelosB\\
			\cite{LOG-a-TEC-testbed}	&	LOG-a-TEC	&	Partially Open Access	&	Web	&	-	&	-	&	Unix, Contiki, Custom	&	VESNA 3.0 (ARM Cortex-M3) \\
			\cite{adjih2015fit}	&	FIT-IOT LAB	&	Partially Open Access	&	Web, REST and CLI	&	2728 Static and 117 Mobile	&	C Language	&	RIOT, OpenWSN, FreeRTOS, Contiki, TinyOS, Linux	&	M3,A8 and WSN430 \\
			\cite{NetsecIoT}	&	NetSecIoT	&	Open Access	&	CLI	&	-	&	Python	&	Raspberry Pi OS  Image and RIOTS	&	Raspberry Pi 2 \\
			\cite{appavoo2018indriya2}	&	INDRIYA2	&	Partially Open Access	&	GUI	&	140	& C, Perl and PHP	&	TinyOS and Contiki	& SensorTag CC2650/ CC1350 and TelosB \\
			\cite{tutornet}	&	TUTORNET	&	-	&	GUI	&	113	&	-	&	-	&	MicaZ, OpenMote and TelosB  \\
			\cite{cci-iot-testbed}	&	CCI IOT Testbed	&	-	&	GUI	&	-	&	-	&	Raspberry Pi & ARM Cortex A53	\\
			\bottomrule
		\end{tabular}
	}
	\label{tbltestbed}
\end{table}
\textcolor{black}{
	\indent INDRIYA2 \cite{appavoo2018indriya2} is a testbed for wireless sensor network deployed in National University of Singapore. INDRIYA2 provides support for various hardware sensor platforms with high data rate capabilities to handle time series data generated at the time of running application. The architecture of INDRIYA2 comprises  Nodes (such as SensorTag cc2560 and TelosB), Gateways and Servers.  Gateways are built from 6 Mac-minis running the Ubuntu 14 OS while Servers are quad core machines with 16 GB of RAM running Ubuntu version 16 OS. The High Influx DB is used to handle the high rate time stamped data in the form of batches of JSON format. The hardware platform uses the ARM Cortex-M3 and TI MSP430 micro-controller with various sensor including magnetometer, barometer, accelerator, gyroscope and microphone. The receiver signal strength, and packet reception rate parameters are used to measure the link quality during the execution.\\
	\indent TUTORNET \cite{tutornet} testbed is deployed at University of Southern California (USC). The overall layout of the testbed has four types of nodes as OpenMote, MicaZ, TelosB and Concentrator nodes. The TUTORNET testbed has three layer of architecture consisting of Central Server, Concentrator  and Sensor Nodes. The Central Server performs duties of testbed reservation implemented by Apache Server in Perl and MySQL running on Ubuntu Linux. The Concentrator node running Ubuntu Linux deployed on Raspberry-Pi hardware having USB hub for connecting devices. There are various datasets available in the form of JavaScript Object Notation (JSON) of running experiments of packet delivery ratio.\\
	\indent CCI IOT \cite{cci-iot-testbed} is a campus wide testbed for Internet of Things deployed at the Center for Cyber-Physical System and IoT at the University of Southern California. The testbed is currently in the development phase. The overall architecture contains the Raspberry-Pi gateway, WiFi access points, server and user application. The Raspberry-Pi version 3 of Model B with 1.2GHZ quad-core ARM Cortex-A53 processor having 1 GB of RAM and 32 GB of flash memory used in the deployment of testbed. The various sensors like temperature, humidity, light, air quality, sound and vibration are being deployed for experimental purpose. Gateways are backed by solar power, covering one mile radius area in the USC campus. \\
	\indent The above discussion related to the existing testbed that can be used for the RPL-security research. However, all the testbeds are critically evaluated on the basis of scalability (numbers of nodes, topology supported), compatibility (hardware nodes, OS supported, programming languages), accessibility (documentation, access type, user interface, execution, energy profiling), environmental factors (physical tempering, interference, node mobility), heterogeneity (constrained nodes) and real-world scenarios (attacks and simulation). The critical evaluation of tested is as follows:  
	\begin{itemize}
		\item WISEBED	\cite{chatzigiannakis2009wisebed}: It doesn't support extendibility, customize-ability and relocation of nodes, provide remote access of testbed facility but not public accessibility, realistic setting for natural disasters and building monitoring and support limited mobility facility among nodes, heterogeneity which offer variety of hardware nodes, various topologies. 
		\item LOG-a-TEC	\cite{LOG-a-TEC-testbed}: It doesn't have public documentation support or scalability. It provides the facility for remote monitoring and reconfiguration of experiments. It has an outdoor implementation which is affected by environmental factors. It is focused on heterogeneous application areas consisting of M2M/MTC/Dense IoT networks which are based on LR-WPAN and LPWAN.  
		\item FIT-IOT LAB	\cite{adjih2015fit}: It is an open source and open access (free of charge) testbed that provides scalability, dynamic topologies, mobility support, re-programmed and reconfigured nodes, heterogeneous hardware nodes that support a larger set of IoT application. It consists of energy profiling modules to measure the energy consumption of nodes as well as monitor the performance metrics such as throughput and end-to-end delay.
		\item NetSecIoT	\cite{NetsecIoT}: It has limited documentation with open access accessibility of the testbed. It doesn't support hardware heterogeneity as well as scalability. It is compatible with only a single operating system and doesn't have an energy consumption module. It supports the remote access of the testbed via remote which gives flexibility to run the experiments.  
		\item INDRIYA2	\cite{appavoo2018indriya2}: It widely supports the scalability of testbed with numerous heterogeneous sensor and wireless technologies. It supports high data rates with real-time monitoring of currently running experiments. It provides a limited access facility to sensor nodes and doesn't provide the mobility of nodes so limited applicability to real-time solutions.    
		\item TUTORNET	\cite{tutornet}: There is no documentation available to experiment on the testbed so there is no accessibility feature incorporated. It supports limited scalability and compatibility of hardware nodes. It is deployed inside the building in an indoor environment with limited mobility support. There is no concept of a real-world scenario incorporated for this testbed.
		\item CCI IOT Testbed \cite{cci-iot-testbed}: It provides a wide range of scalable sensor nodes and wireless support for running and testing IoT experiments. It includes heterogeneous hardware platforms facilitating edge computing and cloud computing. It incorporates advanced technologies such as Software Defined Networks for IoT research applications. It is deployed in outdoor environments which limits environmental factors such as physical tempering, and inference when experiments are performed.    
	\end{itemize}
	While designing the advance IoT testbeds for running experiments some of the points may be consider such as dynamic testing which includes node mobility, dynamic topology for real time network, energy modules which give the insights on power consumption of nodes in terms of battery life, incorporates real world condition such as physical interference/obstacles or noise and support of advanced emerging technologies such as 5G, SDN. The prebuilt attack scenario may be added for batter understanding of security threat detection.}

\subsection{Simulators}
The simulator creates a virtual version of real-life situations for experimentation or training purposes. There are several widely used network simulators for the evaluation and simulation of RPL-based IoT networks by various researchers. Table~\ref{tblsimulator} provides a resemblance to the state-of-the-art simulators used for the RPL protocol simulation. The table gives information about parameters such as the type of simulator, licensing arrangements, compliance of RPL simulators, the programming language used to develop the simulators, the operating system supported by the simulators, and the user interface provided by simulators.\\
\textcolor{black}
{
	\indent More than 90\% of the researchers used Cooja \cite{osterlind2006cross} tool for simulation of RPL networks. Currently the latest Cooja simulator is available with Contiki-NG \cite{Contiki-NG} OS which is regarded as the next generation operating system for low power devices. Cooja is a Java based simulator which supports hardware motes such as Z1 motes \cite{Zolertiaz1} and Tmote Sky \cite{tmotesky} to facilitate the emulation of various sensors. It supports various functions of RPL protocol such as the operating mode and objective functions (such as OF0 and MRHOF). \\
	\indent The next widely used simulator of RPL protocol is NetSim \cite{netsim}. NetSim simulator is a proprietary software of TETCOS. This simulator supports up to 500 sensor nodes in wired or wireless or mobile for simulation purpose. The simulator supports the objective functions (OF0 and MRHOF). An additional feature of this simulator is that it provides the interface for other software packages such as MATLAB \cite{higham2016matlab} and Wireshark \cite{orebaugh2006wireshark}. The NetSim provides the features such as packet animation, output performance matrices and detailed event and packet trace. However, this simulator has limited RPL support and does not provided some the RPL feature like DODAG repair mechanism, power consumption analysis, storage overhead in terms of RAM and ROM usage.\\
	\indent The Network Simulator version 3 (NS-3) is another simulator used used for simulating RPL networks. The NS-3 RPL module is developed by \cite{bartolozzi2012ns}, as a framework to support the IPv6 based routing over low power and lossy devices. This simulator supports the objective function (OF0 and MRHOF). However, NS-3 has limited support of power consumption analysis, development of new objective function and mobility support. The NS-3 has currently in the development phase for 802.15.4 and 6LoWPAN. \\
	\begin{table}[h!]
		\centering
		\scriptsize
		\caption{Simulators for RPL protocol Simulation}
		\renewcommand{\arraystretch}{1.5}
		\resizebox{\linewidth}{!}{
			\begin{tabular}{l p{1cm} p{2cm} p{1.8cm} c p{1.8cm} p{4cm} p{2cm}}
				\toprule
				\textbf{Ref}	& \textbf{Simulator Name}	&	\textbf{Simulator Type} 	&	\textbf{Licensing}	&	\textbf{RPL Compliance}	&	\textbf{Programming Language}	&	\textbf{OS Platform} 	&	\textbf{User Interface}	\\
				\midrule
				\cite{osterlind2006cross}	&	Cooja	&	Discrete-Event	&	Open Source	&	Fully	&	C Language	&	Linux, Contiki and Contiki-NG, MacOSX	&	GUI and CLI	\\
				\cite{netsim}	&	Netsim	&	Research-Based	&	Licensed	&	Moderate	&	C Language	&	Windows/Linux	&	GUI \\
				\cite{agustin2017ipv6}	&	NS-3	&	Discrete-Event	&	Open Source	&	Moderate	&	C++ and Python	&	Windows (Cygwin) /Linux	&	GUI and CLI	\\
				\cite{hosseini2021implementation}	&	OMNet++	&	Discrete-Event	&	Open Source	&	Moderate	&	C++ and NED	&	Windows (Cygwin) /Linux	&	GUI	\\
				\cite{ko2011contikirpl}	&	TinyRPL	&	-	&	Open Source	&	Moderate	&	C Language	&	Tiny OS	&	-	\\
				\bottomrule
		\end{tabular}}
		\label{tblsimulator}
	\end{table} 
	\indent The OMNeT++ supports the limited functionally for the RPL protocol simulation. Recent work in ~\cite{hosseini2021implementation} provides the implementation of RPL for OMNeT++. This implementation considers the existing INET framework available on OMNeT++ for the modification according to RPL protocol specification available in RFC 6550 \cite{winter2012rpl}. Basically it implements various modes of operation along with the three modules which maintain upward routing, source routing table and parent table. The existing implementation supports more than 200 motes which can be advantages for the simulation of larger network.\\
	\indent Finally, TinyRPL used in the early stage for simulating the RPL based network on TinyOS proposed by Stanford. TinyRPL\cite{ko2011contikirpl,ko2011evaluating} is the implementation of IETF RPL draft version of RPL protocol in TinyOS. The IPv6 interfacing is provided by the Berkeley Low-power IP stack (BLIP) module responsible for neighbour discovery, header compression and IP forwarding operation in TinyRPL. The implementation supports both OF0 and MRHOF objective functions for RPL protocol used in parent selection process. The current version of this simulator supports only the upward routes in strong modes and a single instance ID. \\
	The existing simulator discussed above are used for RPL security research. However, every simulator has its own advantage and dis-advantage and it is critically evaluated on the parameters such as accessibility (documentation, licensing type), scalability (number of nodes can be used in simulation task), Usability (user experience when it is used), Compatibility (Specialized module for analysis of packet monitoring, energy consumption), RPL Readiness (How much RPL support in terms of implementation). The critical analysis of existing simulator as follows:
	\begin{itemize}
		\item Cooja\cite{osterlind2006cross}: The Cooja simulator has detailed official documentation and rich set of tutorial for experimental purpose. It is open source simulation tools which support variety of hardware motes, mobility support for dynamic topology, energy consumption monitoring module. It has simple GUI for performing simulation which gives best user experience. Basically in terms of scalability, the performance is reduced when number of nodes increased in simulation. It provide high RPL support and doesn't have pre-built attacks modules.    
		\item Netsim \cite{netsim}: Netsim provides the limited RPL support for RPL functionality and it required licensing fees for the use which is not cost effective. It provides inbuilt traffic monitoring modules, user friendly GUI and no attack modules for RPL-based attack simulation. It has medium scale scalability for RPL nodes in simulation process.    
		\item NS-3 \cite{agustin2017ipv6}: The NS-3 is open source simulator tools with limited RPL readiness. The customized modules are developed for RPL protocol simulation and it is highly scalable (2000 nodes) in terms of number of nodes in simulation. It is complex tools which consist extensive library, advanced log analysis facility and required expertise to use it. It is also no support for pre-built RPL attack modules and energy modules.  
		\item OMNet++ \cite{hosseini2021implementation}: The open source OMNet++ simulator tool is customized for RPL support consisting limited functionality. It is flexible and support large scale topology but sometimes performance is degraded. It required the manual configuration for implementation attack scenario. The current version is unstable and future extension will have physical layer support and neighbor detection.   
		\item TinyRPL \cite{ko2011contikirpl}: This simulator has focused implementation of RPL protocol with limited features of RPL with small amount of scalability for sensor nodes. There is no proper documentation available for attack scenario implementation. As can be accessed through CLI interface so it is less user friendly and required expertise users. The performance is degraded when network size is increased.
	\end{itemize}
	The future network simulators are designed for RPL-security research should have support of modular attack library, hybrid scale network for smaller and larger network topologies, energy and performance metrics module, and real world customization of emerging technologies.  
}
\section{Research Challenges} \label{sec_research_challenges}
Many defense solutions have been developed in response to the range of RPL-based routing attacks described in this survey. However, because of RPL's resource constraints, the developed solutions should not hinder the protocol's performance. We will now examine possible future research directions that can be explored in designing a novel defense mechanism against RPL-based routing attacks.

\subsection{Unaddressed Security Features of RPL}
The RPL standard has three security features: unsecured, preinstalled, authenticated, and treated as optional, as stated in RFC 6550 \cite{winter2012rpl}. However, as far as we know, none of the in-built security features of RPL has been implemented in full in any research work with real-time applications. Therefore,  the RPL security feature implementation is essential for future research while minimizing performance degradation.
\subsection{Security in RPL-based Mobile Networks}
In this survey, we have reviewed numerous defense mechanisms in terms of mitigation schemes and intrusion detection systems. In their evaluation, very few proposed mechanisms consider the mobility of nodes. Mobile nodes introduce additional challenges due to link disconnection, lower packet delivery ratio, and collision among nodes, making detecting attacks more difficult.     
\subsection{Approaches related to Intrusion Detection}
We have observed the following aspects regarding intrusion detection for RPL-based routing attacks. 
\begin{itemize}
	\item Extensible IDS: The IDS-based solutions proposed in the literature have only been tested on small-scale networks. As a result, there is an opportunity to investigate the scalability of IDS by evaluating large-scale networks with heterogeneous IoT devices.
	\item Collaborative IDS: The majority of IDS solutions that have been developed are based on hybrid mechanisms for node monitoring with a single border router and nodes to detect routing attacks. However, rapid detection of attacker nodes requires collaboration between the border router and the other nodes. As a result, there is a need to develop secure collaborative IDS \cite{arshad2018colide} solutions for routing attacks. 
\end{itemize}

\subsection{Security Solutions based on Cutting-Edge Hardware}
Most defense solutions have been evaluated on the Cooja simulator with ContikiOS ~\cite{kim2017challenging}. However, ContikiOS uses only the TelosB\cite{telosB} and Zolertia Z1\cite{Zolertiaz1} hardware platforms for simulation purposes. Furthermore, TelosB has become obsolete, and the manufacturer has discontinued Zolertia. Due to their high resource consumption, these platforms are unsuitable for developing efficient security solutions. Hence, there is a compelling need for innovative security measures that can support recent advances in the hardware, such as Cortex-M3\cite{Cortex-M3} and Raspberry-Pi\cite{Raspberry-Pi-4} having the excellent capability of processing speed and higher memory storage.  
\subsection{Emerging Technology-Based Security Solutions}
Numerous new technologies, including machine learning, software-defined networks, and Blockchain, are increasingly used in developing security solutions. Only a few such solutions deploy these technologies against RPL-based routing attacks, which are discussed below.
\begin{itemize}
	\item Machine Learning: A machine learning model poisons datasets to launch an instant AI attack. Adversarial machine learning techniques enable diverse AI attacks \cite{he2023adversarial}. AI attacks primarily involve input attacks (adding patterns) and data poisoning/false injection (altering datasets). Future research will create a GAN-based self-supervised defence against AI attacks \cite{wu2023self}. Developing IoT security solutions requires research on Explainable AI \cite{kalutharage2023explainable}. Explainable AI addresses datasets, machine learning models, and algorithm trustworthiness. 
	\item Software Defined Network (SDN): Violettas et al. \cite{violettas2021softwarized} proposed an SDN-based IDS for routing attacks named as ASSET. The ASSET has an novel attacker identification and mitigation procedures against the RPL-based and anomaly based attacks. 
	\item Blockchain: Sahay et al. \cite{alsirhani2022securing} \cite{sahay2022holistic},  designed a framework based on Blockchain and Deep learning for detection of RPL attacks. The main idea was to design a smart contract which stores the features and use of deep learning to detect the attack. 
	\item 6G-based RPL network: The RPL protocol can be adopted in 6G-based network to enable communication among the low power devices \cite{alotaibi2023securing}. 
	The 6G network requires high data rate, reliability, and low latency, so a new technological solution is needed to seamlessly integrate RPL in 6G-based wireless communication. In low-power devices, massive data transfer may leak data and consume a lot of energy.Research on RPL-based 6G communication may focus on secure data exchange and energy harvesting. 
	\item Metaverse: Metaverse built the 3-D virtual world where a user can experience the power of the internet in the physical world using digital objects, i.e., avatars \cite{li2022internet}. Numerous IoT devices and sensors are used to create digital avatars to enable the virtual experience in front of users. As RPL is the most popular routing protocol to connect IoT devices and sensors, it has a more significant role in the Metaverse applications. 
	The attacker can create fake digital avatars using Sybil attacks and disrupt communication. As a future aspect, RPL attacks and defense strategies may be explored in the context of Metaverse application.
	\item Smart Factories: The Industry 4.0 revolution introduces the concept of smart factories, where real-time monitoring, data storage in the cloud, machine utilization, and process efficiency can be achieved through the use of smart devices \cite{tange2020systematic}. 
	The RPL protocol is suitable for communicating among the sensor nodes. The vulnerabilities against RPL can significantly disrupt the factory operation and cause downtime and unexpected equipment behavior. 
	The substantial aspect of RPL protocol in smart factories scenario will be explored as a future challenge to protect against RPL-based vulnerabilities for reducing the risk of failure.
\end{itemize}

\subsection{New Dimensions in RPL Routing Attacks}
Here we briefly outline some additional dimensions of routing attacks and their defense mechanisms that are worth further exploration.
\begin{itemize}
	\item Unaddressed Attacks: There is a compelling need for innovative security measures that can handle  additional attacks such as Induced Blackhole \cite{chen2018analysis}, DIO Flooding, Hatchetman \cite{pu2018Hatchetman}, DIO Suppression \cite{perazzo2017dio}, DAO Insider \cite{ghaleb2018addressing} and Routing Table Falsification \cite{mayzaud2016taxonomy}.
	\item Coordinated Attacks: The limited number of coordinated attacks \cite{essaadi2021detection} presented in literature. This type of attacks where a network of bad actors collaborates to carry out a single attack. This type of attacks are more destructive in the nature. Hence, the effect of such attack and defense solution can be explored in future.
	\item Hybrid Attacks: The copycat \cite{verma2020addressing} and sink-clone \cite{mirshahjafari2019sinkhole+} are the only two hybrid attacks that have been considered so far. There is an opportunity to explore new possibilities of hybrid attacks by combining two or more attacks and their mitigation.
	\item Cross-layer Attacks: Cross-layer attacks attack multiple layers of a protocol's stack, each of which might severely influence performance. Researchers have not paid much attention to such cross-layer attacks, despite their increasing prominence in the RPL domain. Asati \textit{et al.} \cite{asati2018rmdd} is the only work addressing a cross-layer attack involving rank manipulation and drop delay for RPL.
\end{itemize} 

\subsection*{\textcolor{black}{Prominent Future Research Directions in the Context of RPL}}
\textcolor{black}{This section introduces a prioritized list of possible research directions in the context of RPL. The directions consider the latest challenges in RPL that push further on what could be achieved with low-power and loss-prone networks efficiently, reliably, and adaptively, thereby being useful for almost any IoT and sensor-based applications.}
	\begin{itemize}
		\color{black}
		\item \textbf{Energy Efficiency and Resource Optimization}: Energy efficiency is highly important in RPL for IoT and sensor networks because they have limited sources of power and are battery-run. Energy-aware routing metrics help in reducing power consumption and increases the performance of the network. Decisions on the RPL routing should pay attention to such factors as node battery levels, density of the network, and traffic patterns. Design of an efficient radio duty cycling mechanism allows nodes to sleep during idle periods; this will, in turn, reduce energy use, whereas optimizing Trickle timers will balance energy efficiency with responsiveness in resource-constrained networks.
		\item \textbf{Security Enhancements for RPL}: RPL security should guarantee data integrity and confidentiality for IoT networks. Low-power devices are vulnerable to malicious attacks like DODAG inconsistency and rank spoofing. Research must be done in lightweight encryption and trust-based routing to detect the malicious activity. Mechanisms need to be developed that prevent rank and control message spoofing, which would ensure the reliability of RPL in hostile environments.
		\item \textbf{Mobility Support in RPL}: As mobility is gaining more popularity in IoT applications, RPL must be able to adapt and manage mobile nodes such as drones and robots. Research areas include mobility-aware routing metrics, stable routes, and fast DODAG reconfiguration. In RPL, efficient handoff mechanisms will ensure seamless connectivity for applications like autonomous vehicles, mobile healthcare, and disaster response.
		\item \textbf{RPL in 6G and Next-Gen IoT Networks}: Ultra-low-latency, massive device connectivity, and high data rates will need to be supported by RPL in 6G networks. The other integration of 6G network slices will enable the management of a wide variety of IoT services with diverse performance and power requirements. Hybrid RPL designs can further enhance data forwarding efficiency in dynamic environments supported with edge computing. 
		\item \textbf{Context-Aware and Adaptive RPL}: Context-aware routing enables the decision-making of RPL through actual factors such as the real-time energy of nodes, congestion of traffic, and outer conditions. Using machine learning, RPL could alter the routing according to network conditions, ensuring optimally high performance. Some enhancements may be made regarding DODAG adjustments or through adaptive metrics based on network load and resource availability.
		\item \textbf{Integration of Renewable Energy Sources}: The lifetime of energy-constrained IoT devices may be increased through the integration of renewable energy sources, including solar or wind energy. RPL can be tailored to favor nodes that are energy harvesting enabled and optimize routing based on renewable energy usage. Predictive algorithms may predict the availability of energy, thereby allowing for proactive routing adjustments in order to improve network sustainability.
		\item \textbf{RPL in Time-Sensitive Applications}: Time-sensitive applications like industrial automation and healthcare require real-time communication. RPL must be adapted to meet strict latency requirements, with enhancements like deterministic routing to guaranteed latency. Multi-channel communication can reduce congestion and improve performance for time-critical applications.
		\item \textbf{Data Aggregation and Compression in RPL}: Data aggregation and compression reduce the overhead and energy consumption in the RPL networks, mostly when nodes are reporting the same data. Aggregation-aware routing combines data from various nodes before transmission to reduce messages and conserve bandwidth and energy. The inclusion of data compression will further enhance network efficiency and extend the lifetime of the network.

	\end{itemize}
	
	\section{Conclusion} \label{sec_conclusion}
	\textcolor{black}{
		This survey gives the comprehensive groundwork for improving the RPL-based security solutions in IoT networks by novel attack and defense taxonomy, in-depth discussion of RPL-based routing attacks and defense mechanisms, impact of routing attacks in real life situations, evaluations of existing attack datasets, testbeds, simulators, potential research challenges and directions. The main idea for classifying attacks on the basis of nature means how it exploits the RPL and defense mechanisms and focuses on type of strategy.  As per the observation, the version number attack is the most destructive one because of the high impact on energy consumption, routing decisions whereas increasing rank is least because of low impact of packet delivery ratio. Similarly, most of the defense solutions used RPL specification to overcome the effect of routing attacks and least used the cryptographic based solution due to high overhead and memory constraints. The top three performance metrics are packet delivery ratio, energy consumption and detection accuracy. It is also observed that, the most prominent dataset IRAD, RADAR and RPL-NIDDS17 were used to evaluate their ML/DL based solutions. Two most significant testbeds that can be used for RPL study are CICIOT and FIT-IOT LAB as widely available and rich functionalities. Consequently, the COOJA simulator is mostly used by the authors for evaluations of defense mechanisms as it is open source and provides fully RPL compliance. The major focused attacks are new dimensions on attacks such as hybrid attacks, cross-layer attacks, some of attacks have unaddressed or limited solutions. The artificial intelligence based secure solutions for RPL is recommended for real-time threat modeling in the future.  Further,  research challenges and future research direction will provide a substantial foundation for researchers to design and develop more effective defense solutions for emerging RPL routing in IoT networks. Our survey is limiting the practical implementation of RPL-based routing attacks and defense which could be addressed as a future work.}

	\printcredits

	
	
	
	
\end{document}